\documentclass[utf8,bibauthoryear]{aa}  
\usepackage{silence}
\usepackage{graphicx}
\usepackage{txfonts}
\usepackage{siunitx}
\usepackage[utf8]{inputenc}
\usepackage{natbib}
\usepackage{caption}
\usepackage{subcaption}
\usepackage{amsmath}
\usepackage{grffile}
\usepackage{array}
\usepackage{bm}
\usepackage{xcolor}
\usepackage[pdftex, hidelinks]{hyperref}
\addto\extrasenglish{  
    
}
\defcitealias{2017ApJ...841...30T}{PT17}
\defcitealias{10.1111/j.1365-2966.2011.18315.x}{H11}
\bibpunct{(}{)}{;}{a}{}{,}
\addto\extrasenglish{
}

\begin{document} 

   \title{Flows, Circulations, and Energy Transport in the Outer and Deep Atmospheres of Synchronous and Non-synchronous Hot Jupiters}

   \author{F. Sainsbury-Martinez\inst{1,2}\thanks{\email{f.sainsbury-martinez@leeds.ac.uk}}\and Pascal Tremblin\inst{1}  \fnmsep}

   \institute{Université Paris-Saclay, UVSQ, CNRS, CEA, Maison de la Simulation, 91191, Gif-sur-Yvette, France, \\ School of Physics and Astronomy, University of Leeds, Leeds LS2 9JT, UK
}

   \date{Received \#\#\# \#\#\, 2024; accepted \#\#\# \#\#, 2024}

  \abstract
   {}
   {Recent studies have shown that vertical enthalpy transport can explain the inflated radii of highly irradiated gaseous exoplanets. Simultaneously, they have also shown that rotation can influence this transport, leading to highly irradiated, rapidly rotating, objects that are uninflated. Here we explore the flows which underpin this transport, including the impact of synchronous/non-synchronous rotation. }
  {We use DYNAMICO to run a series of long-timescale, HD209458b-like, atmospheric models at various rotation rates. For models that are tidally-locked, { we consider rotation rates between $1/16^\mathrm{th}$ and $40$ times the rotation rate of HD209458b, whilst for non-synchronous models we consider the range $1/8^\mathrm{th}$ to $4$ times HD209458b.}}
   {We find that our synchronous models fall into one of three $\Omega$-dependent regimes: at low $\Omega$, we find that the outer atmosphere dynamics are driven by a divergent day-night wind, { which drives weak vertical transport} and can lead to the formation of a night-side hot-spot. At intermediate $\Omega$, we find classical hot Jupiter dynamics, whilst at high $\Omega$ we find a strong Coriolis effect that suppresses off-equator dynamics, including the jet-driving standing waves, thus also reducing vertical transport. 
   As for non-synchronicity, when small, { like \citet{Showman_2009}}, we find that it has little effect on the dynamics. However as it grows, we find that temporal variations prevents the formation of the persistent structures that drive large scale dynamics and transport. }
   {We find that rotation can significantly impact the atmospheric dynamics of irradiated exoplanets, including vertical enthalpy advection, which may help to explain the scatter in the hot Jupiter radius-irradiation relation. We have also identified a seemingly robust atmospheric feature at slow rotation: a night-side hot-spot. As this may have important implications for both the phase curve and atmospheric chemistry, we suggest that this study be followed up with next-generation GCMs that robustly models radiation and chemistry.  }

   \keywords{Planets and satellites: interiors - Planets and satellites: atmospheres - Planets and satellites: fundamental parameters - Planets: HD209458b - Hydrodynamics}
   \titlerunning{Flows, Circulations, and Energy Transport in Hot Jupiters}
   \maketitle
%

\section{Introduction} \label{sec:introduction}
Thanks to the continuing efforts of the observational and instrumental astrophysics community, the parameter regime in which { highly irradiated Jupiter-like planets} can be found has expanded greatly (see, for example, \citealt{2017AJ....154..188S,2019MNRAS.484.3522H,2020ApJ...892L...7H,2020AJ....160..275S,2021ApJ...920L..16D,2022A&A...662A.101S,2022arXiv220703911U}). In particular, there has been a significant increase in the range of orbital radii in which { irradiated Jupiter-mass planets can be observed, leading to an increase in the number of known `warm' Jupiters}. Simultaneously, numerical studies of hot Jupiter, and hot Brown Dwarf, atmospheres has revealed that rotation can play a significant role on both the horizontal and vertical dynamics \citep{2014arXiv1411.4731S,10.1093/mnras/stz3050,2021A&A...656A.128S}. For example, \citet{2021A&A...656A.128S}, found that, in addition to the link between radius inflation and planetary irradiation, a link might also exist between the level of radius inflation observed and the rotation rate of the exoplanet in question (see \autoref{fig:rad_rot}). Specifically, they suggested that the lack of observed radius inflation for the ultra-short-orbit hot Brown Dwarf SDSS1411b could be explained by the influence that rotation has on the atmospheric dynamics. { A similar dynamical story might also explain the radius of the ultra-short-orbit hot Brown Dwarf WD0137-349b, with the atmospheric models of \citet{2020MNRAS.496.4674L} also suggesting that the dynamics of this hot Brown Dwarf are highly influenced by rotation}. \\

Examples of how the Coriolis effect can influence atmospheric dynamics can be found in our own solar system. Generally, planetary atmospheric dynamics depend upon the influence of both planetary irradiation and rotation, with the latter effect leading to the formation of three distinct atmospheric regimes.\\
At slow rotation rates, e.g. Venus and Titan, atmospheric dynamics are dominated by planetary scale Hadley cells \citep{1735RSPT...39...58H,doi:10.1146/annurev.earth.34.031405.125144,2013cctp.book..277S} which mix temperature horizontally, resulting in weak latitudinal temperature gradients. This is known as Cyclostrophic circulation, and leads to significant super-rotation of the winds, as well as strong, circumpolar, vortexes, { the latter} of which is described well by Thermal Wind Balance \citep{Flasar2010}. {Note that, in the slow rotation regime, strong drag can also play a role in shaping the atmospheric wind, with a force balance between pressure-gradients and drag leading to winds to just flowing down pressure gradients rather than developing Cyclostrophic circulations \citep{2013ApJ...762...24S}}. \\
Increasing the rotation rate, we come to the dynamical regime occupied by both Earth and Mars: At low latitudes, the Coriolis effect remains relatively weak, and Hadley cells develop (the tropics), However as we move to higher latitudes (i.e. the extra-tropics - $\theta>30^\circ$ for Earth), the regime changes and rotational effects start to play a much more significant role. This results in the wind changing from being Cyclostrophic to Geostrophic, which in turn is paired with a change in thermal/angular-momentum transport from Hadley circulations to baroclinic instabilities (which drive the extra-tropical cyclones/anti-cyclones that dominate much of Europe's and the US's weather - \citealt{OntheExistenceofStormTracks,StormTrackDynamics,wcd-2-1111-2021}). As a consequence of this change in dynamics, meridional heat transport has weakened, resulting in latitudinal temperature gradients that are comparable to vertical gradients. \\
The strength of the meridional heat transport continues to decrease as we move into the rapid rotation regime, as seen on Jupiter, Saturn, and the other gaseous solar-system giants. Here the Coriolis effect plays a significant role at all latitudes, leading to a Geostrophic atmosphere dominated by the infamous banded jet structure \citep{ingersoll_vasavada_1998} (with baroclinic instabilities { being proposed as a mechanism to} drive mixing/angular-momentum-transport between bands; \citealt{YOUNG2019225}).\\

A similar change in dynamics is expected to occur in exoplanetary atmospheres, albeit over a significantly wider parameter space thanks to the much wider variety of planetary type/orbital radii combinations found outside of our solar system. One such combination found only outside of our solar-system is the hot Jupiter, a Jupiter-like, gaseous planet which is in a very short orbit and hence is very highly irradiated. Whilst a number of recent studies have explored how atmospheric dynamics changes with orbital radii (\citealt{2014arXiv1411.4731S,2023ApJ...958...68S}), these come with the proviso that the models link effects caused by changes in both planetary irradiation and rotation, making it difficult to explore the effects that rotation alone has on the dynamics. Further, even studies which have explored the isolated effects of rotation alone \citep[e.g;][]{10.1093/mnras/stz3050} have only done so for a few rotation rates, although even this limited study showed how rotation can shape the horizontal and vertical transport.  \\

As such, and in order to isolate how rotation affects the atmospheric dynamics (with a particular focus on vertical { potential-temperature/enthalpy} transport, { which has the potential to explain the radius inflation problem for both hot Jupiters and hot Brown Dwarfs}; \citealt{2017ApJ...841...30T,2019A&A...632A.114S,2021A&A...656A.128S,2023MNRAS.524.1316S}), here we present a series of deep, HD209458b-like atmospheric models in which everything but the rotation rate is held constant. Specifically the rotation rate will be varied between $\Omega_{\mathrm{rot}}=0.0625\Omega_{0}$ and $\Omega_{\mathrm{rot}}=40\Omega_{0}$, where { $\Omega_{0}=2.1\times10^{-5}\mathrm{s^{-1}}$} is the canonical angular rotation rate of HD209458b, and the atmospheric models will be run until the deep atmosphere (with $P_{max}=200\si{\bar}$) has equilibrated. These values where chosen such that our models would span the rotation regimes found in our own solar system, and we will investigate cases in which the rotation is both synchronous and non-synchronous.   \\

The structure of this work is as follows: In \autoref{sec:model_setup} we introduce the model we use to explore the effects of rotation on atmospheric dynamics and observables: This includes both a brief description of our 3D global circulation model (GCM) DYNAMICO, as well as a more indepth discussion of the parametrisations we use. Then, in \autoref{sec:synchronous}, we discuss how rotation influences the atmospheric dynamics of synchronously rotating hot Jupiters, starting with zonal-mean dynamics before moving onto the horizontal and vertical winds and heat transport, which we explore via both the Helmholtz wind decomposition as well as the enthalpy flux. Together these allow us to investigate how rotation influences deep atmospheric heating, and hence observed radius inflation. We then, in \autoref{sec:NS}, perform a similar analysis for a set of hot Jupiter atmospheric models in which the orbital period and planetary rotation period are allowed to diverge (i.e. non-synchronous rotation), investigating how sensitive the standing-wave driven dynamics, and hence the vertical heat transport, are to longitudinal shifts in the planetary irradiation profile. We finish, in \autoref{sec:conclusion}, with concluding remarks, discussing the implications of our results and suggestions for future studies, with next-generation GCMs, which will investigate the robustness of our results. In particular, we are very interested in confirming if the formation of a night-side hot-spot at slow rotation rates is just a feature of our atmospheric models, or if it is a robust feature that may have important implications for both the chemistry and observations of more slowly rotating hot Jupiters (i.e. hot Jupiters in longer orbits around massive/bright stars).  \\

\section{Model Setup}\label{sec:model_setup}
In order to investigate how rotation impacts atmospheric dynamics in both the observable outer atmosphere, and the more quiescent (but still dynamically important) deep atmosphere, we adopt the simplified atmospheric setup \citet{2019A&A...632A.114S} used to model the deep atmosphere of HD209458b. This is necessitated by both the high computational cost of evolving the quiescent deep atmosphere to steady state (even when we initialise the model with an adiabat close to the expected equilibrium - \citet{Rauscher_2010,Mayne_2014,2021A&A...656A.128S}) as well as the large range of rotation rates considered: from $\Omega_{\mathrm{rot}}=0.0625\Omega_{0}$ and $\Omega_{\mathrm{rot}}=40\Omega_{0}$, where $\Omega_{0}=\Omega_{\textrm{HD209458b}}=2.1\times10^{-5}\mathrm{s^{-1}}$. \\
Here, we give a very brief overview of the highly computationally efficient GCM DYNAMICO (\autoref{sec:dynamico_NS}), which uses a Newtonian Cooling approach to model the radiative forcing (\autoref{sec:newtonian_cooling}), which we use to run a large array of atmospheric models to steady state at all pressures, and from which a smaller subset of models which represent the different dynamical regimes observed are selected (\autoref{sec:model_parameters}).

\subsection{DYNAMICO} \label{sec:dynamico_NS}
DYNAMICO is one of a number of highly computationally efficient global circulation models which has been developed to run on the next-generation of Exascale supercomputers, enabling atmospheric models of resolutions and { regimes} that where previously out of reach. To achieve this, DYNAMICO solves the primitive equations of meteorology (\citealt{2014JAtS...71.4621D,Vallis17}) on a spherical icosahedral grid \citep{gmd-8-3131-2015}. Here we configure this horizontal-grid such that the angular resolution of each grid-cell is approximately $2.5^\circ$. Vertically, the levels of the pressure-grid are linearly spaced in $\log\left(P\right)$ space with a maximum pressure of $200\si{\bar}$ and a minimum pressure of $7 \times 10^{-3}\si{\bar}$. This pressure range was selected such that we model both the highly irradiated outer atmosphere where irradiation always dominates over transport as well as a sufficiently large region of the deep atmosphere such that the effects of differing vertical heat transport rates are clearly visible. Note that the vertical boundaries of our model are closed and stress-free such that the only means of energy transport into and out-of the system are the numerical-dissipation (implemented as a hyperdiffusion filter with a fixed dissipation time-scale) required to stabilise the system against grid-scale noise, and the Newtonian cooling relaxation scheme which we use to model the stellar irradiation (\autoref{sec:newtonian_cooling}).  \\

Following the calibration procedures of \citet{2019A&A...632A.114S} and \citet{2021A&A...656A.128S}, we select a numerical dissipation time-scale of $\tau_{dissip}=2500\,\mathrm{s}$ in order to ensure the stability of the equatorially confined jets that develop in our most rapidly rotating models (e.g. \autoref{fig:Zonal_Wind_15}). These narrow jets with their high latitudinal velocity gradients also result in us setting a 90 second model time-step, which balances the need to correctly reproduce model physics, whilst also minimising the computational cost modelling equilibrium deep atmospheres. 

\subsection{Radiative forcing via Newtonian cooling} \label{sec:newtonian_cooling}
As discussed above, for computational efficiency reasons, we do not directly model either the incident stellar irradiation or resulting night-side thermal emission. Instead, as discussed in \citet{2019A&A...632A.114S}, we take a Newtonian cooling approach in which the outer atmosphere is relaxed onto a spatially varying equilibrium temperature profile $T_{eq}$ over a time-scale based upon the local radiative time-scale $\tau_{rad}$. In DYNAMICO, this is achieved by adding a source term to the temperature evolution equation:
\begin{equation}
\frac{H}{c_{p}}=\frac{\partial T\left(P,\theta,\phi\right)}{\partial t} = - \frac{T\left(P,\theta,\phi\right)-T_{eq}\left(P,\theta,\phi\right)}{\tau_{rad}\left(P\right)} \,,
\end{equation}
{ where $H$ is the local heating rate and $c_{p}$ is the specific heat capacity.}
For numerical reasons, we express the equilibrium temperature profile ($T_{eq}\left(P,\theta,\phi\right)$) onto which the atmosphere is relaxed, in terms of the night-side temperature profile $T_{n}\left(P\right)$ \citep{2005A&A...436..719I} and a pressure dependent day-night temperature difference ($\Delta{T}(P)$) which has been calibrated such as to reproduce observations { of the hot Jupiter HD209458b}:
\begin{align}
T_{eq}\left(P,\theta,\phi\right) &= T_{n}\left(P\right) \notag \\
&+ max|0, \Delta{T}\left(P\right) \cdot cos\left(\theta\right) \cdot cos\left(\phi - 180\right)| \notag \\
&+ \Delta{T}(P) \cos\left(\theta\right)\max \left[ 0, \cos (\phi - \pi) \right], \label{eq:teq}
\end{align}
where the night-side temperature profile is parametrised as series of linear in $\log(P)$ space interpolations between the following T-P points
\begin{equation}
  T_{night}=\left(800, 1100, 1800\right)\,@\,P=\left(10^{-6}, 1, (\ge)10 \right)\textrm{ bar},  
\end{equation}
and the day-night temperature difference takes the form
\begin{align}
\Delta T (P) &=\left\{ \begin{array}{ll}
  600 \si{\kelvin} & \textrm{if } P<0.01 \si{\bar} \\
 600 \si{\kelvin} \log (P/(0.01)) & \textrm{if } 0.01\si{\bar}  < P < 10 \si{\bar} \\
 0 & \textrm{if } P > 10 \si{\bar}  \end{array}
\right. \label{eq:deltaT}
\end{align}

The radiative time-scale over which this relaxation occurs is also based upon the work of \citet{2005A&A...436..719I}, with the linear in $\log(P)$ space interpolation taking the form:
\begin{equation}
  \log(\tau_{rad})=\left(2.5, 5.0, 7.5, \log(\tau_{rad}^d)\right)\,@\,P=\left(10^{-6}, 1, 10, 200 \right)\textrm{ bar}.  
\end{equation}
Note that, for the majority of models discussed here, we set the deep radiative time-scale to infinity $\tau_{rad}^d = \infty$, thus disabling radiative forcing in the deep atmosphere, { between 20 and 200 bar,} where it is expected to be significantly slower than advection/transport \citep{2019A&A...632A.114S}. 

\subsection{HD209458b-like Atmospheric Models} \label{sec:model_parameters}
In order to explore how rotation impacts the outer and deep atmospheric dynamics and observables of highly irradiated, gaseous, exoplanets, we have explored synchronously rotating atmospheres at 28 different rotation rates (between $\Omega_{\mathrm{rot}}=0.0625\Omega_{0}$ and $\Omega_{\mathrm{rot}}=40\Omega_{0}$ - discussed in \autoref{sec:synchronous}) and non-synchronously rotating atmospheres at 13 different planetary rotation rates (between $\Omega_{\mathrm{rot}}=0.25\Omega_{0}$ and $\Omega_{\mathrm{rot}}=4\Omega_{0}$, with the orbital period held constant at that of HD209458b - discussed in \autoref{sec:NS}). { Here each model was run for over 200 Earth-years of simulation time to ensure that the radiatively active layers ($P<10$ bar) were equilibrated. Note that the deep atmospheres in all the models presented here should also be close to equilibrium, however the deep heating rate asymptotically slows as we approach steady-state and hence modelling the last few kelvin of heating would require significant additional computation for little change in the atmospheric dynamics (relative to the difference between a deep isotherm and adiabat; \citealt{2019A&A...632A.114S}).}   \\
The parameters for these models are given in \autoref{tab:lr_params} and are the same for both synchronous and non-synchronous models, with one major difference. In the synchronous models, the Newtonian Cooling profile is fixed in both latitude and longitude, representing the irradiation of a tidally-locked exoplanet. On the other hand, in our non-synchronous models, the longitude of the Newtonian Cooling (forcing) profile is updated every time-step based upon the difference between the planetary and orbital rotation rates:
\begin{align}
  \Delta\Omega &= \Omega_{\mathrm{rot}} - \Omega_{\mathrm{orb}} \\
  \phi_{net} &= \left(t * \Delta\Omega\right)\bmod(360).
\end{align}

\begin{table}[t]
  \centering
  \def\arraystretch{1.5}
  \begin{tabular}{c|c|c}
    Quantity (units) & Description & Value \\
    \hline \hline
    dt (\si{\second}) & Time-step & 90 \\
    $N^\circ$ & Angular Resolution & $~2.7^\circ$\\
    $P_{top}$ (\si{\bar}) & Pressure at Top  & $7 \times 10^{-3}$  \\
    $P_{bottom}$ (\si{\bar}) & Pressure at Bottom & 200  \\
    $g$ (\si{\meter\per\second\squared}) & Gravity & 8.0 \\
    $R_{HJ}$ (\si{\meter}) & HJ Radius  & $10^8$ \\
    $\Omega_{0}$ (\si{\per\second}) & Base HJ Angular Rotation & $2.1 \times 10^{-5}$ \\
    $c_p$ (\si{\joule\per\kilo\gram\per\kelvin}) & Specific Heat & 13226.5 \\
    $\mathcal{R}$ (\si{\joule\per\kilo\gram\per\kelvin}) & Ideal Gas Constant & 3779.0
  \end{tabular}
  \caption{Parameters for Simulations}
  \label{tab:lr_params}
\end{table}
\begin{figure*}[tbp] %
\begin{centering}
\begin{subfigure}{0.3\textwidth}
\begin{centering}
\includegraphics[width=0.99\columnwidth]{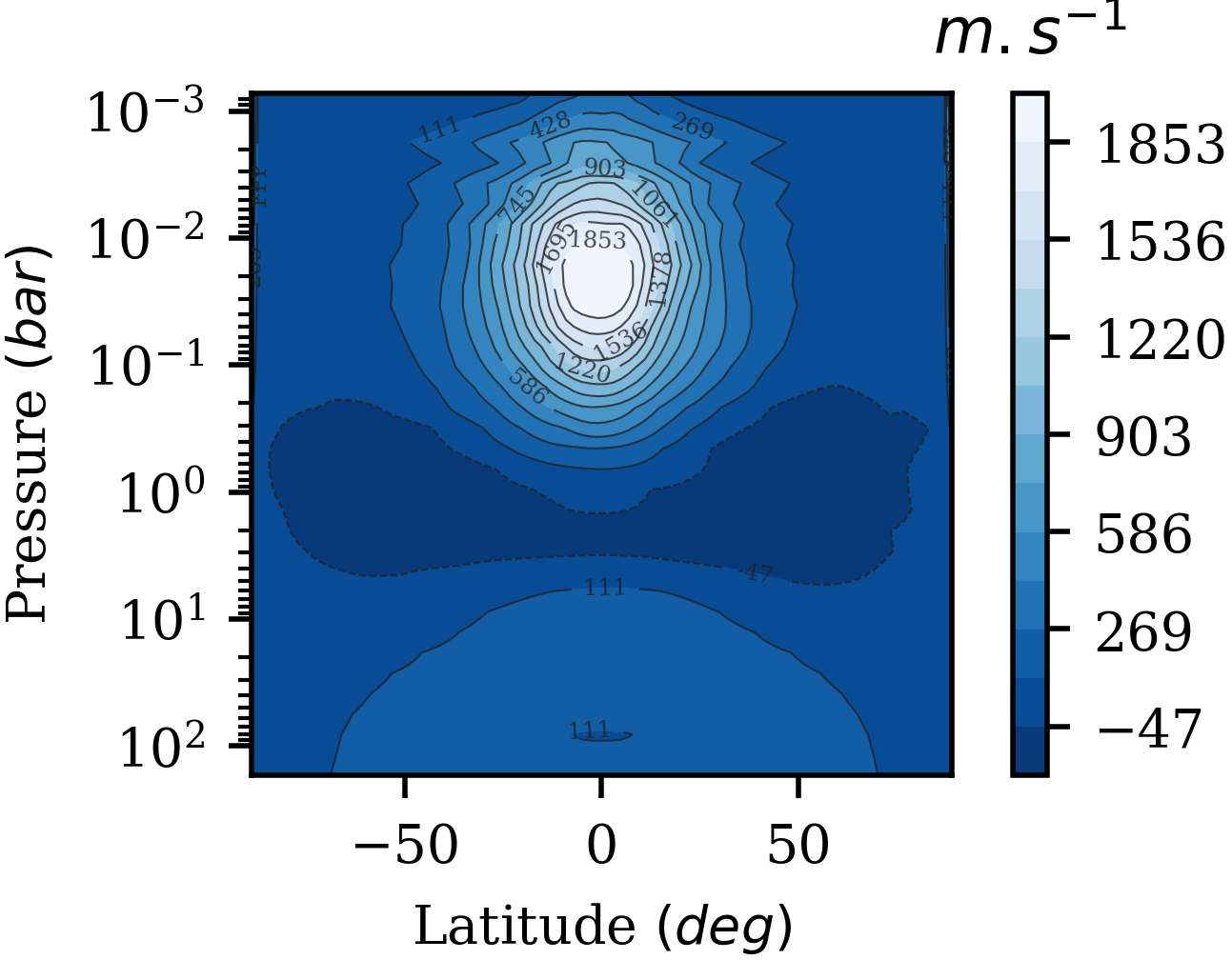}
\caption[]{$0.125\Omega_{0}$  \label{fig:Zonal_Wind_0125} }
\end{centering}
\end{subfigure}
\begin{subfigure}{0.3\textwidth}
\begin{centering}
\includegraphics[width=0.99\columnwidth]{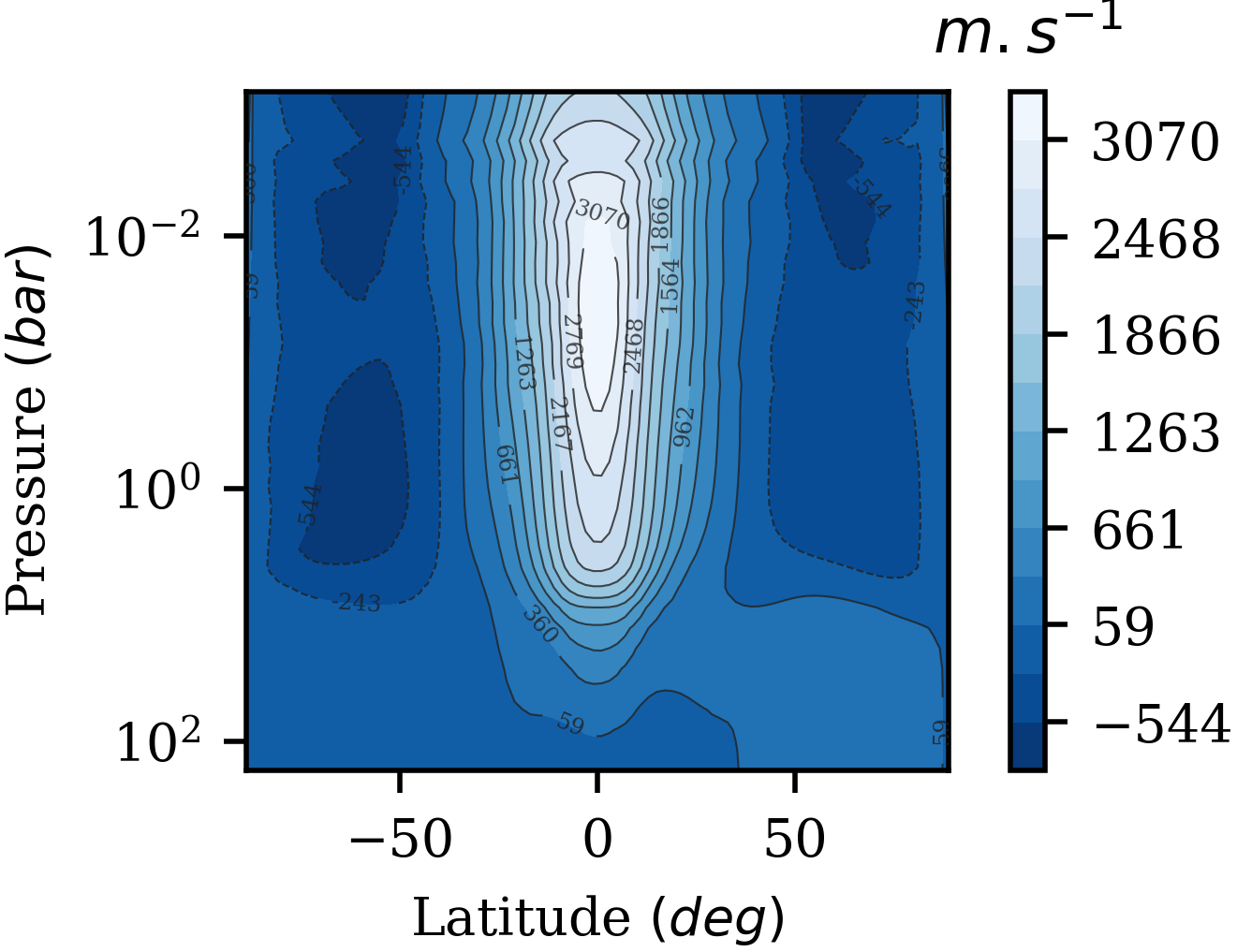}
\caption[]{$1\Omega_{0}$   \label{fig:Zonal_Wind_1} }
\end{centering}
\end{subfigure}
\begin{subfigure}{0.3\textwidth}
\begin{centering}
\includegraphics[width=0.99\columnwidth]{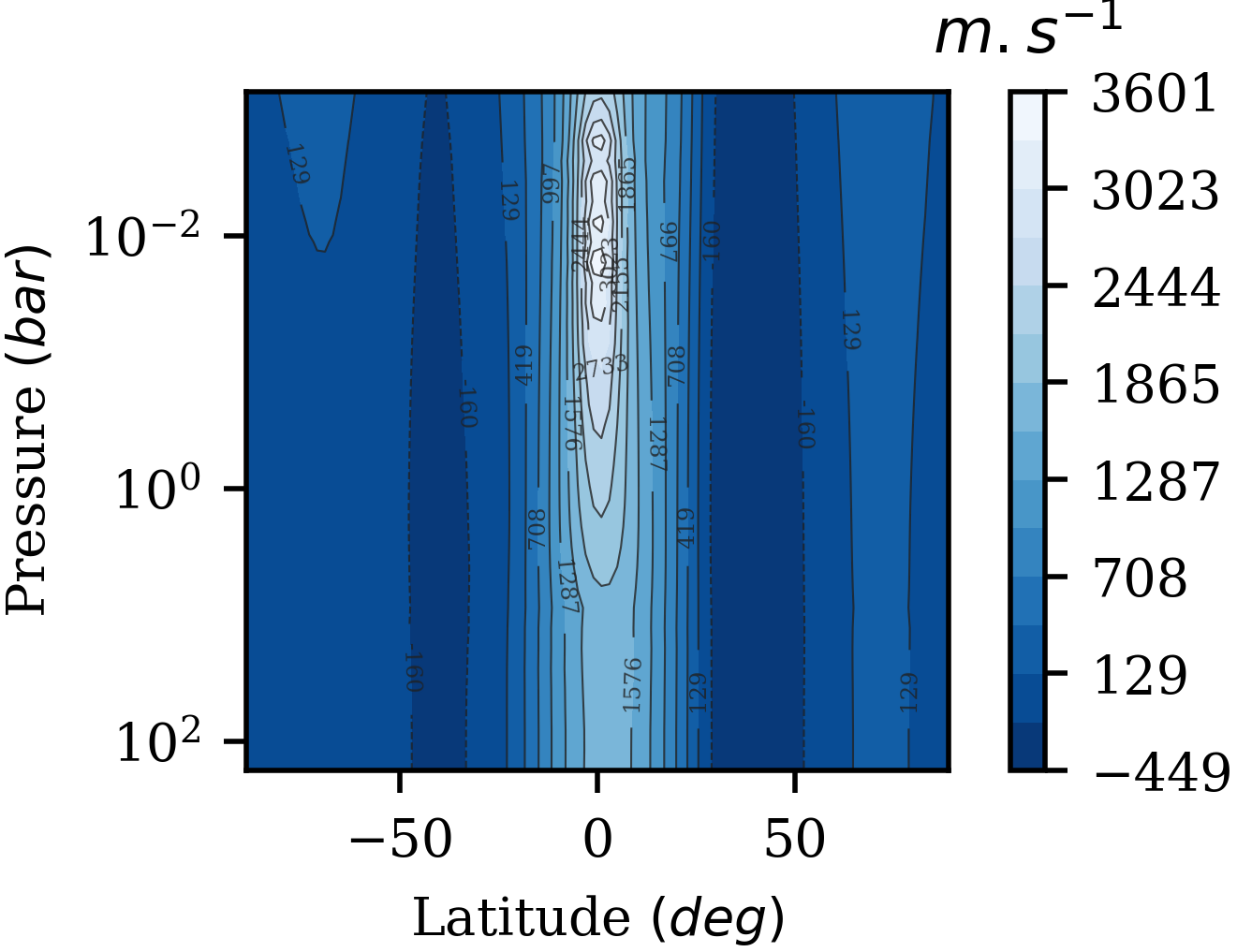}
\caption[]{$15\Omega_{0}$    \label{fig:Zonal_Wind_15} }
\end{centering}
\end{subfigure}
\caption[Zonal wind profiles for three HD209458b-like models ]{Zonally and temporally averaged, { over five Earth-years of simulation time,} zonal wind profiles for three HD209458b-like models with rotation rates of; $0.125\Omega_{0}$ - top, $\Omega_{0}$ - middle, and $15\Omega_{0}$ - bottom.  \label{fig:Zonal_wind} }
\end{centering}
\end{figure*} 
\begin{figure}[tbp] %
\begin{centering}
\begin{subfigure}{0.99\columnwidth}
\begin{centering}
\includegraphics[width=0.7\columnwidth]{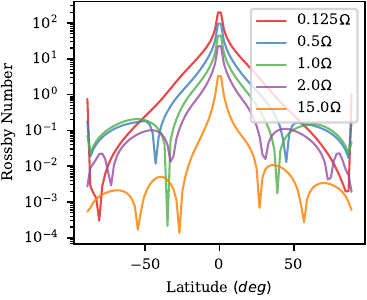}
\caption[]{$R_{0}$  \label{fig:Latitudinal_Rossby} }
\end{centering}
\end{subfigure}
\begin{subfigure}{0.99\columnwidth}
\begin{centering}
\includegraphics[width=0.7\columnwidth]{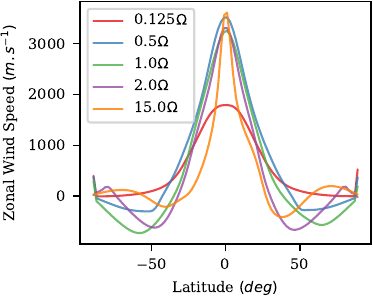}
\caption[]{$u$  \label{fig:Latitudinal_Wind} }
\end{centering}
\end{subfigure}
\caption[Latitudinal Rossby Number profiles at 0.016 bar]{Latitudinal Rossby number, $R_{0}$ - top, and zonal-mean zonal-wind, $u$ - bottom, profiles in the outer atmospheres ($P\simeq0.016\textrm{bar}$) of five HD209458b-like models at different rotation rates spanning the range $0.125\Omega_{0}\rightarrow15\Omega_{0}$. \label{fig:Latitudinal_Rossby_Wind} }
\end{centering}
\end{figure}
\begin{figure*}[tbp] %
\begin{centering}
\begin{subfigure}{0.3\textwidth}
\begin{centering}
\includegraphics[width=0.99\columnwidth]{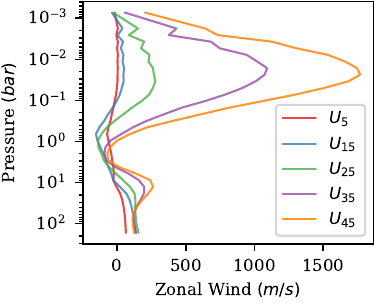}
\caption[]{$0.125\Omega_{0}$  \label{fig:Zonal_wind_slices_0125} }
\end{centering}
\end{subfigure}
\begin{subfigure}{0.3\textwidth}
\begin{centering}
\includegraphics[width=0.99\columnwidth]{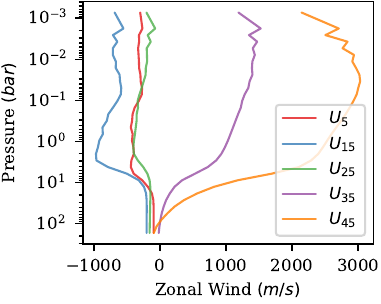}
\caption[]{$1\Omega_{0}$   \label{fig:Zonal_wind_slices_1} }
\end{centering}
\end{subfigure}
\begin{subfigure}{0.3\textwidth}
\begin{centering}
\includegraphics[width=0.99\columnwidth]{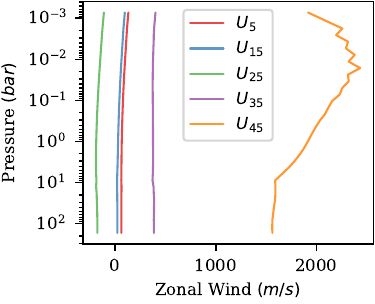}
\caption[]{$15\Omega_{0}$    \label{fig:Zonal_wind_slices_15} }
\end{centering}
\end{subfigure}
\caption[Slices of the longitudinally averaged zonal wind plotted against pressure at five different latitudes for three different rotation rates ]{Longitudinally averaged profiles of the zonal wind at five different latitudes, spanning the region from the equator to the northern pole ($90^{\circ}$), for three HD209458b-like models at different rotation rates: $0.125\Omega_{0}$ - left, $\Omega_{0}$ - centre, and $15\Omega_{0}$ - right. These three rotation rates have been chosen to emphasise the change in zonal flow dynamics with rotation regime. Note: Positive wind speeds indicate eastwards flows and for each wind profile we average over a $20^{\circ}$ latitudinal band centred at the labelled latitude. \label{fig:Zonal_wind_slices} }
\end{centering}
\end{figure*} 
\section{Synchronous Rotation} \label{sec:synchronous}
We start by exploring how rotation affects the atmospheric dynamics of synchronously rotating exoplanets, focusing our analysis on three rotation rates ($0.125\Omega_{0}$, $\Omega_{0}$, $15\Omega_{0}$) which are typical of the three dynamical regions observed. We return to our full sample in \autoref{sec:enthalpy_transport} where we investigate the influence of rotation on vertical heat transport. And we finish, in \autoref{sec:length_scales}, with a brief discussion of non-dimensional length scales, rotational regimes, and the resolution of our models. 

\subsection{Zonal-Mean Zonal-Winds} \label{sec:zonal_winds}
We begin our study by exploring how the zonal-mean zonal-winds change with rotation rate. Specifically, as previously alluded to, we find that our models all fit into one of three dynamical regimes (or the transition between regimes), as shown in \autoref{fig:Zonal_wind}. \\

At slower rotation rates ($\Omega_{\mathrm{rot}}<0.5\Omega_{0}$ - \autoref{fig:Zonal_Wind_0125}), we find that a relatively slow, very shallow, and latitudinally extended `jet' appears to dominate the zonal-mean zonal-wind (this can also be seen in vertical slices of the zonal-mean zonal-wind: \autoref{fig:Zonal_wind_slices_0125}). This is paired with a rather quiescent mid to deep atmosphere ($\left|U\right|_{max} < 150 \si{\metre\per\second}$ for $P > 0.6\si{\bar}$), which implies that deep mixing may be suppressed in this regime. Note that this weakening of the equatorial jet is similar to that found in the ten day rotational period models of \citet{10.1093/mnras/stz3050}.  \\
As $\Omega_{\mathrm{rot}}$ increases (between $0.5\Omega_{0}<\Omega_{\mathrm{rot}}\lesssim5\Omega_{0}$ for our models), we find that the zonal-mean zonal wind (\autoref{fig:Zonal_Wind_1}) { resembles} that found in the classical hot Jupiter regime: a strong zonal jet that extends all the way from the irradiated outer atmosphere to the advective deep atmosphere ($P>10\si{\bar}$), flanked by weaker high latitude counter-flows (see \autoref{fig:Zonal_wind_slices_1}) which are typically associated with the { zonal jet driving} standing wave pattern (see \autoref{sec:helmholtz_wind} and \citealt{2011ApJ...738...71S}).
Note that this jet is slightly narrower than the `jet' that forms in the slow rotation regime. As shown in \autoref{fig:Latitudinal_Wind}, the latitudinal extent of the jet is highly sensitive to the rotation rate, likely due to the influence that rotation plays, via the Coriolis effect, on off-equator dynamics (see \autoref{fig:Latitudinal_Rossby}).\\
This can be seen in our $\Omega_{\mathrm{rot}}=15\Omega_{0}$ model where we find (\autoref{fig:Zonal_Wind_15}) an even narrower zonal jet which extends all the way from the outer atmosphere to the very bottom of the simulation domain (\autoref{fig:Zonal_wind_slices_15}). Once again we also find that this jet is braced by weaker counter flows, however due to the Coriolis effect, these counter flows have become more latitudinally confined, revealing even weaker easterly flows near the poles. This pattern of alternating easterly and westerly flows is reminiscent of the wind structure found for; gas giants in our solar system, the weakly irradiated and rapidly rotating hot Jupiter models of \citet{2014arXiv1411.4731S}, and the rapidly rotating brown Dwarf SDSS1411b \citep{2021A&A...656A.128S}. \\
This narrowing of the jet and suppression of off-equator flows continues as we move to even higher rotation rates. However here another effect also becomes apparent: { a reduction in the total zonal transport by the jet, which can be seen in \autoref{fig:Rossby_vs_U_mean} as a reduction in the mean zonal wind speed with increasing $Ro$ (an effect we discuss in more detail in \autoref{sec:helmholtz_wind}). As we discuss in \autoref{sec:helmholtz_wind}, this is likely due to the Coriolis suppression of the jet-driving off-equator standing Rossby and Kelvin waves. }

\begin{figure*}[tbp] %
\begin{centering}
\begin{subfigure}{0.3\textwidth}
\begin{centering}
\includegraphics[width=0.99\columnwidth]{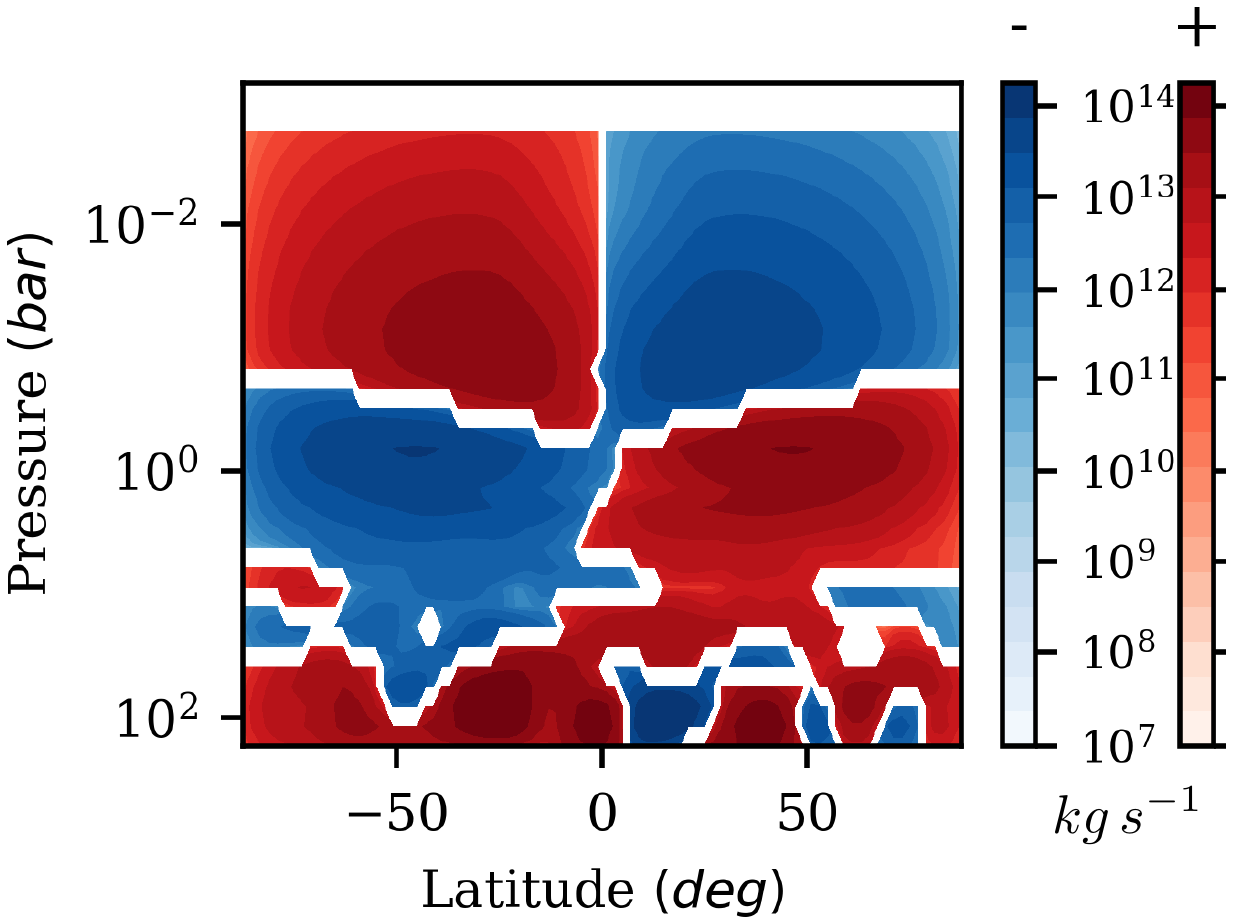}
\caption[]{$0.125\Omega_{0}$  \label{fig:Streamfunction_0125} }
\end{centering}
\end{subfigure}
\begin{subfigure}{0.3\textwidth}
\begin{centering}
\includegraphics[width=0.99\columnwidth]{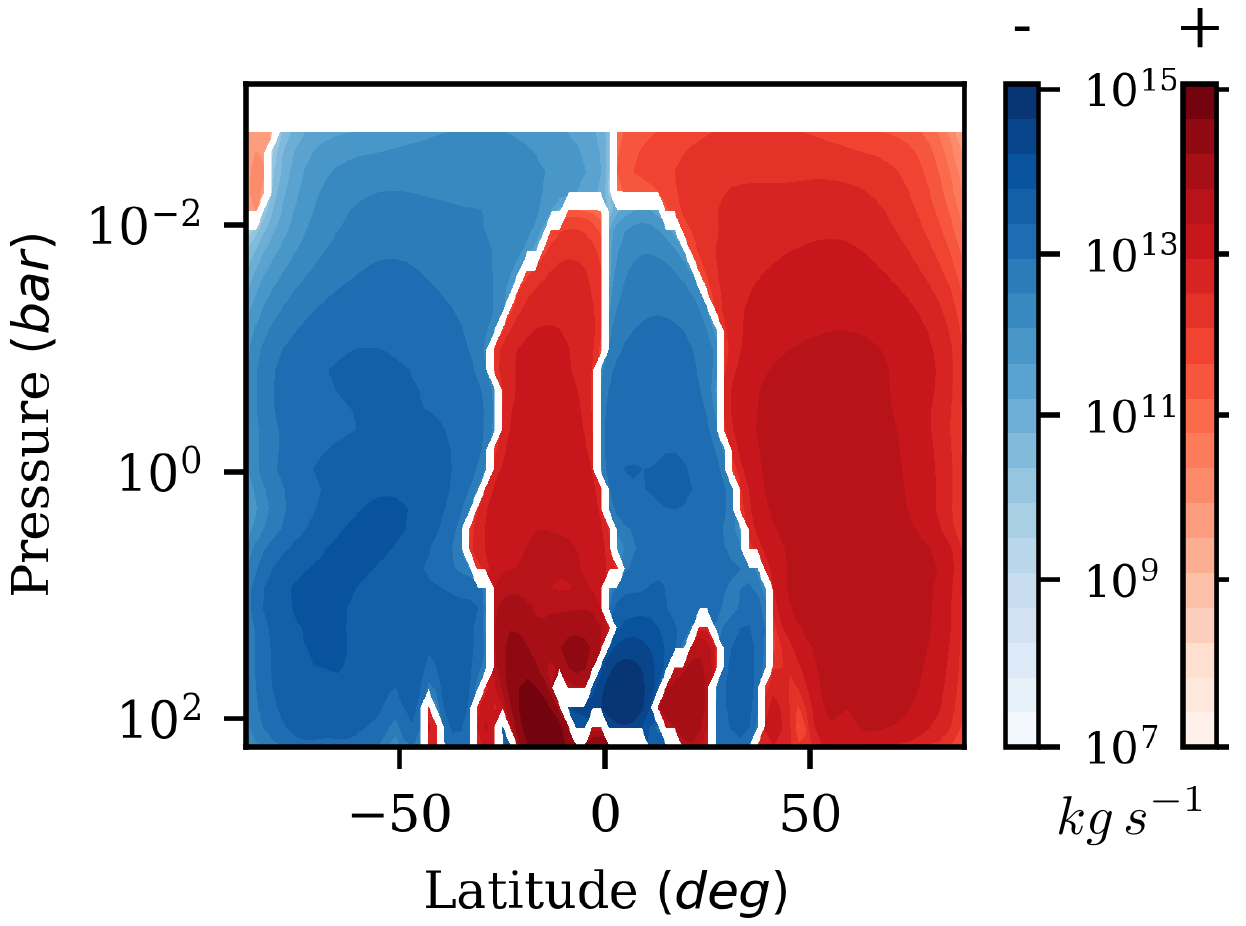}
\caption[]{$1\Omega_{0}$   \label{fig:Streamfunction_1} }
\end{centering}
\end{subfigure}
\begin{subfigure}{0.3\textwidth}
\begin{centering}
\includegraphics[width=0.99\columnwidth]{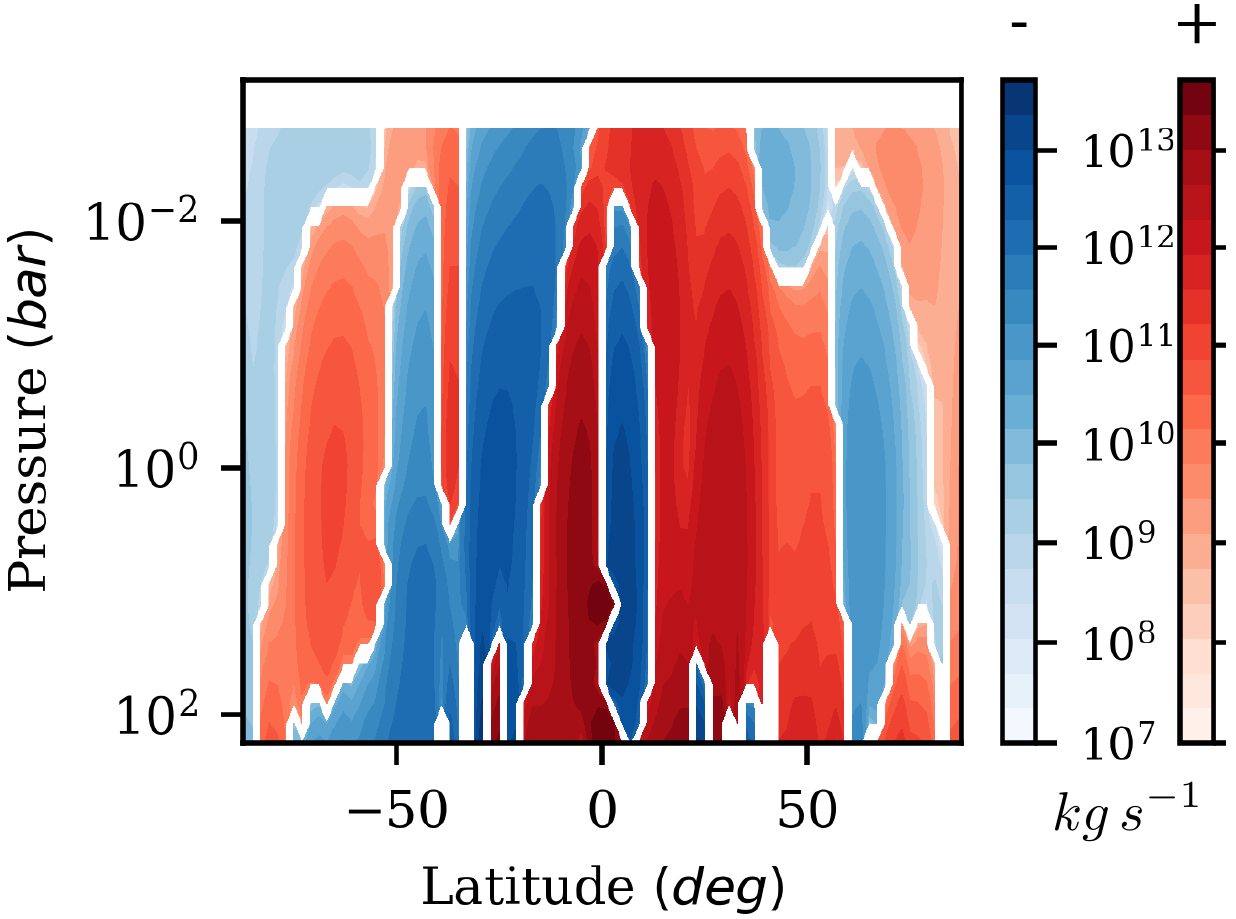}
\caption[]{$15\Omega_{0}$    \label{fig:Streamfunction_15} }
\end{centering}
\end{subfigure}
\caption[Streamfunctions for three HD209458b-like models ]{Zonally and temporally, { over five Earth-years of simulation time,} averaged meridional circulation streamfunctions for three HD209458b-like models with rotation rates of; $0.125\Omega_{0}$ - top, $\Omega_{0}$ - middle, and $15\Omega_{0}$ - bottom. Note that the meridional circulation profile is plotted on a log scale with clockwise circulations shown in red and anti-clockwise circulations shown in blue. At the base rotation rate, for example, these circulations combine at the equator to drive an equatorial downflow at almost all pressures. \label{fig:Streamfunction} }
\end{centering}
\end{figure*}

\subsection{Meridional Circulation Streamfunction} \label{sec:meridional_circ}
Moving onto the meridional circulation profile, specifically the streamfunction of the mass-flux on the meridional (i.e. latitudinal-pressure) plane, we again find that our models fall into three distinct regimes (\autoref{fig:Streamfunction}). This is in agreement with the results of \citet{2019A&A...632A.114S}, who suggests that the meridional circulation, and more generally the vertical dynamics, are driven by the horizontal winds.\\
Starting in the slowly rotating regime, we find (\autoref{fig:Streamfunction_0125}) a relatively weak meridional circulation profile that generally consists of a single circulation cell in each hemisphere. The only exception to this is the very deepest regions of the simulation domain which has not yet equilibrated due to the low local wind speeds (\autoref{fig:Zonal_Wind_0125}) and hence long dynamical time-scales. 
{ In the outer atmosphere, where the influence of rotation on the global overturning circulation is strongest, these circulations are driven by the net-eastward zonal wind, and as a result we find clockwise/anti-clockwise circulation cells in the southern/northern hemisphere respectively, cells that combine at the equator to drive a weak, net, downflow. However, as we move deeper ($P>0.5\si{\bar}$) and the net zonal-wind fades to near zero(see \autoref{fig:Zonal_wind_slices_0125}), the sense of these circulations is reversed, leading to an equatorial upwelling that is typical of a global overturning circulation that is barely influenced by rotation \citep{2021PNAS..11822705H}.} \\
Moving towards the classical hot Jupiter rotation regime, we find a meridional circulation profile { (\autoref{fig:Streamfunction_1})} similar to that found by \citealt{2019A&A...632A.114S}: a pair of meridional circulation cells in each hemisphere which combine together to drive an equatorial downflow from the outer atmosphere to the bottom of the simulation domain (albeit with a slight equatorial upflow at very low pressures where $\tau_{rad}\ll \tau_{adv}$ and hence no jet has formed: \autoref{fig:Zonal_wind_slices_1}), a mass conserving upflow at mid-latitudes, and finally weak polar downflows. Further, this circulation is almost an order of magnitude stronger than that found in the other rotation regimes, which may help { to explain this regimes strong vertical enthalpy advection} (see \autoref{sec:enthalpy_transport}). \\
Finally, in the rapidly rotating regime, { we find a meridional circulation profile with multiple circulation cells per hemisphere} (\autoref{fig:Streamfunction_15}), much like what was found for SDSS1411b by \citet{2021A&A...656A.128S}. These circulation cells combine at the equator to drive a relatively weak downflow, and, much like the zonal wind, the circulation cells become significantly weaker as we move towards higher latitudes. Furthermore, both this effect, as well as the number of circulation cells in each hemisphere, increases with $\Omega$, an effect which is likely linked to banded zonal-wind structure that also develops (\autoref{fig:Zonal_wind_slices_15}). 

\begin{figure*}[tbp] %
\begin{centering}
\begin{subfigure}{0.33\textwidth}
\begin{centering}
\includegraphics[width=0.8\columnwidth]{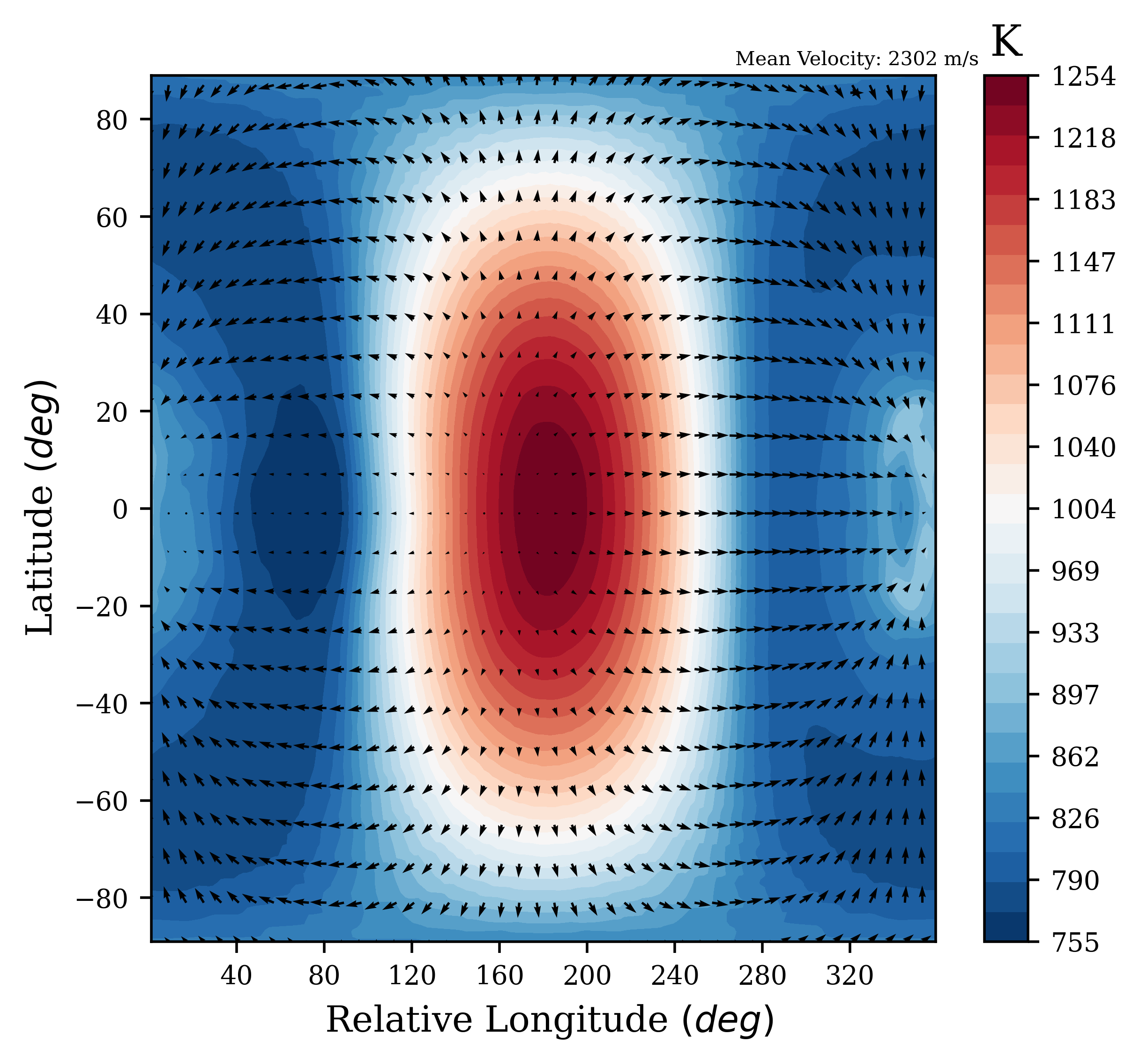}
\caption[]{$0.125\Omega_{0}$ - 0.0026 bar  \label{fig:Wind_Temp_0.125_00026} }
\end{centering}
\end{subfigure}
\begin{subfigure}{0.33\textwidth}
\begin{centering}
\includegraphics[width=0.8\columnwidth]{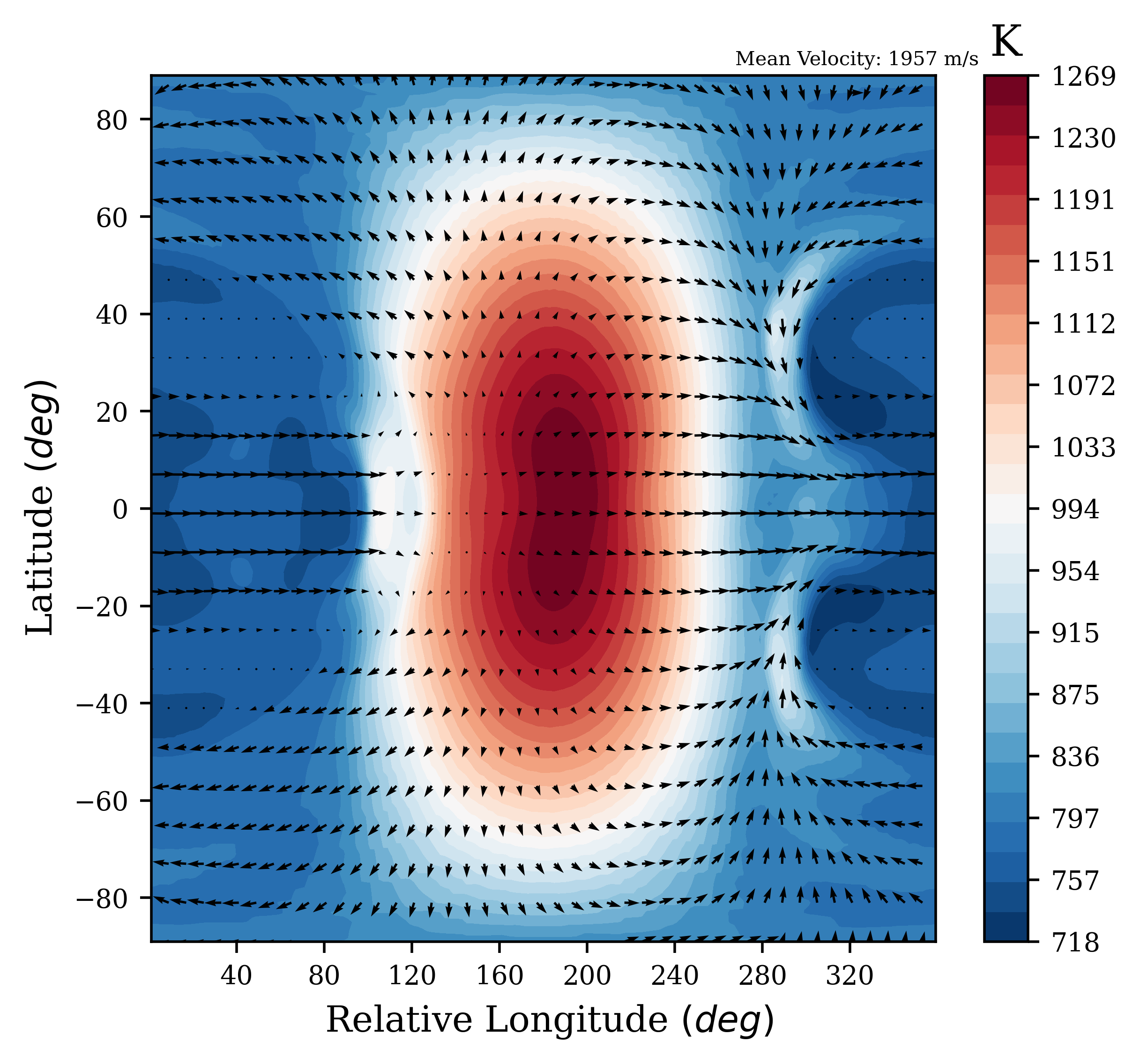}
\caption[]{$1\Omega_{0}$ - 0.0026 bar   \label{fig:Wind_Temp_1_00026} }
\end{centering}
\end{subfigure}
\begin{subfigure}{0.33\textwidth}
\begin{centering}
\includegraphics[width=0.8\columnwidth]{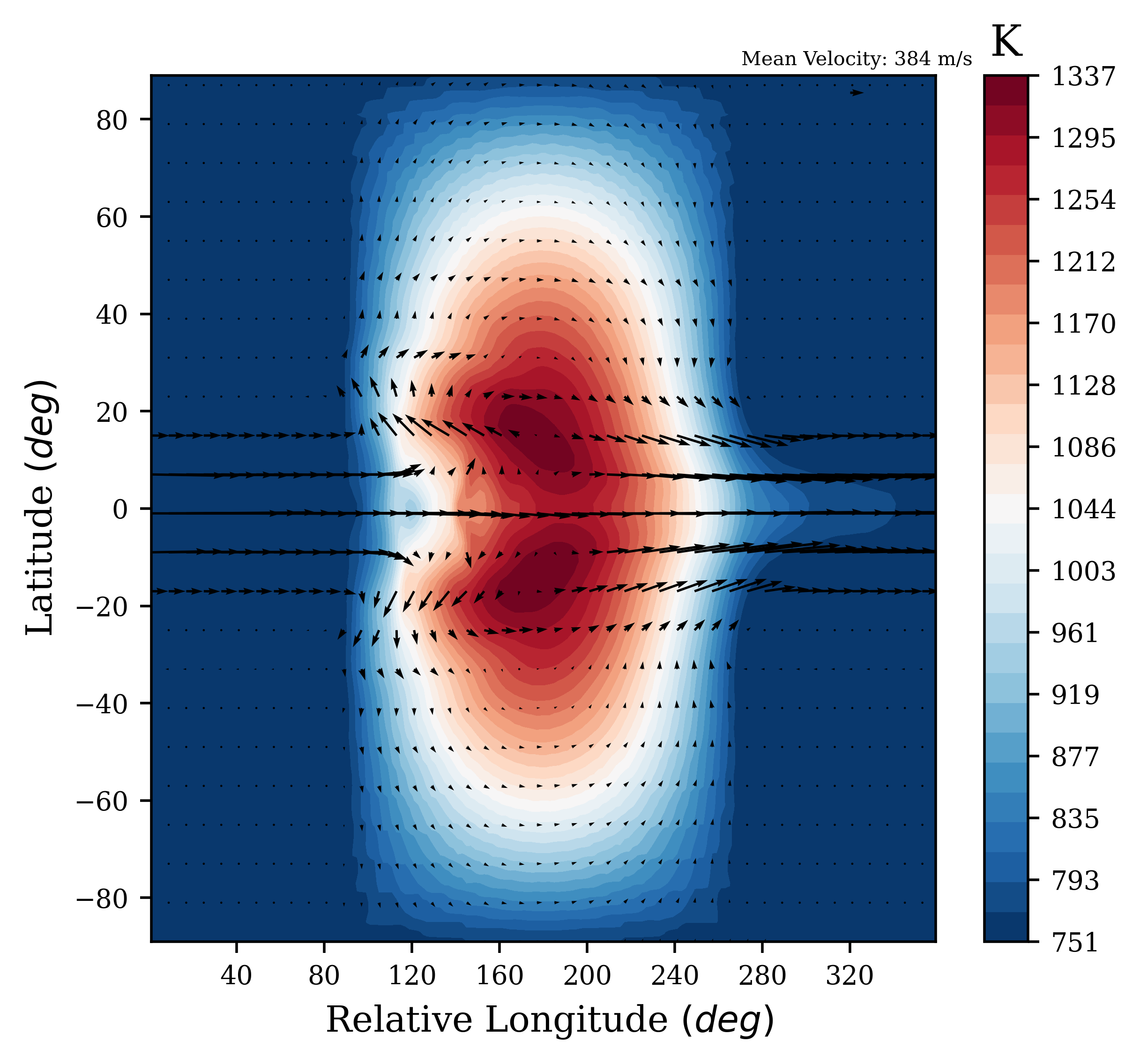}
\caption[]{$15\Omega_{0}$ - 0.0026 bar   \label{fig:Wind_Temp_15_00026} }
\end{centering}
\end{subfigure}
\begin{subfigure}{0.33\textwidth}
\begin{centering}
\includegraphics[width=0.8\columnwidth]{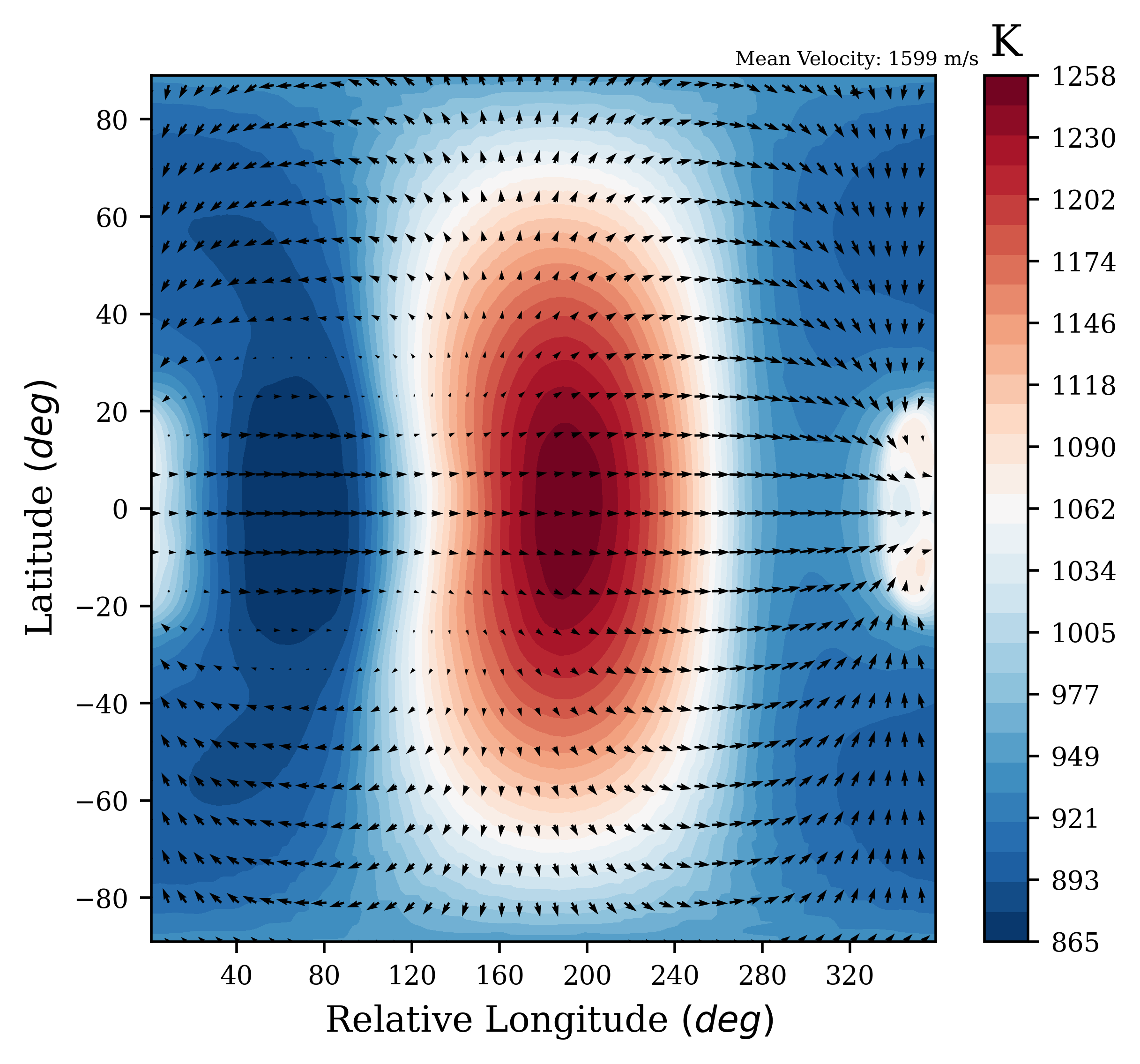}
\caption[]{$0.125\Omega_{0}$ - 0.016 bar  \label{fig:Wind_Temp_0.125_0016} }
\end{centering}
\end{subfigure}
\begin{subfigure}{0.33\textwidth}
\begin{centering}
\includegraphics[width=0.8\columnwidth]{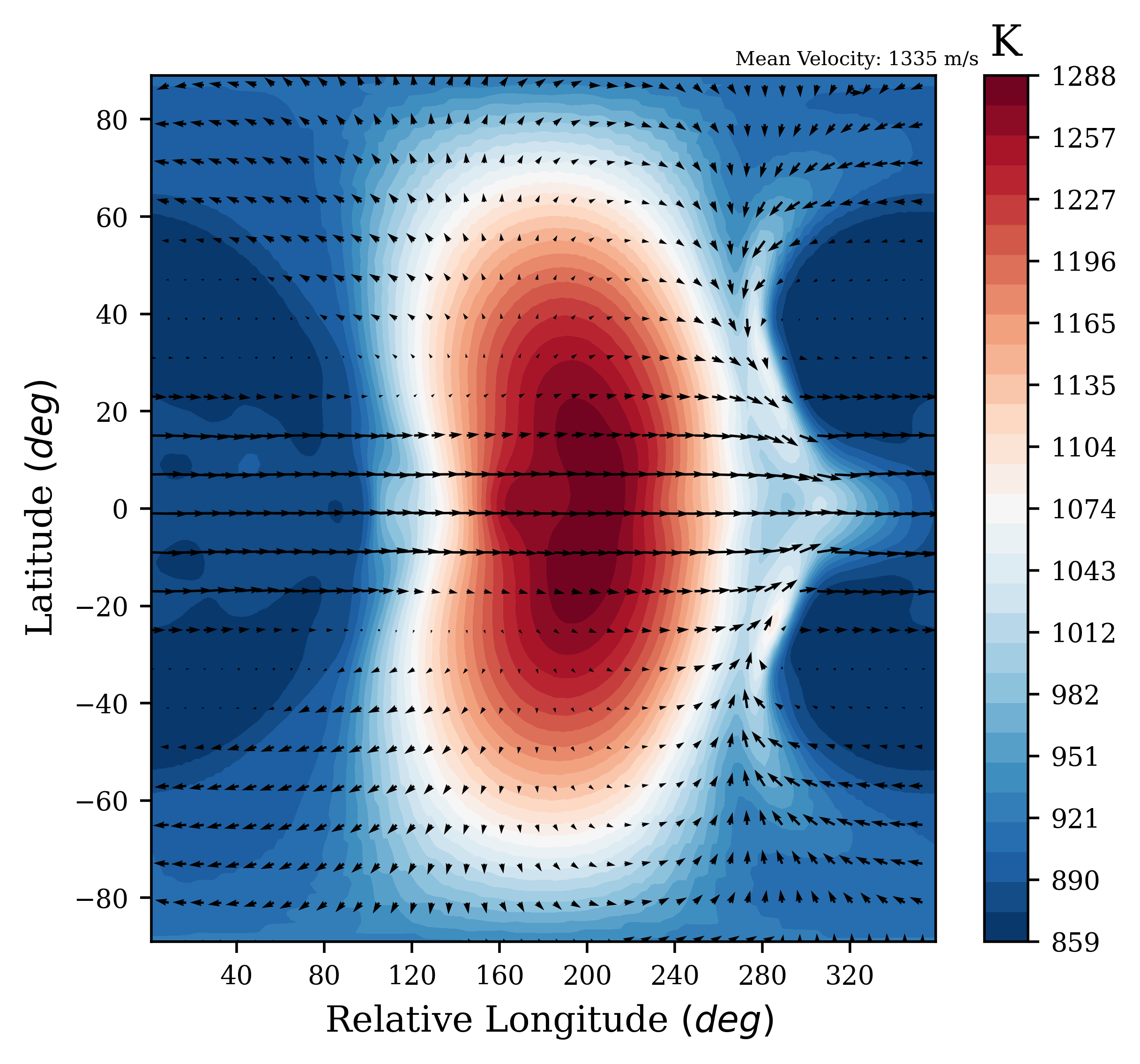}
\caption[]{$1\Omega_{0}$ - 0.016 bar   \label{fig:Wind_Temp_1_0016} }
\end{centering}
\end{subfigure}
\begin{subfigure}{0.33\textwidth}
\begin{centering}
\includegraphics[width=0.8\columnwidth]{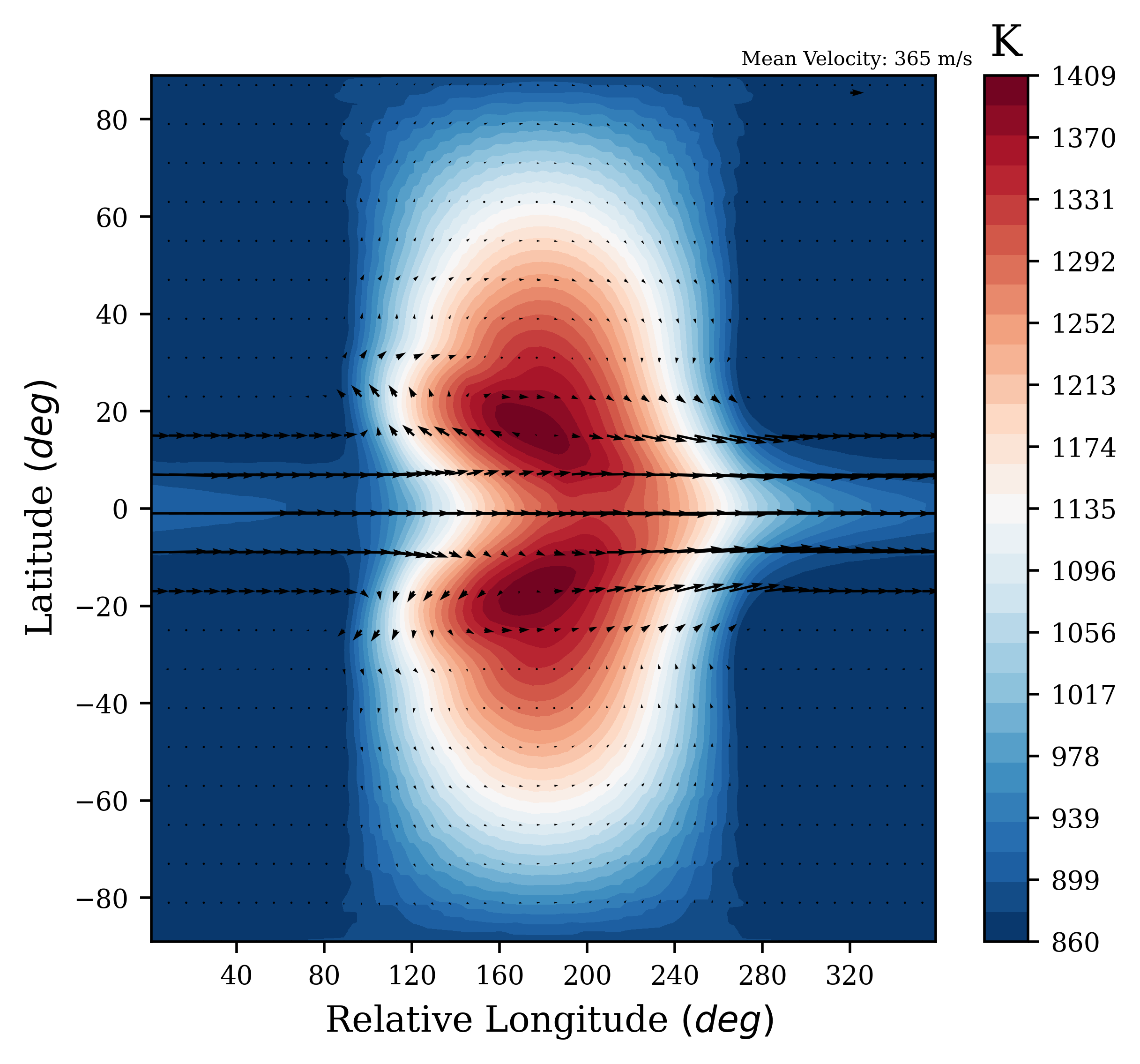}
\caption[]{$15\Omega_{0}$ - 0.016 bar   \label{fig:Wind_Temp_15_0016} }
\end{centering}
\end{subfigure}
\begin{subfigure}{0.33\textwidth}
\begin{centering}
\includegraphics[width=0.8\columnwidth]{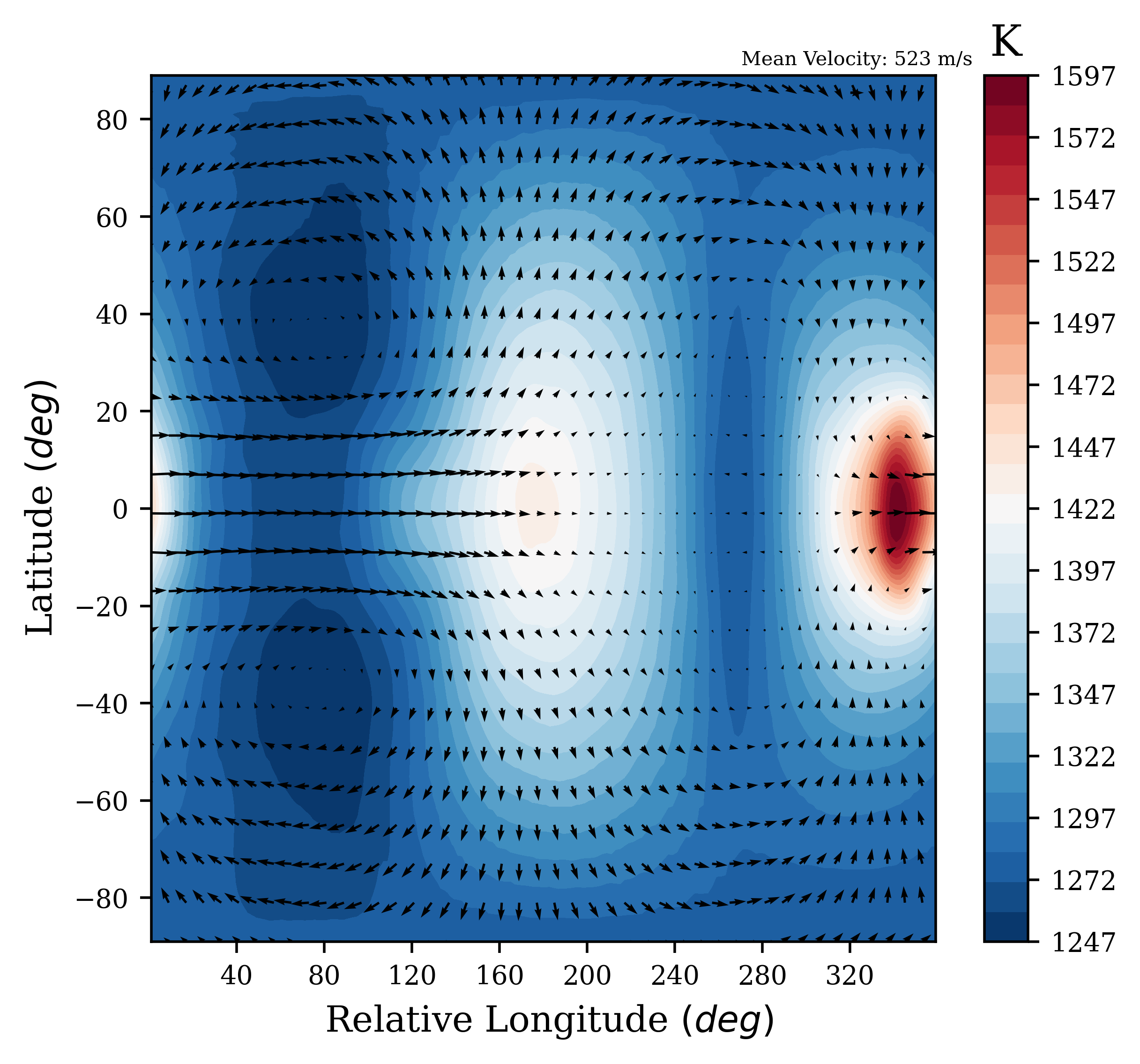}
\caption[]{$0.125\Omega_{0}$ - 0.2 bar  \label{fig:Wind_Temp_0.125_02} }
\end{centering}
\end{subfigure}
\begin{subfigure}{0.33\textwidth}
\begin{centering}
\includegraphics[width=0.8\columnwidth]{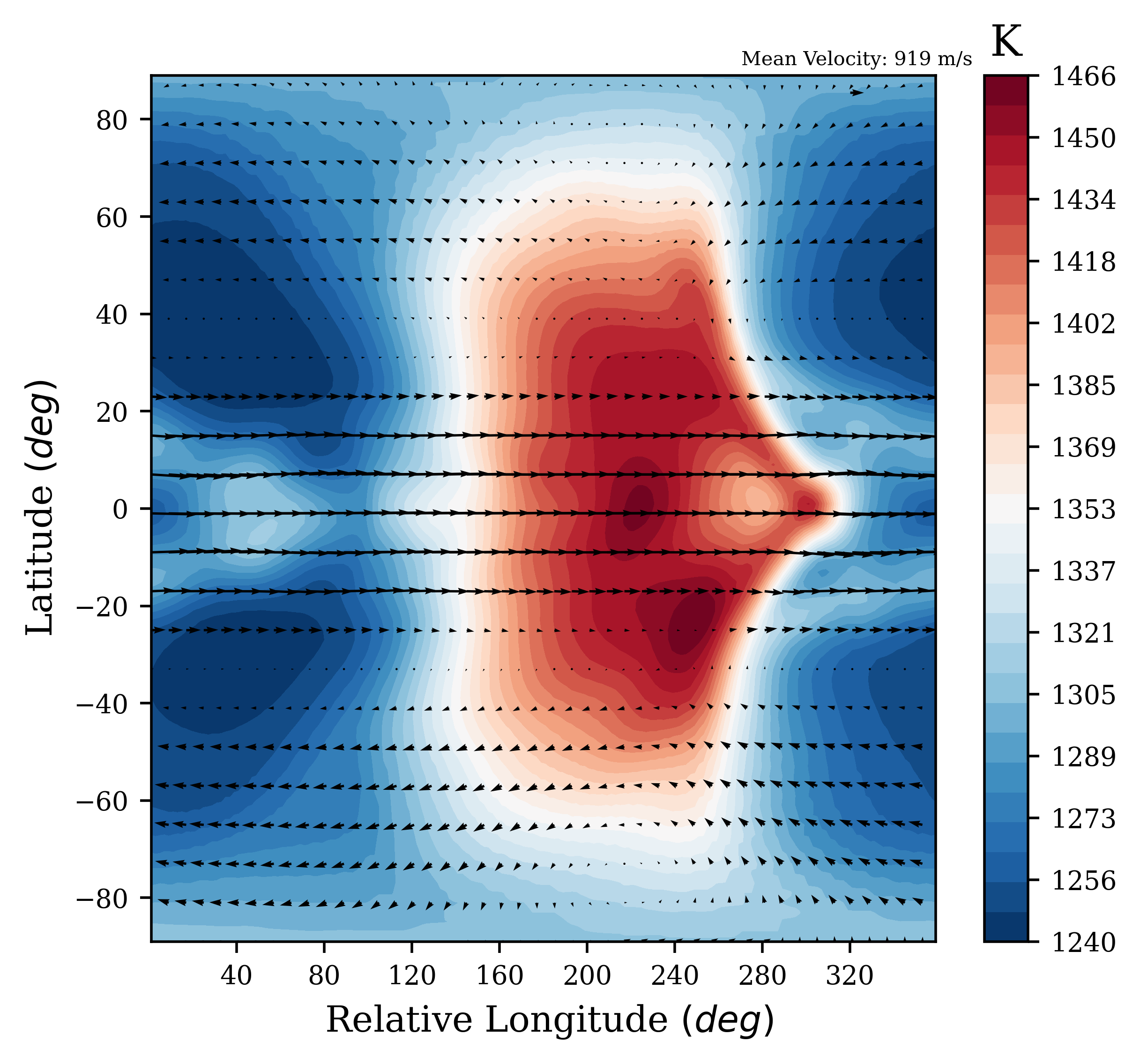}
\caption[]{$1\Omega_{0}$ - 0.2 bar   \label{fig:Wind_Temp_1_02} }
\end{centering}
\end{subfigure}
\begin{subfigure}{0.33\textwidth}
\begin{centering}
\includegraphics[width=0.8\columnwidth]{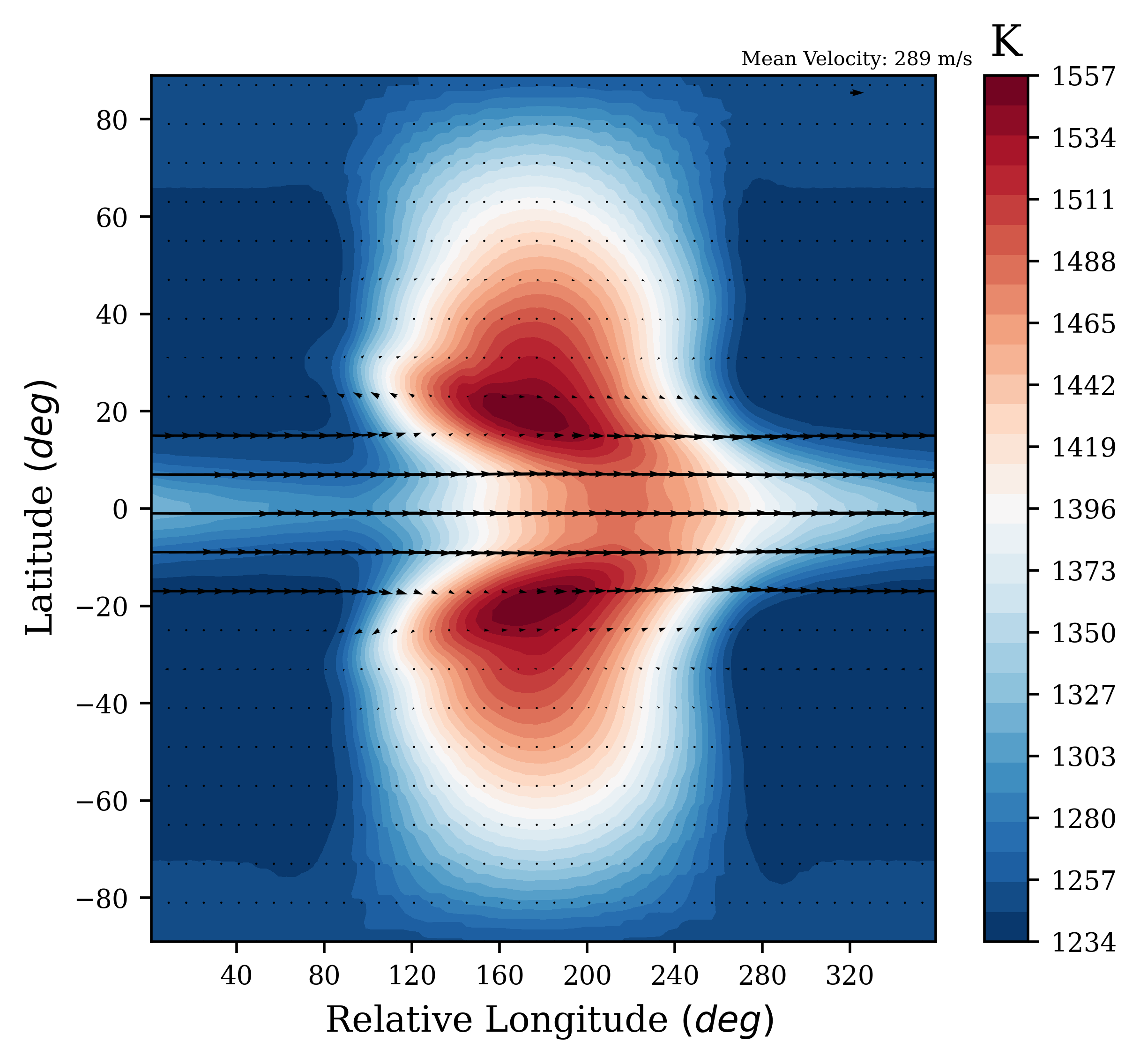}
\caption[]{$15\Omega_{0}$ - 0.2 bar   \label{fig:Wind_Temp_15_02} }
\end{centering}
\end{subfigure}
\caption[Temperature-wind maps at four different pressures for three HD209458b-like models at different rotation rates]{ Temporally averaged, { over five Earth-years of simulation time,} zonal wind (quivers) and temperature maps at three different pressures (0.0026 bar - top - to 0.2 bar - bottom) for three HD209458b-like models at different rotation rates ranging from $0.125\Omega_{0}$ - left - to $15\Omega_{0}$ - right. { For each pressure level and rotation rate, the { RMS} wind speed is shown in the top right.} \label{fig:Wind_Temp} }
\end{centering}
\end{figure*} 

\subsection{Horizontal-Wind and Temperature Maps} \label{sec:horizontal_wind_temp}
{ In order to better understand the dynamics discussed above, we next investigate the horizontal wind and temperature { maps} at select pressure levels for three HD209458b-like models which fall into our three regimes of interest, as shown in \autoref{fig:Wind_Temp}. Note that pressure levels have been chosen so as to emphasise the difference between the dynamical regimes (i.e. the differences associated with varying both rotation rate and relative radiative/advective dynamical strength).} \\

We start our analysis with the slowly rotating regime (left column of \autoref{fig:Wind_Temp}). Here, in the outer atmosphere, \autoref{fig:Wind_Temp_0.125_00026}, we find that the horizontal wind primarily flows from the day-side to the night-side with almost no asymmetry - i.e. an almost isotropic divergence of the wind on the hot day-side away from the substellar point and convergence of wind on the cold night-side. However almost isotropic does not mean completely: we find that the easterly flow from the day to night-side is faster than its westerly equivalent, and that the convergence point of the winds on the night-side is west of the anti-stellar point. Both of these effects can be linked to the, weak, influence of rotation on the atmosphere, either in the form of standing waves attempting to drive an equatorial jet (and hence introducing a preferred direction to flows), or the preferential enhancement/suppression of flows with/against the direction of rotation.  \\
{ Moving deeper into the atmosphere( \autoref{fig:Wind_Temp_0.125_0016}) we start} to find evidence that horizontal advection is shaping the thermal structure as the influence of radiation wanes.
Of particular note are the equatorial winds, which are now { eastwards} at all longitudes, although the influence of divergent day-night winds remains with the equatorial winds being significantly slower just west of the substellar point { (see \autoref{sec:helmholtz_wind} for more details)}. 
We also catch our first glimpse of a rather unusual feature that is exclusive to the slowly rotating regime: a slight hot-spot on the night-side, where the divergent winds converge (as we discuss in \autoref{fig:Enthalpy_Wind}, this is not a coincidence). \\
{ This night-side hot-spot grows in strength as we move towards mid-pressures} ($0.2\si{\bar}$ - \autoref{fig:Wind_Temp_0.125_02}), { resulting in a} rather unusual wind and temperature structure: the hottest region of the atmosphere has moved from the day-side to just west of the anti-stellar point, where the divergent day-night winds converge. Further, this night-side hot-spot, and the resulting temperature gradients, { drive} a significant equatorial wind from the night-side to the day-side, which appears in the zonal-mean zonal-wind profile as part of the jet, despite it not forming in the same way as the classical hot Jupiter zonal-mean equatorial jet. { In part, these differences may occur due to differences in the behaviour of Rossby waves, and barotropic and baroclinic instabilities, as the Rossby deformation radius approaches the planetary radius \citep{2010JGRE..11512008M}.} \\
If robust, the formation of this night-side hot spot, which occurs for all of our HD209458b-like models that fall into the slowly rotating regime, may have important implications for our understanding of the dynamics of such objects. For example, as discussed in \autoref{sec:enthalpy_transport} and shown in \autoref{fig:Longitudinal_T_0125x}, the night-side hot-spot is associated with a thermal inversion, the presence of which might have strong implications for both the observed phase curve (depending upon the wavelength/molecule observed, and hence the pressure probed) as well as the chemistry, since the inversion can act as a cold-trap, leading to condensation/rain-out and hence changing the, observable, composition of the atmosphere (which in turn may effect the thermal structure due to changes in absorption/emission). Note that, as discussed above, this feature is common to all our slowly rotating regime models, and, in fact, its presence is almost a defining feature of the slowly rotating regime. This is because, as $\Omega$ increases, rotational influences shape the day-night flow, shifting the hot-spot even further westwards, away from the anti-stellar point and towards the terminator, until it eventually combines with the { eastwards} advected day-side hot-spot, leading to the formation of the thermal butterfly that is emblematic of the classical hot-Jupiter regime { at a transitional rotation rate of around} $\Omega=0.5\rightarrow0.6\Omega_{0}$. { This shift in the location of the hot-spot is in agreement with the Spitzer phase curve observations analysed by \citet{2022AJ....163..256M}, which revealed (as shown in their Figure 10) an increasing trend of hot-spot off set with increasing orbital period (i.e. decreasing synchronous rotation rate). They even found a slight westward shift at the most rapid rotation rates, a shift which might correspond to the off-equator westwards advection that we find in the rapidly rotating regime (see below).  }  \\

{ Moving into the classical hot Jupiter regime, which we plot on the middle column of \autoref{fig:Wind_Temp}, we find dynamics that are in good agreement with previous studies of hot Jupiters, such as \citet{2019A&A...632A.114S}.}\\
{ For example, in the outer atmosphere (\autoref{fig:Wind_Temp_1_00026}),} we find slight advection from the day-side towards the eastern terminator, forming `wings' that: a) trace the off-equator divergent/convergent flows, which are increasingly rotationally influenced, and hence have become longitudinally asymmetric, with eastern flows compressed and, higher latitude, western flows stretched longitudinally; and b) will eventually form, deeper in the atmosphere, a key part of the thermal butterfly structure that is synonymous with the classical regime. { Note that this structure is also known as a global `chevron' pattern \citep{2011ApJ...738...71S}.}
We also find that, unlike in the slowly rotating regime, super-rotating flows are present at all longitudes on the equator, albeit with the proviso that winds west of the sub-stellar point are slightly suppressed/weakened by the divergent component of the wind (see \autoref{sec:helmholtz_wind}). \\
{ Evidence for the formation of foreshadowed thermal butterfly increases as we move deeper}, \autoref{fig:Wind_Temp_1_0016}, with the influence that rotation has on the dynamics also becoming increasingly clear. For example, the high latitude flows have become increasingly asymmetric, shifting the `wings' associated with convergent flows closer towards the terminator and the day-side hot-spot. Together { with} the growing strength of the equatorial jet at all longitudes (although it does remain slightly weaker just west of the sub-stellar point) which drives the eastward advection (by approximately $20^{\circ}$) of the near equator day-side hot-spot, we are starting to see the basis of the structure that will form the thermal butterfly. \\
A structure that becomes apparent as we move into the mid atmosphere ($0.2\si{\bar}$ - \autoref{fig:Wind_Temp_1_02}), where we find that the day-side hot-spot is now located $\sim50^{\circ}$ east of the sub-stellar point. Interestingly, we also find a secondary peak in the temperature even further east, across the terminator and on the night-side - this is a remnant of the `wings' seen at lower pressures and can also be considered as a sign that the off-equator heat transport was sufficiently rotationally influenced as to prevent the formation of the night-side hot-spot found in the slowly rotating regime.
Further, whilst the off-equator flows are mostly hidden by the equatorial jet, the visible remnants suggest that, at high latitudes, the easterly circulation that previously had been confined to the day-side has been almost completely suppressed, leaving only a westerly day-night flow, which appears in the zonal-mean zonal-wind as a counterflow (\autoref{fig:Zonal_Wind_1}).  \\
Note that, as we increase the rotation rate in the classical regime, the influence of rotation grows, with off-equator flows playing an increasing role in shaping the circulation, leading to westward and poleward advection at mid-latitudes and more distinct `wings' to the thermal butterfly. \\

This rotational influence becomes increasingly apparent as we move into the rapidly rotating regime, which we plot on the right-hand column of \autoref{fig:Wind_Temp}. \\
Here, in the outer atmosphere (\autoref{fig:Wind_Temp_15_00026}) we find that the off-equator wind dynamics are very strongly suppressed {(by the Coriolis effect)}, particularly on the cold night-side, resulting in only a weak day-side circulation from west of the sub-stellar point to the east via mid-latitudes. { This leaves the strong super-rotating equatorial jet as the dominant flow, but even this is influenced by rotation, with the jet being visibly narrower than its classical regime counterpart. }
Consequently, this highly rotationally-confined wind structure drives thermal advection that is limited to equatorial regions, with weak eastwards advection at the equator (weak due to the reduced jet speed that latitudinally confined standing waves can drive), weak westwards advection at low latitudes, and basically no heat transport at high latitudes (i.e. above $\pm\sim30^{\circ}$). { This weak low-latitude westwards advection may explain the slight westward phase curve offset found for short orbital period hot Jupiters by \citet{2022AJ....163..256M}}. \\
Moving deeper, \autoref{fig:Wind_Temp_15_0016} and \autoref{fig:Wind_Temp_15_02}, we find a very similar structure but with one significant difference: As the radiative forcing weakens, the eastwards equatorial and westwards low-latitude advection strengthens, leading to stronger zonal mixing, such as the band of equatorial day-night heating found at 0.2 bar (\autoref{fig:Wind_Temp_15_02}). { Note that as we move even deeper into the atmosphere, this zonal mixing grows, leading to an equatorial warm band between 5 and $20\si{\bar}$. } \\
Note that, at even faster rotation rates, we find that both the equatorial jet, as well as the low-latitude counterflows become increasingly latitudinally confined, which has implications for not only the thermal structure, but also the strength of the zonal winds (as we discuss in \autoref{sec:helmholtz_wind}), and hence the vertical transport of potential-temperature/enthalpy (as we discuss in \autoref{sec:enthalpy_transport}). \\

\begin{figure*}[tbp] %
\begin{centering}
\begin{subfigure}{0.33\textwidth}
\begin{centering}
\includegraphics[width=0.99\columnwidth]{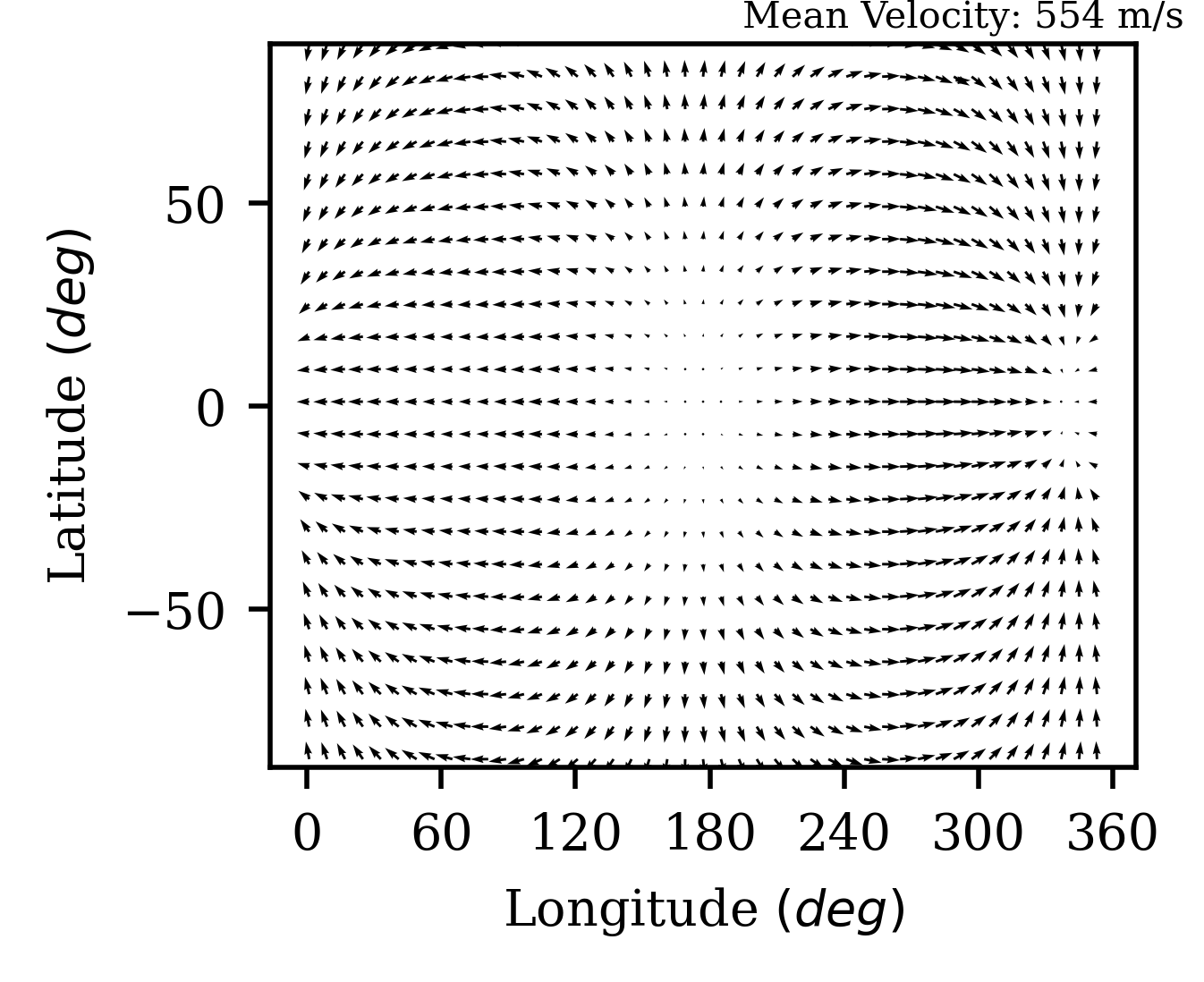}
\caption[]{$0.125\Omega_{0}$ - Divergent  \label{fig:Helmholtz_0.125_div} }
\end{centering}
\end{subfigure}
\begin{subfigure}{0.33\textwidth}
\begin{centering}
\includegraphics[width=0.99\columnwidth]{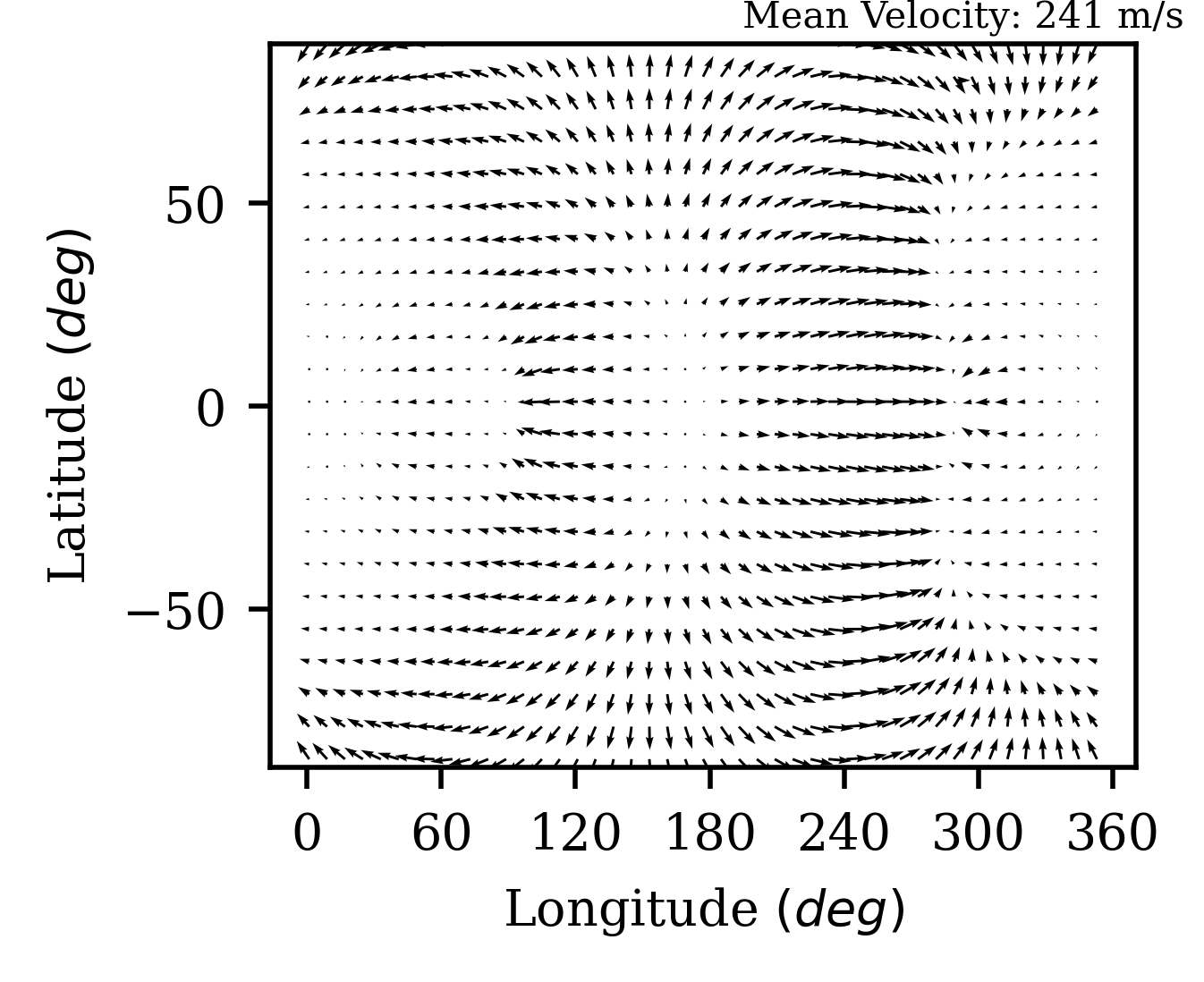}
\caption[]{$1\Omega_{0}$ - Divergent   \label{fig:Helmholtz_1_div} }
\end{centering}
\end{subfigure}
\begin{subfigure}{0.33\textwidth}
\begin{centering}
\includegraphics[width=0.99\columnwidth]{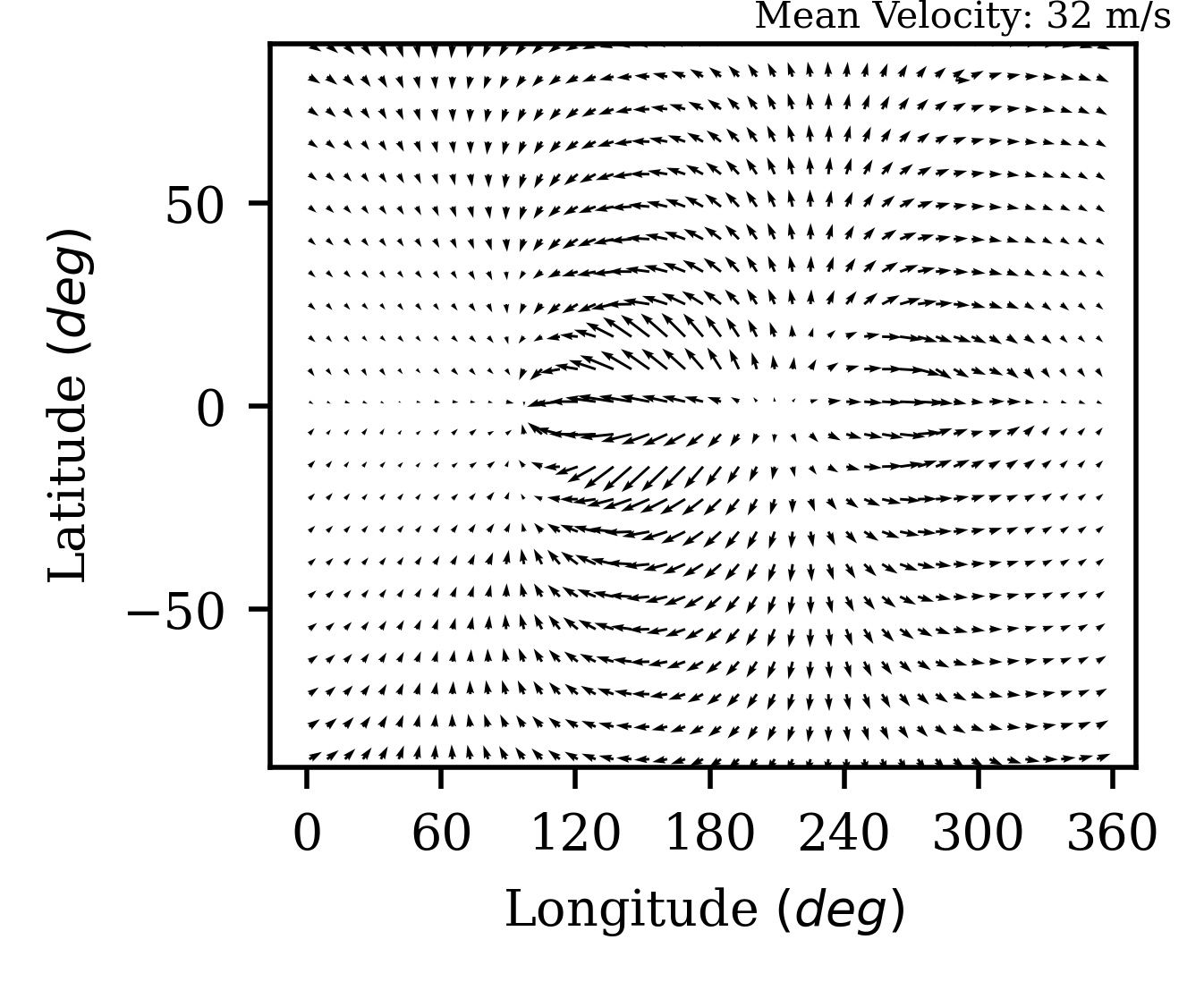}
\caption[]{$15\Omega_{0}$ - Divergent   \label{fig:Helmholtz_15_div} }
\end{centering}
\end{subfigure}
\begin{subfigure}{0.33\textwidth}
\begin{centering}
\includegraphics[width=0.99\columnwidth]{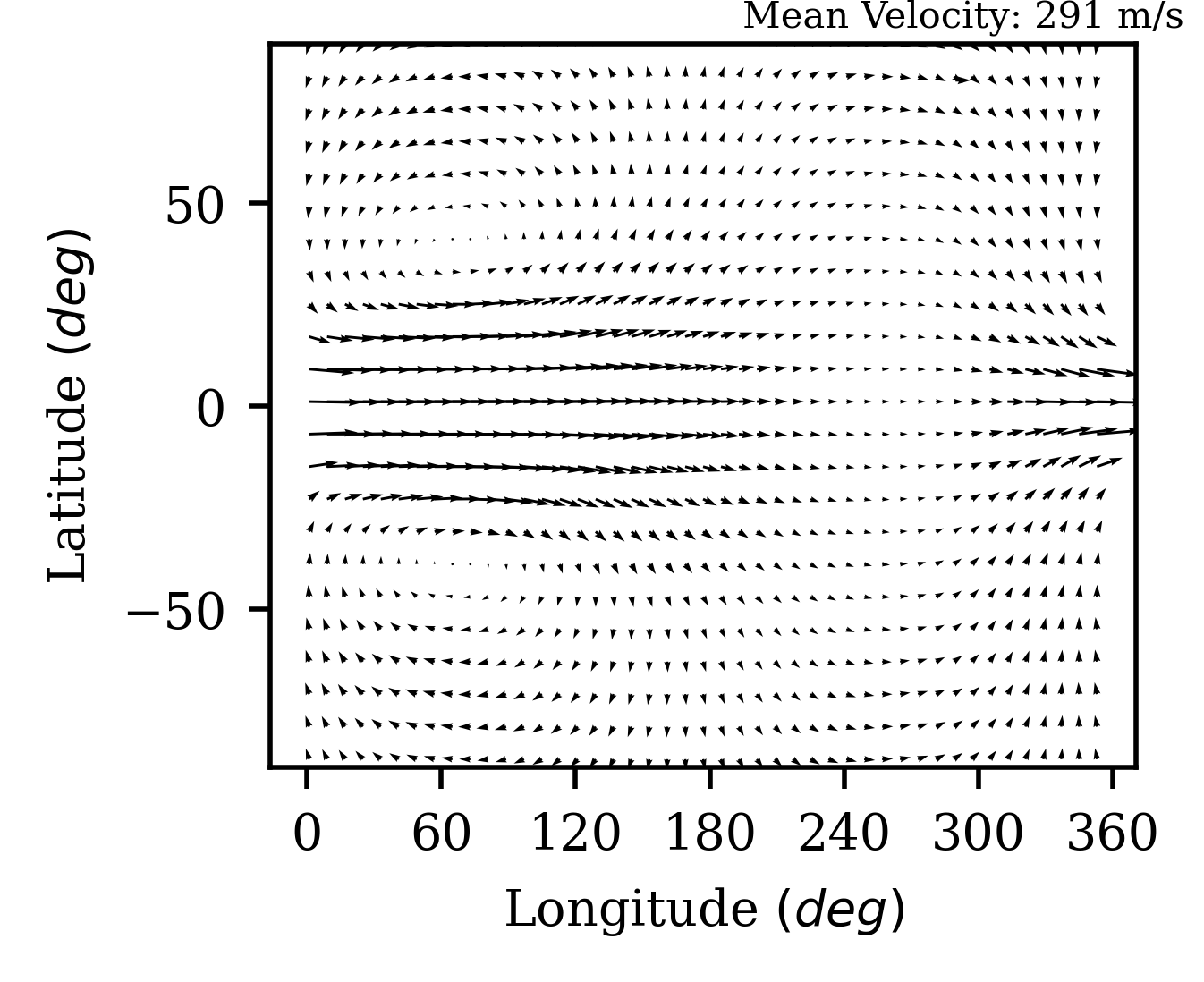}
\caption[]{$0.125\Omega_{0}$ - Rotational  \label{fig:Helmholtz_0.125_rot} }
\end{centering}
\end{subfigure}
\begin{subfigure}{0.33\textwidth}
\begin{centering}
\includegraphics[width=0.99\columnwidth]{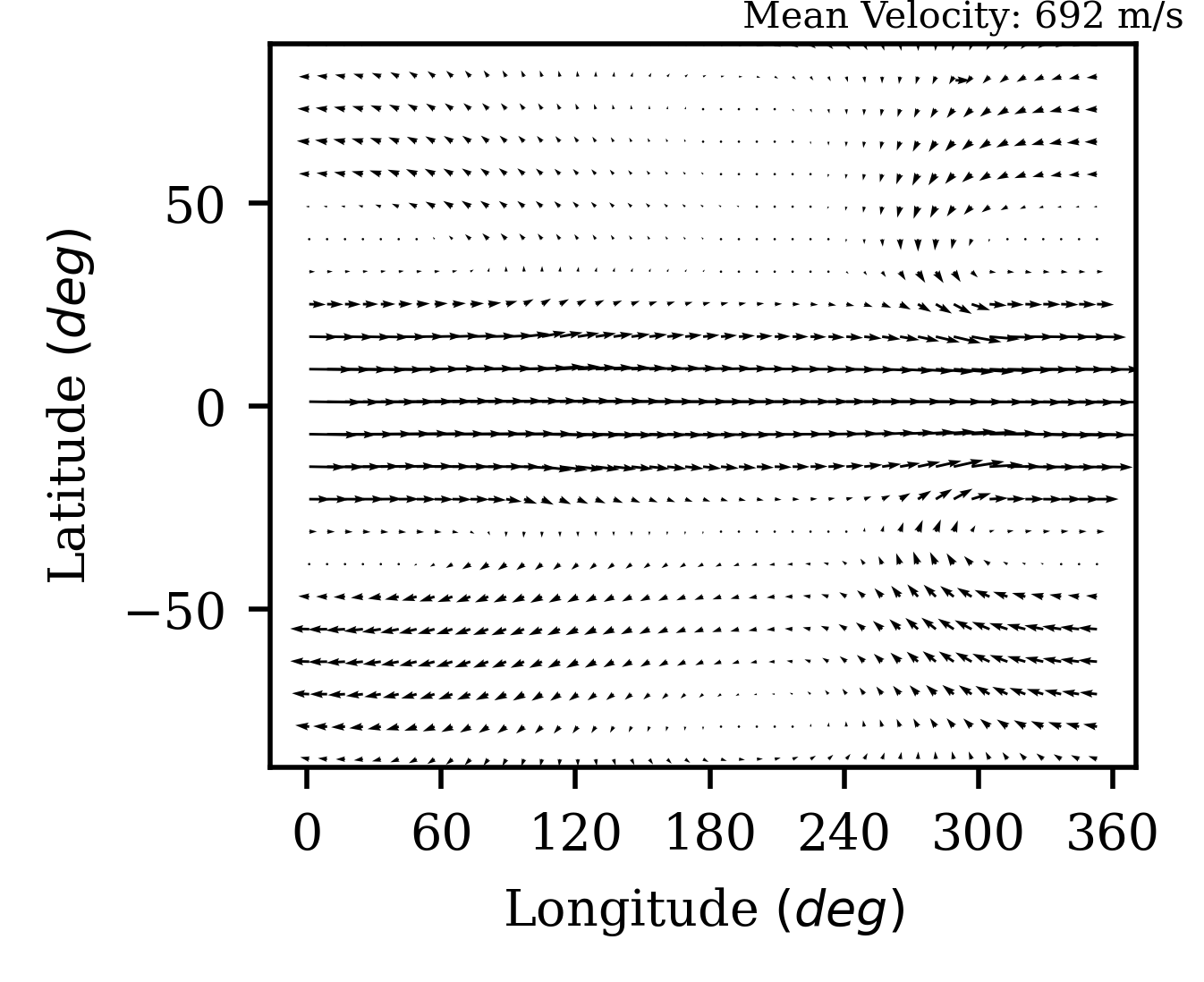}
\caption[]{$1\Omega_{0}$ - Rotational   \label{fig:Helmholtz_1_rot} }
\end{centering}
\end{subfigure}
\begin{subfigure}{0.33\textwidth}
\begin{centering}
\includegraphics[width=0.99\columnwidth]{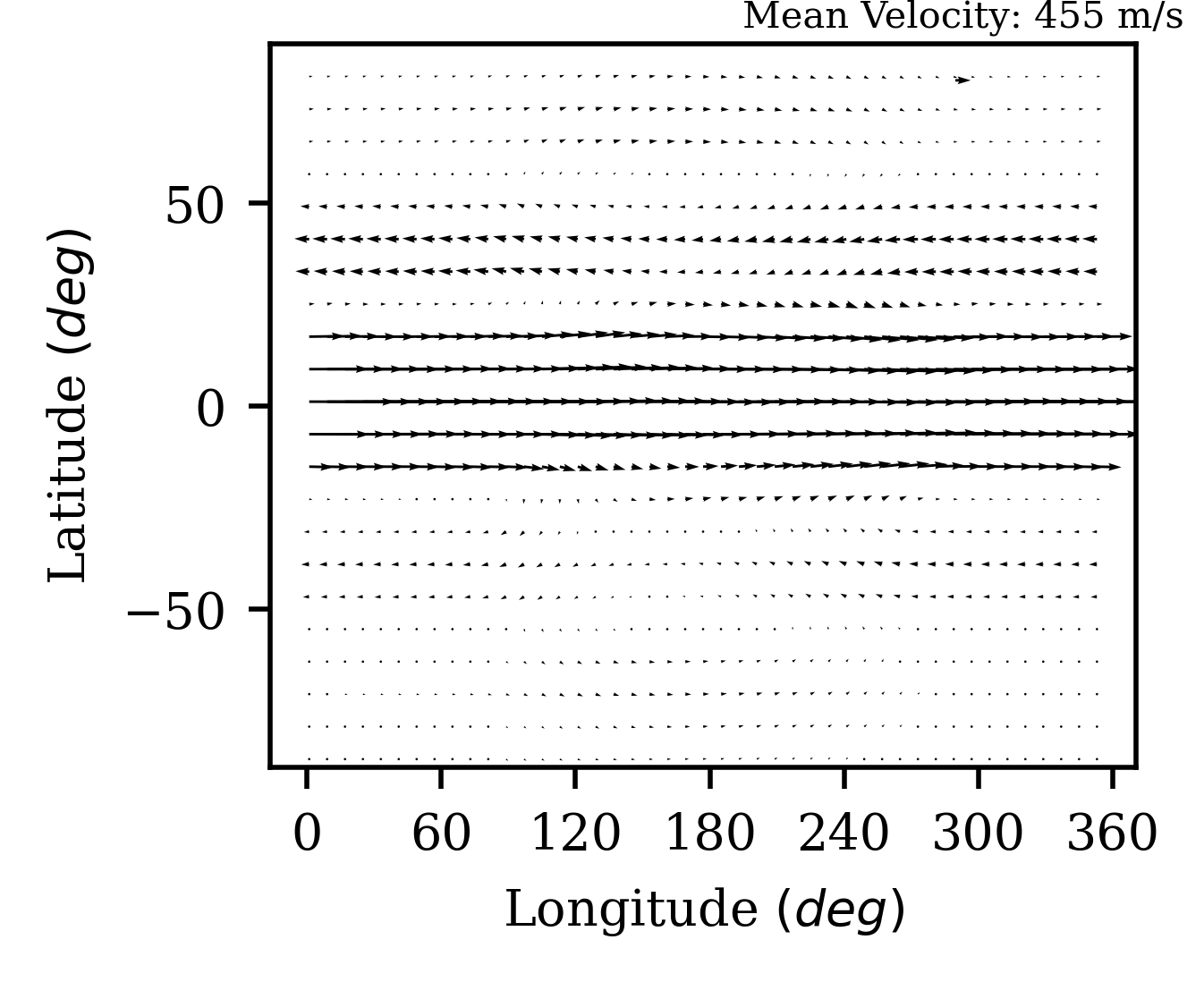}
\caption[]{$15\Omega_{0}$ - Rotational   \label{fig:Helmholtz_15_rot} }
\end{centering}
\end{subfigure}
\begin{subfigure}{0.33\textwidth}
\begin{centering}
\includegraphics[width=0.99\columnwidth]{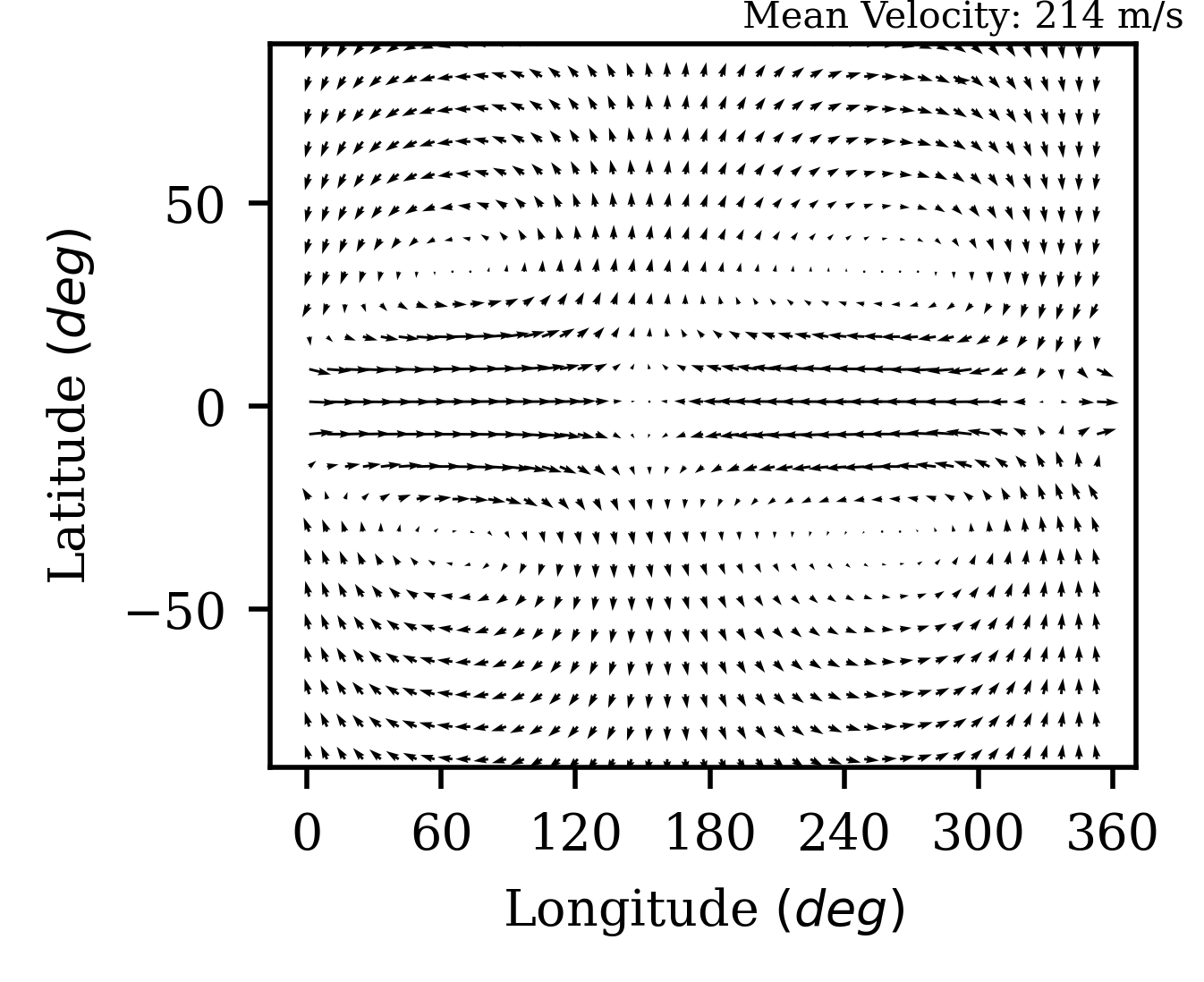}
\caption[]{$0.125\Omega_{0}$ - Eddy  \label{fig:Helmholtz_0.125_eddy} }
\end{centering}
\end{subfigure}
\begin{subfigure}{0.33\textwidth}
\begin{centering}
\includegraphics[width=0.99\columnwidth]{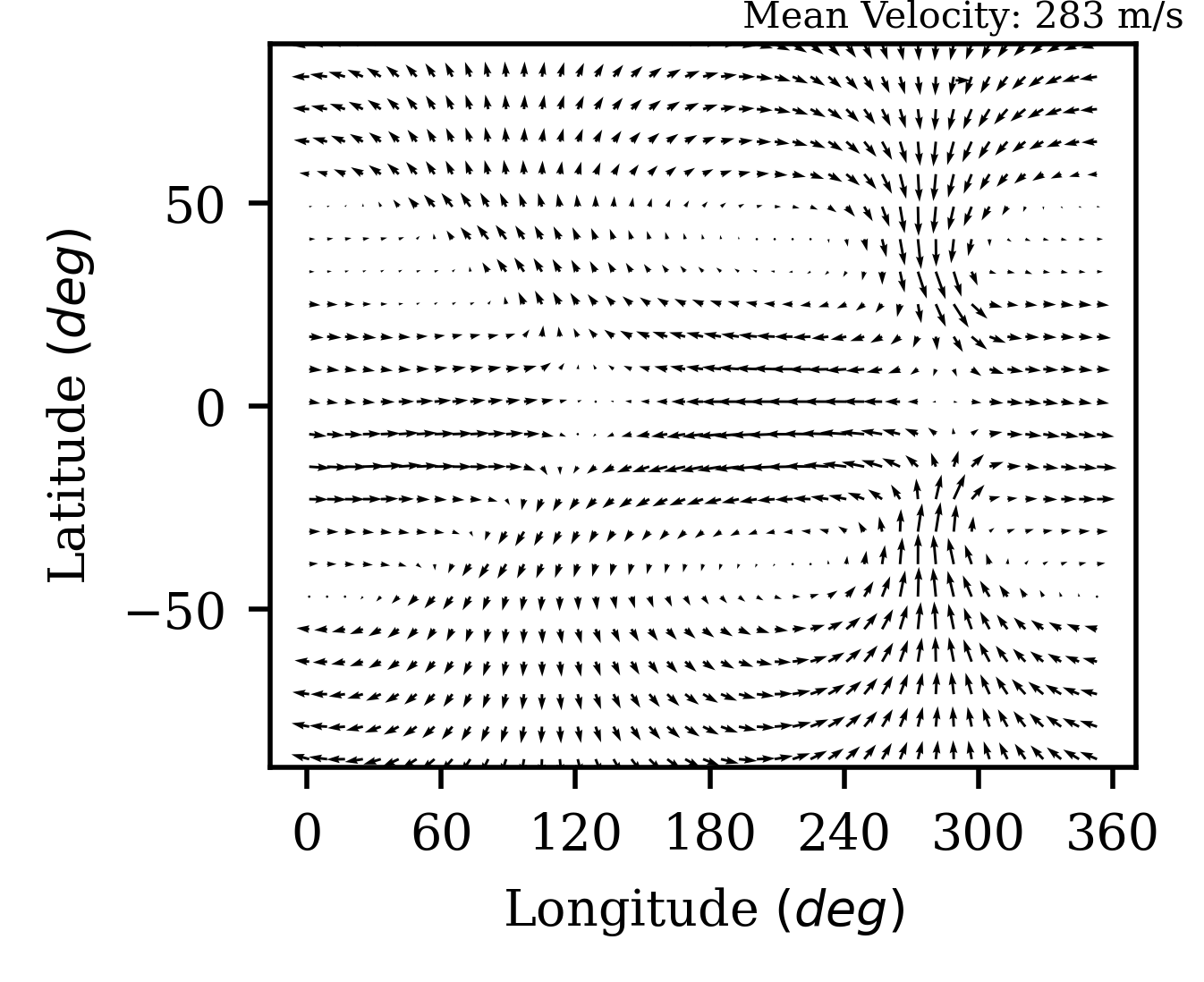}
\caption[]{$1\Omega_{0}$ - Eddy   \label{fig:Helmholtz_1_eddy} }
\end{centering}
\end{subfigure}
\begin{subfigure}{0.33\textwidth}
\begin{centering}
\includegraphics[width=0.99\columnwidth]{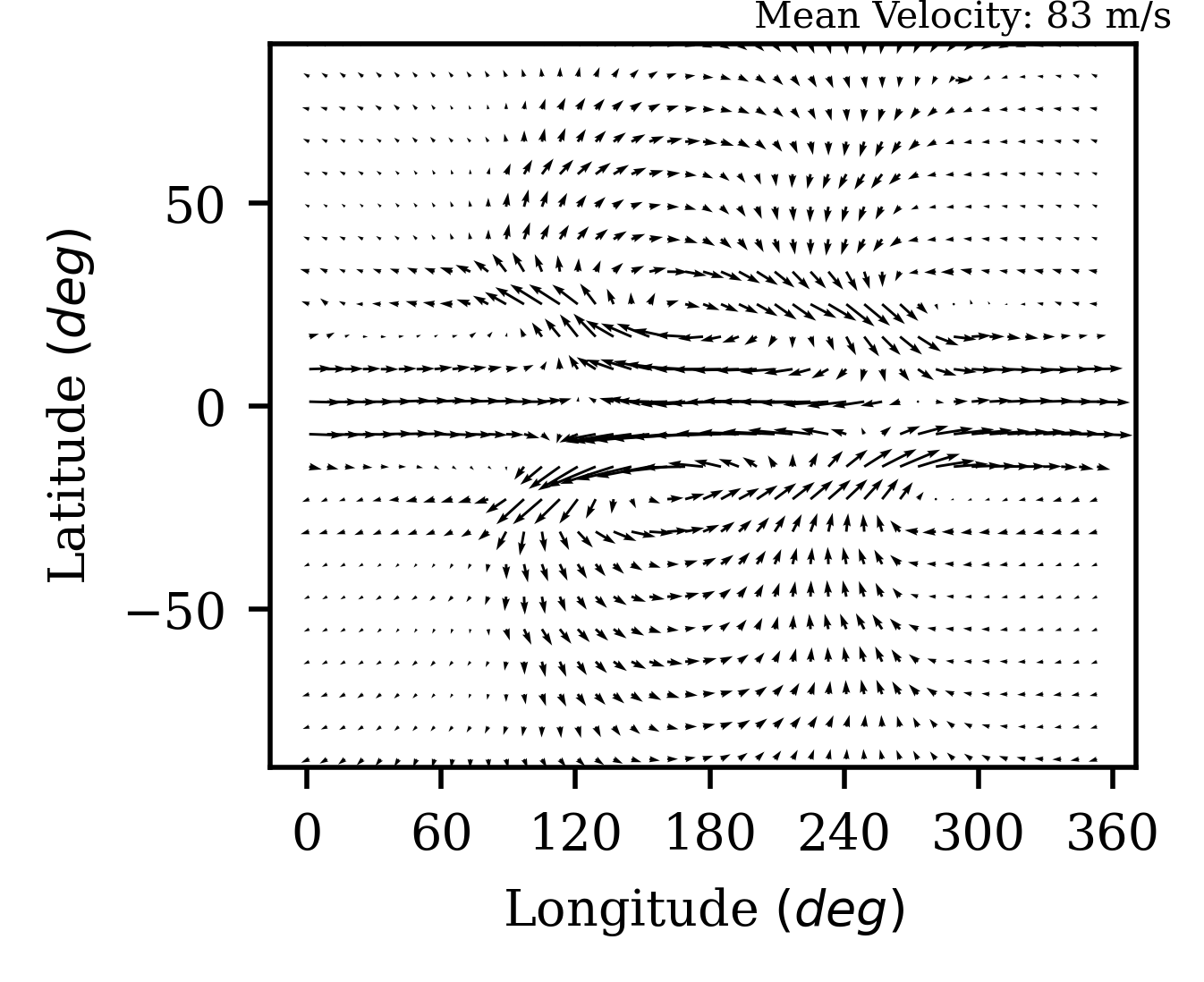}
\caption[]{$15\Omega_{0}$ - Eddy   \label{fig:Helmholtz_15_eddy} }
\end{centering}
\end{subfigure}

\caption[Helmholtz decomposition of the horizontal velocity for three synchronised models of HD209]{Helmholtz decomposition of the radially averaged (over the outer and mid atmosphere) horizontal wind of three, synchronous, HD209458b-like atmospheric models with rotation rates ranging from 0.125$\Omega_{0}$ to 15$\Omega_{0}$. The top row plots the divergent component ($H_{div}$) of the Helmholtz decomposition, the middle row plots the rotational component ($H_{rot}$) of the decomposition, and the bottom row shows the eddy component ($H_{eddy} = H_{rot} - \left<H_{rot}\right>$) of the decomposition. Note the change (decrease) in { RMS mean} velocity as the rotation rate increases.  \label{fig:Helmholtz} }
\end{centering}
\end{figure*} 

\subsection{Helmholtz Wind Decomposition} \label{sec:helmholtz_wind}
Our above analysis has revealed that the horizontal winds are highly sensitive to both rotation rate and pressure (which acts as a proxy for the relative influence of radiative forcing to advective transport). In order to disentangle these effects, we turn to the Helmholtz wind decomposition, which has long been a staple of Earth atmospheric studies \citep{dutton1986ceaseless}, and which has recently been applied to both hot Jupiters \citep{2021PNAS..11822705H} and tidally-locked rocky exoplanets \citep{2022PSJ.....3..214S}. \\
Briefly, a Helmholtz decomposition can be used to split the horizontal wind at a select pressure level (or over an averaged pressure range), $\bm{u} = (u,v)$, into divergent (i.e. `vorticity free') and rotational (i.e. `divergence free') components \citep{dutton1986ceaseless}: 
\begin{align}
  \bm{u} &= \bm{u}_{d} + \bm{u}_{r} \\
  &= -\bm{\nabla}\chi + \bm{k}\times\bm{\nabla}\psi,
\end{align}
where $\bm{k}$ denotes the unit vector in the (eastward) zonal direction, $\chi$ is the velocity potential function, $\psi$ is the velocity stream-function, and both $\chi$ and $\psi$ can be linked to the divergence $\delta$ / vorticity $\mathcal{W}$ directly:
\begin{align}
  \nabla^{2}\chi &= \delta\\
  \nabla^{2}\psi &= \mathcal{W}.
\end{align}
Additionally, in order to isolate the equatorial super-rotating jet from any other rotational wind dynamics it may be masking, we further split the rotational component, $\bm{u}_r$ into a zonal-mean component $\bm{u}_{z}$ and an eddy-wind component $\bm{u}_{e}$:
\begin{align}
  \bm{u}_{z} &= \left<\bm{u}_{r}\right>\\
  \bm{u}_{e} &= \bm{u}_{r} - \bm{u}_{z},
\end{align}
where $\left<\right>$ indicates the zonal-mean.\\
As for exactly what these components represent for a tidally-locked atmosphere: $\bm{u}_d$ represents flows that diverge from the hot-spot on the day-side and converge somewhere on the cold night-side, forming a closed cycle when combined with an upwelling below the day-side hot-spot and the resulting downwelling on the nightside; $\bm{u}_r$ represents dynamics driven by angular momentum transport via stationary Rossby and Kelvin waves - in typical hot Jupiters these standing waves transport angular momentum from mid-latitudes to the equator, resulting in slight westward flows at mid-latitudes and a super-rotating flow (jet) at the equator \citep{2011ApJ...738...71S}; and finally, as mentioned above, $\bm{u}_e$ and $\bm{u}_z$ represent a further decomposition of $\bm{u}_r$, allowing us to explore the transport by standing waves in cases where the presence of a super-rotating equatorial jet would completely dominate the dynamics. \\

In \autoref{fig:Helmholtz}, we plot $\bm{u}_d$ (top), $\bm{u}_r$ (middle), and $\bm{u}_e$ (bottom), for the three different rotation rates discussed in \autoref{sec:zonal_winds}, in order to investigate how the balance between different components of the radially-averaged { (over the mid/outer atmosphere where $P<1$ bar)} horizontal winds changes with rotation rate. \\

We start by analysing the winds in the slowly rotating regime, which is plotted in the left column of \autoref{fig:Helmholtz}. Here, our analysis reveals that, in the outer and mid atmosphere, the primary component which make up the total horizontal wind is the divergent component { ($\left|\bm{u}_{d}\right|=554\mathrm{\,ms^{-1}}$)}, which is, on average, a factor of two faster than the rotational component { ($\left|\bm{u}_{r}\right|=279\mathrm{\,ms^{-1}}$)}. This makes sense, and is in good agreement with the wind structure discussed in \autoref{sec:zonal_winds}. To start, off-equator, we find flows that trace the dominant, divergent, day-side to night-side flow (compare \autoref{fig:Helmholtz_0.125_div} and \autoref{fig:Wind_Temp_0.125_00026}), and which are driven by the strong radiative forcing and day-night temperature gradient. Whereas at the equator, we find that the zonal wind associated with rotational dynamics opposes the divergent wind west of the sub-stellar point whilst also reinforcing the flow to the east, leading to to observed asymmetry in equatorial wind speed. 
Decomposing the rotational wind further reveals the driving force behind this equatorial jet: as shown in \autoref{fig:Helmholtz_0.125_eddy}, the eddy component of the wind consists of an $m=1$ circulation structure (i.e. two circulation cells per hemisphere), a structure which is associated with Kelvin and Rossby standing waves and which is commonly attributed as being the driving mechanism behind super-rotating jets on hot Jupiters \citep{2011ApJ...738...71S}. Briefly, this mechanism works as follows:
longitudinal variations in the radiative forcing (i.e. the strong day/night heating/temperature contrast) trigger the formation of standing, planetary-scale, Rossby and Kelvin waves, with the Kelvin waves straddling the equator, and the Rossby waves lying at mid latitudes (with their exact location depending upon the rotation rate). { Rotational effects (i.e. Coriolis forces) then cause this standing wave pattern to tilt, with a northwest-southwest tilt in the northern hemisphere and a southwest-northeast tilt in the southern hemisphere. These tilted standing waves pump eastwards angular momentum from mid-latitudes to the equator, inducing equatorial superrotation and mid-latitude counterflows. Note that thus tilt is essential to this angular momentum transporting mechanism and hence an equatorial jet is typically} only found at more rapid rotation rates (\autoref{fig:Helmholtz_1_eddy}-l). \\

As we increase the rotation rate and move into the classical hot Jupiter regime, shown in the middle column of \autoref{fig:Helmholtz}, we find that the balance has shifted in favour of the rotational component { ($\left|\bm{u}_{r}\right|=692\mathrm{\,ms^{-1}}$, { where $\left|\bm{x}\right|$ indicates the RMS mean of $\bm{x}$})} over the divergent component { ($\left|\bm{u}_{d}\right|=241\mathrm{\,ms^{-1}}$)}, with the eddy component { (\autoref{fig:Helmholtz_1_eddy} - $\left|\bm{u}_{e}\right|=283\mathrm{\,ms^{-1}}$)} starting to exhibit signs of the titled standing wave pattern associated with a strong { rotationally influenced} zonal jet, as seen in \autoref{fig:Helmholtz_1_rot} and \autoref{sec:zonal_winds}. This dominance of rotation driven dynamics is also reflected in the thermal/wind structures shown in \autoref{fig:Wind_Temp}.
For example, comparing \autoref{fig:Helmholtz_1_rot} with \autoref{fig:Wind_Temp_1_00026}/f/g reveals the thermal butterfly formation mechanism: a strong jet drives equatorial advection whilst the $m=1$ circulations drives the formation of the off-equator temperature wings east of the sub-stellar point as well as the slightly cooler point found on the equator just behind the tip of the butterfly. Both of these features are associated with the rotational wind, with the divergent wind (\autoref{fig:Helmholtz_1_div}) only acting to slightly slowdown/speedup the jet west/east of the sub-stellar point. \\

Moving into the rapidly rotating regime, as shown on right-hand column of \autoref{fig:Helmholtz}, we start to see strong evidence of the Coriolis driven reshaping of the winds. { For example, we find that the rotational wind} { ($\left|\bm{u}_{r}\right|=377\mathrm{\,ms^{-1}}$)} completely dominates the divergent wind { ($\left|\bm{u}_{d}\right|=40\mathrm{\,ms^{-1}}$)}, leading to a divergent wind whose contributions are insignificant at all but the lowest of pressures where radiative forcing dominates.  
However this change has not come about because of an enhancement of the rotational wind (which is slower here than either slower rotation regime), rather we find that Coriolis effects have started to seriously suppress off-equator flows, essentially destroying the divergent wind and leaving a highly { latitudinally} compressed and extremely titled eddy wind { (\autoref{fig:Helmholtz_15_eddy} - $\left|\bm{u}_{e}\right|=98\mathrm{\,ms^{-1}}$)} that struggles to generate a { broad} zonal jet, but does still manage to drive at least some off-equator advection east of the sub-stellar point (e.g. \autoref{fig:Wind_Temp_15_00026}). 
Evidence for this compression can be seen in \autoref{fig:Zonal_Wind_15}, where we find that the off-equator counter flows associated with the standing wave pattern have moved to low latitudes. 
{ We also find that, off-equator,} both components of the wind are notably weaker on the night-side than the day-side, an effect which we attribute to the tilt of the eddy winds and the standing wave pattern leading to a preferred direction for circulations and essentially zero night-side flow. \\
It is worth mentioning that the compression/suppression of the off-equator flows only grows in strength as we increase the rotation rate further. This has devastating consequences { for overall zonal advection/transport}: as the standing-wave structure gets further compressed/suppressed by Coriolis forces, its ability to pump off-axis, eastwards, angular momentum towards the equator is reduced, resulting in a { slight slowdown in equatorial jet speeds, as shown in \autoref{fig:Rossby_vs_U_mean}, as well as a narrowing of the jet, as shown in \autoref{fig:Latitudinal_Wind}. Together this leads to a significant reduction is zonal transport except at the equator, the effects of which can be seen in \autoref{fig:Wind_Temp}.}
As we discuss below, this can have significant consequences for the vertical advection of enthalpy in rapidly rotating atmospheres, and may help to explain the lack of inflation observed in such objects. \\

{ 
\begin{figure*}[tbp] %
\begin{centering}
\begin{subfigure}{0.3\textwidth}
\begin{centering}
\includegraphics[width=0.99\columnwidth]{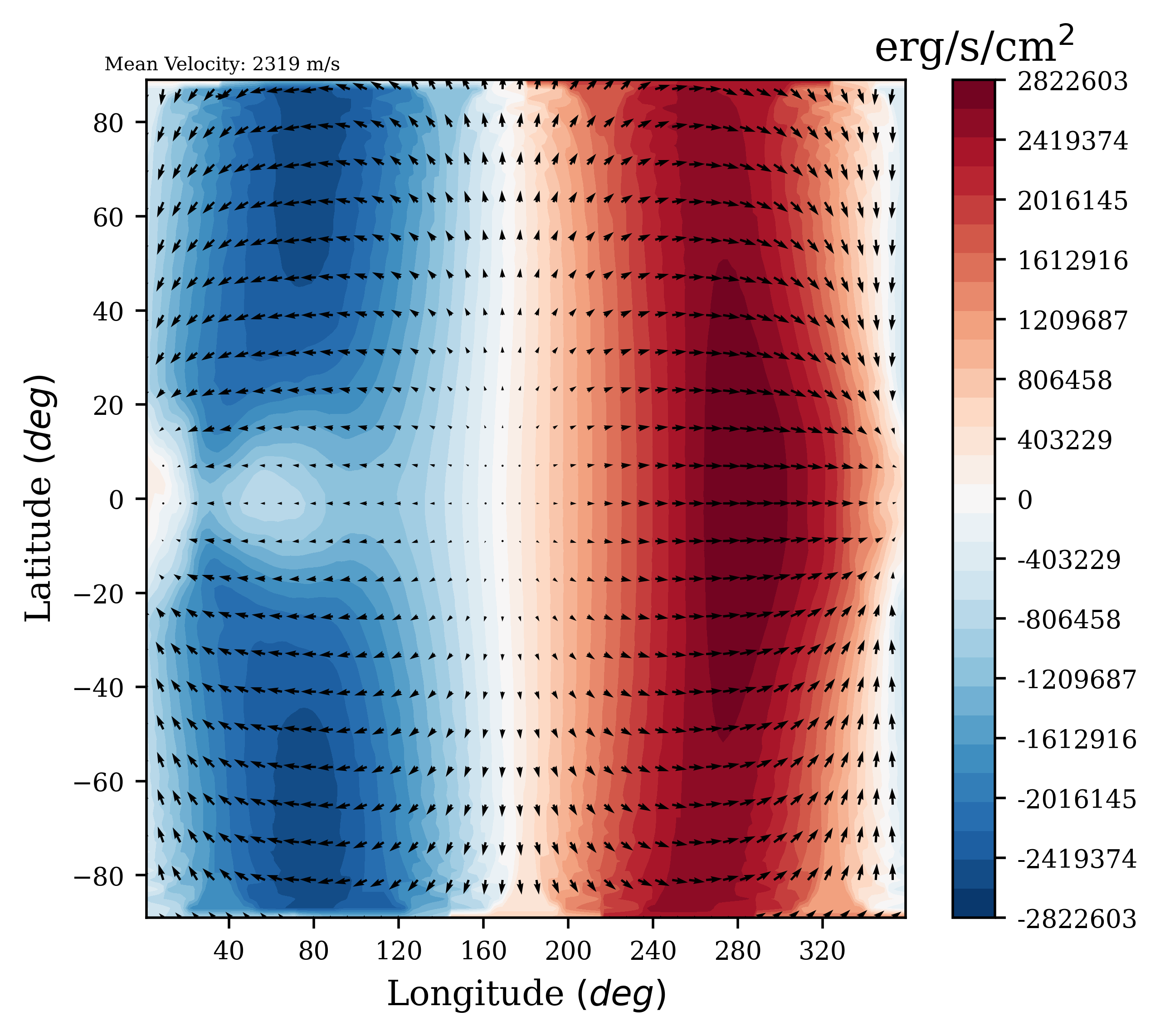}
\caption[]{$0.125\Omega_{0}$ - Zonal - 0.0026 bar  \label{fig:Enthalpy_Wind_zonal_0016_0.125} }
\end{centering}
\end{subfigure}
\begin{subfigure}{0.3\textwidth}
\begin{centering}
\includegraphics[width=0.99\columnwidth]{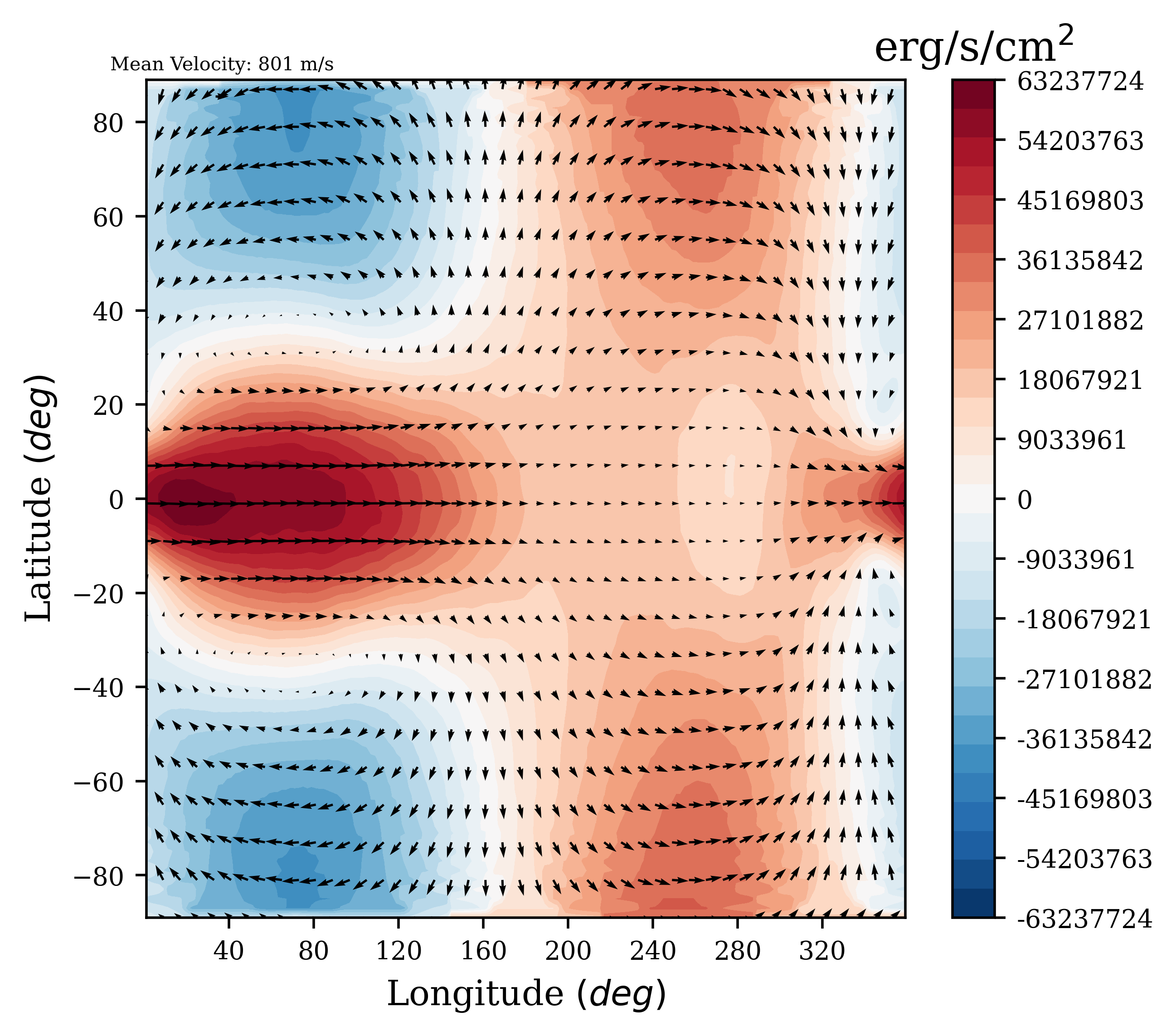}
\caption[]{$0.125\Omega_{0}$ - Zonal - 0.1 bar   \label{fig:Enthalpy_Wind_zonal_02_0.125} }
\end{centering}
\end{subfigure}
\begin{subfigure}{0.3\textwidth}
\begin{centering}
\includegraphics[width=0.99\columnwidth]{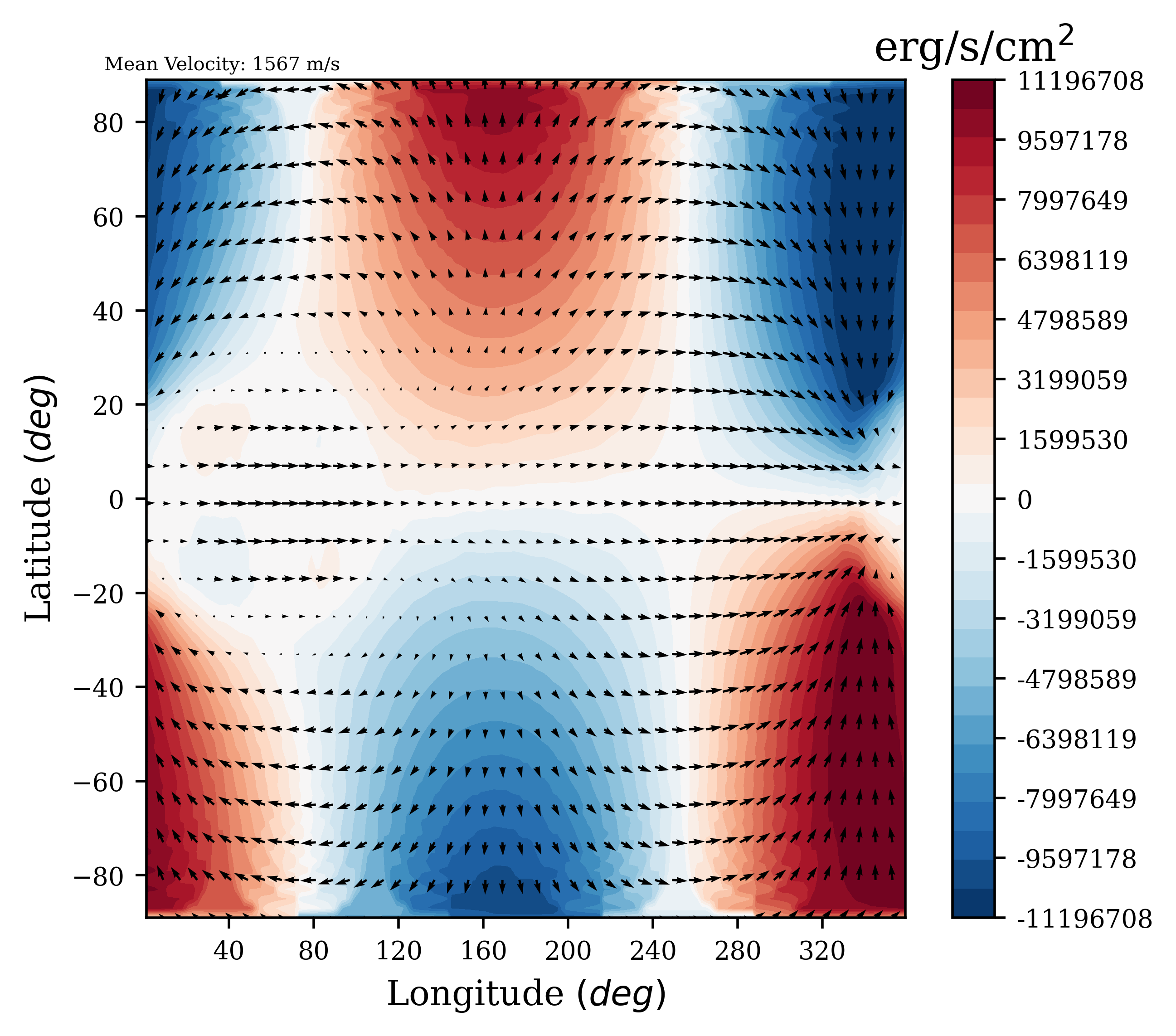}
\caption[]{$0.125\Omega_{0}$ - Meridional - 0.016 bar  \label{fig:Enthalpy_Wind_merid_0.125} }
\end{centering}
\end{subfigure}

\begin{subfigure}{0.3\textwidth}
\begin{centering}
\includegraphics[width=0.99\columnwidth]{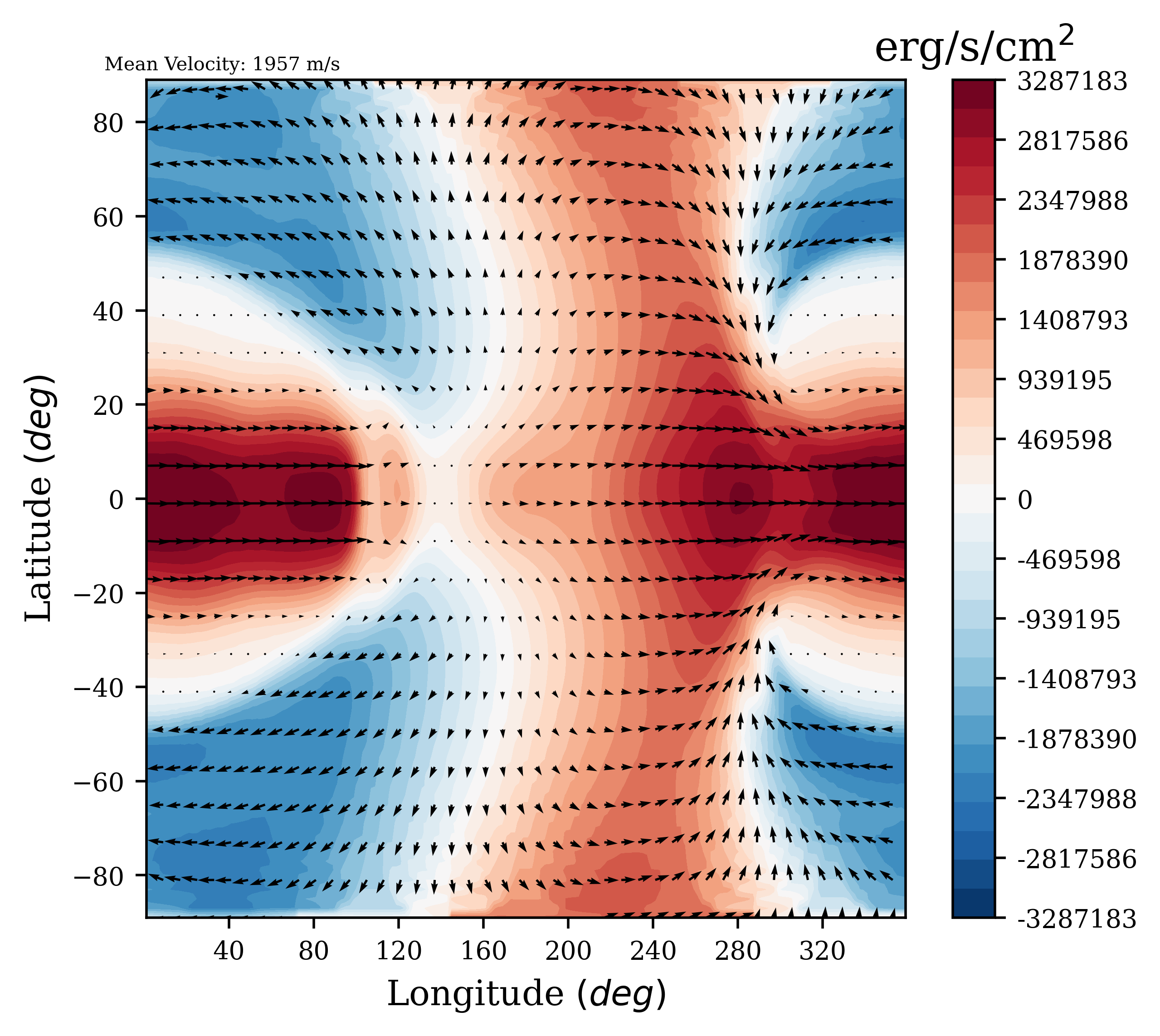}
\caption[]{$1\Omega_{0}$ - Zonal - 0.0026 bar  \label{fig:Enthalpy_Wind_zonal_0016_1} }
\end{centering}
\end{subfigure}
\begin{subfigure}{0.3\textwidth}
\begin{centering}
\includegraphics[width=0.99\columnwidth]{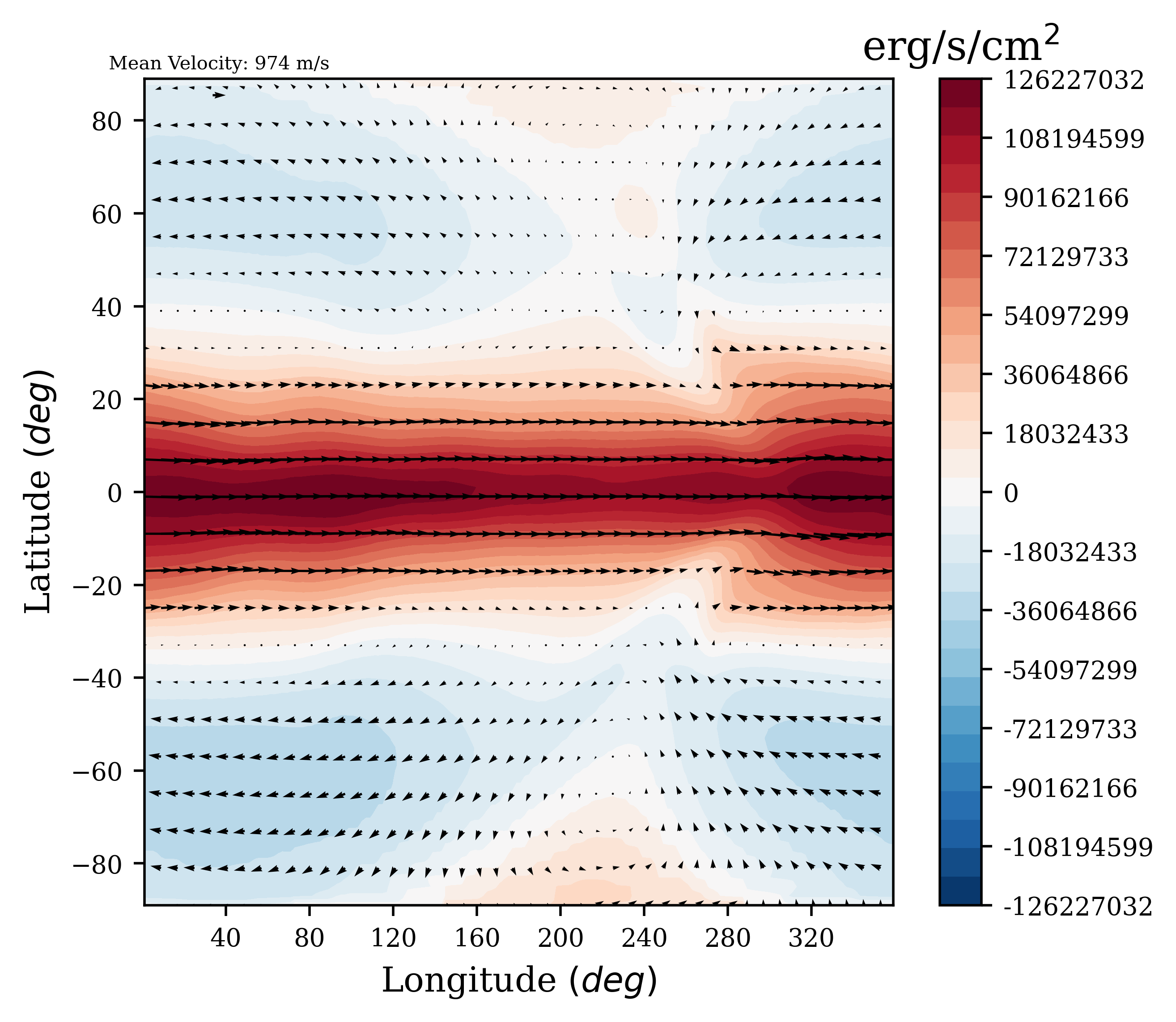}
\caption[]{$1\Omega_{0}$ - Zonal - 0.1 bar   \label{fig:Enthalpy_Wind_zonal_02_1} }
\end{centering}
\end{subfigure}
\begin{subfigure}{0.3\textwidth}
\begin{centering}
\includegraphics[width=0.99\columnwidth]{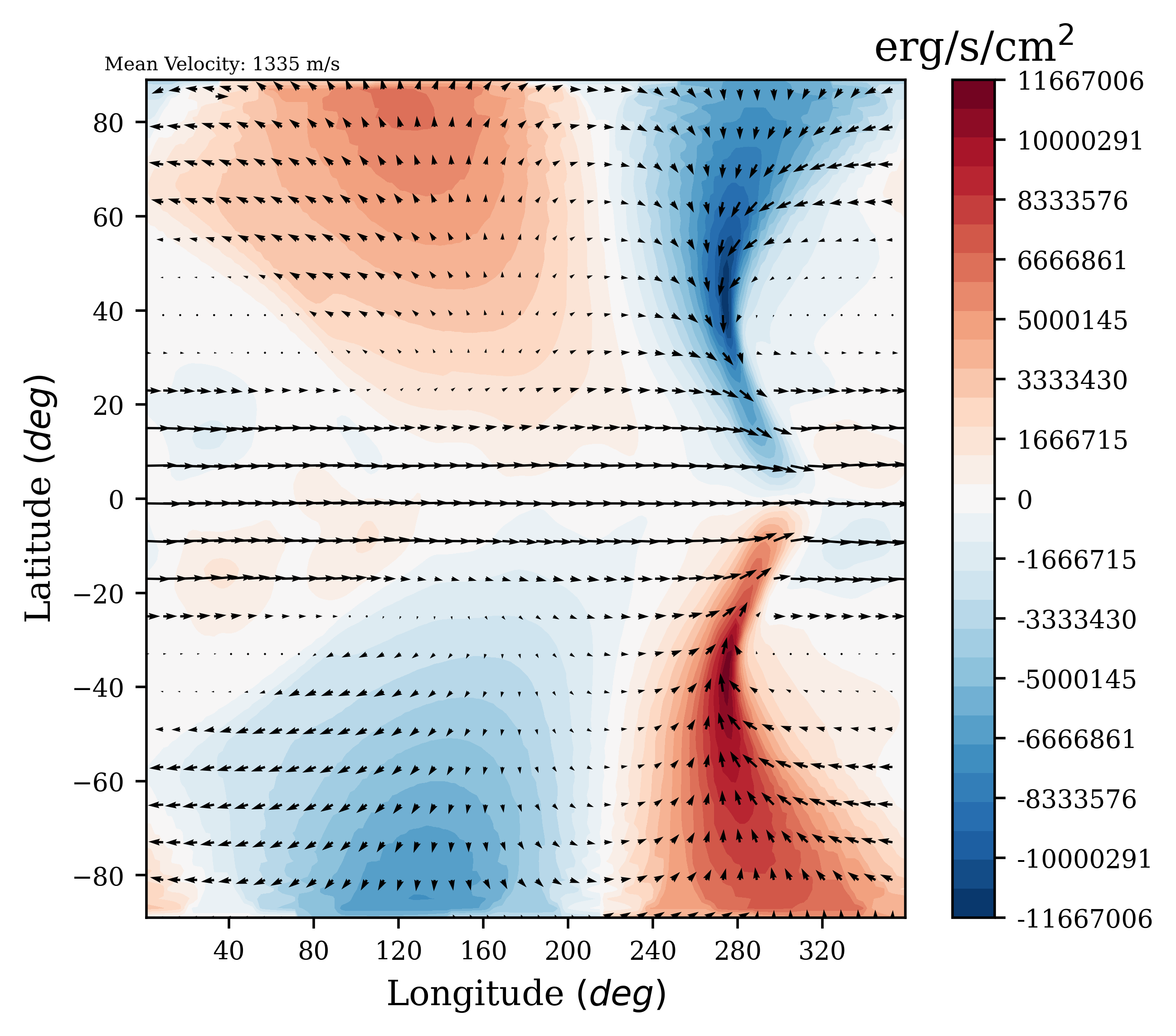}
\caption[]{$1\Omega_{0}$ - Meridional - 0.016 bar  \label{fig:Enthalpy_Wind_merid_1} }
\end{centering}
\end{subfigure}
\begin{subfigure}{0.3\textwidth}
\begin{centering}
\includegraphics[width=0.99\columnwidth]{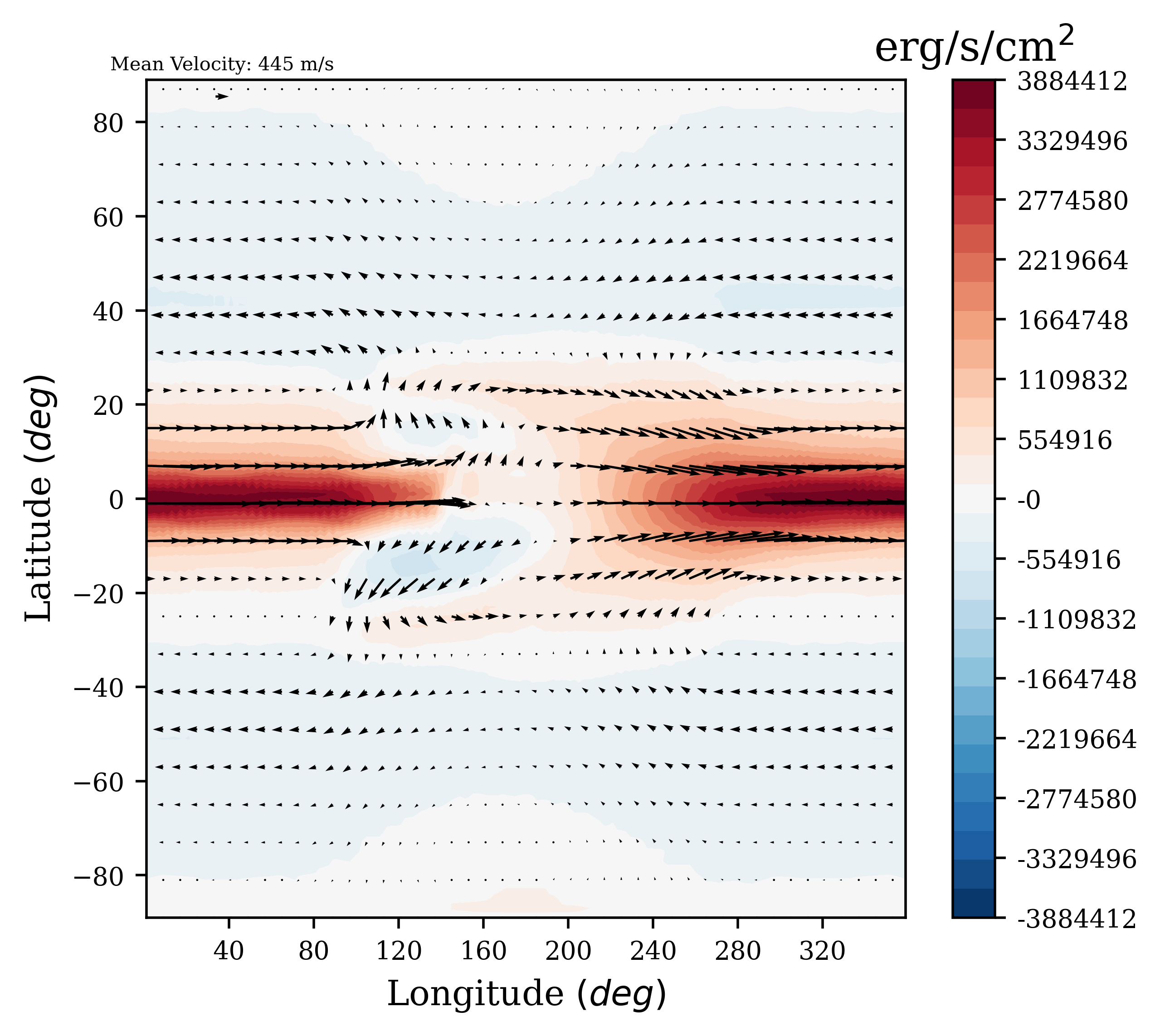}
\caption[]{$15\Omega_{0}$ - Zonal - 0.0026 bar  \label{fig:Enthalpy_Wind_zonal_0016_15} }
\end{centering}
\end{subfigure}
\begin{subfigure}{0.3\textwidth}
\begin{centering}
\includegraphics[width=0.99\columnwidth]{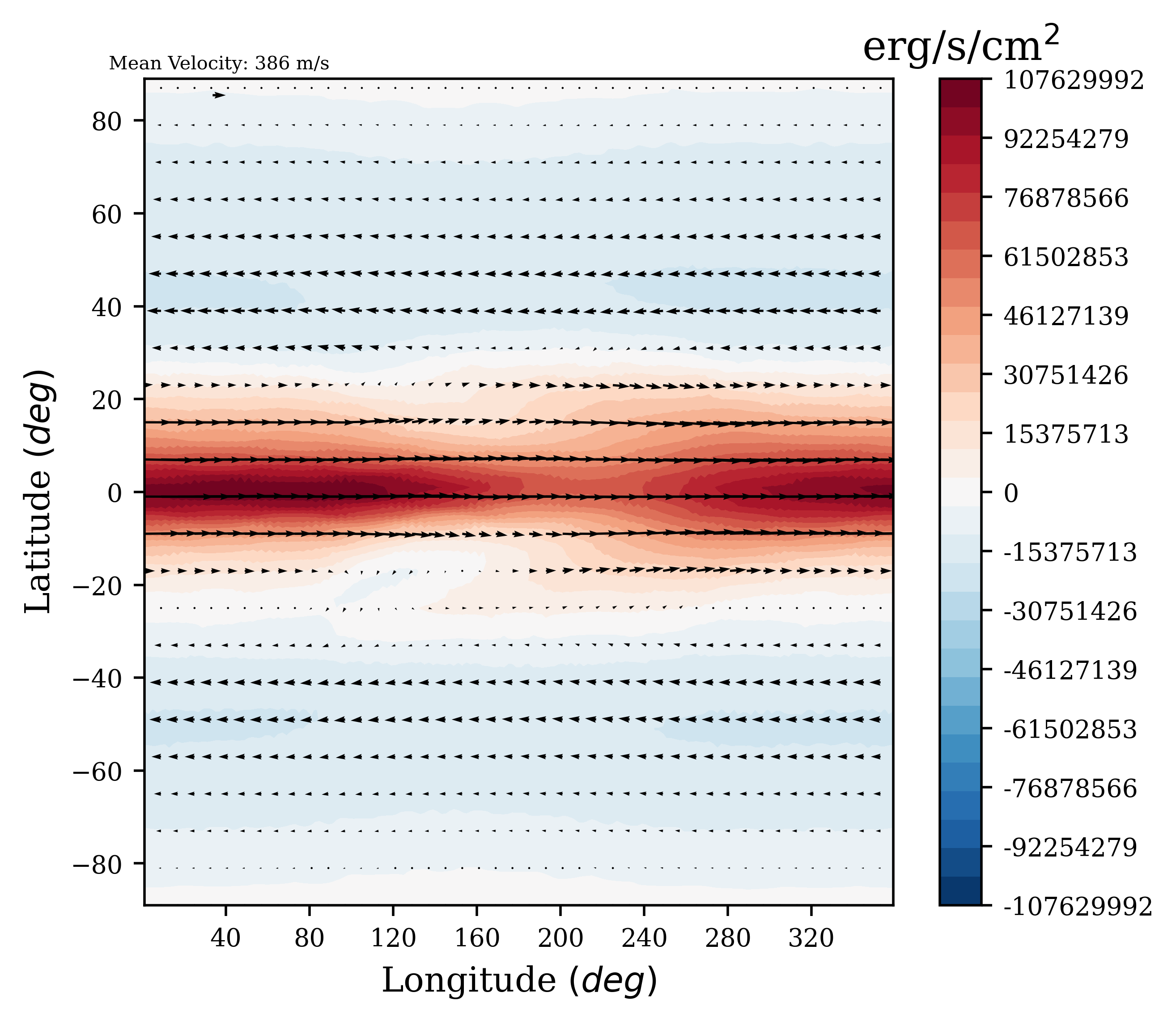}
\caption[]{$15\Omega_{0}$ - Zonal - 0.1 bar   \label{fig:Enthalpy_Wind_zonal_02_15} }
\end{centering}
\end{subfigure}
\begin{subfigure}{0.3\textwidth}
\begin{centering}
\includegraphics[width=0.99\columnwidth]{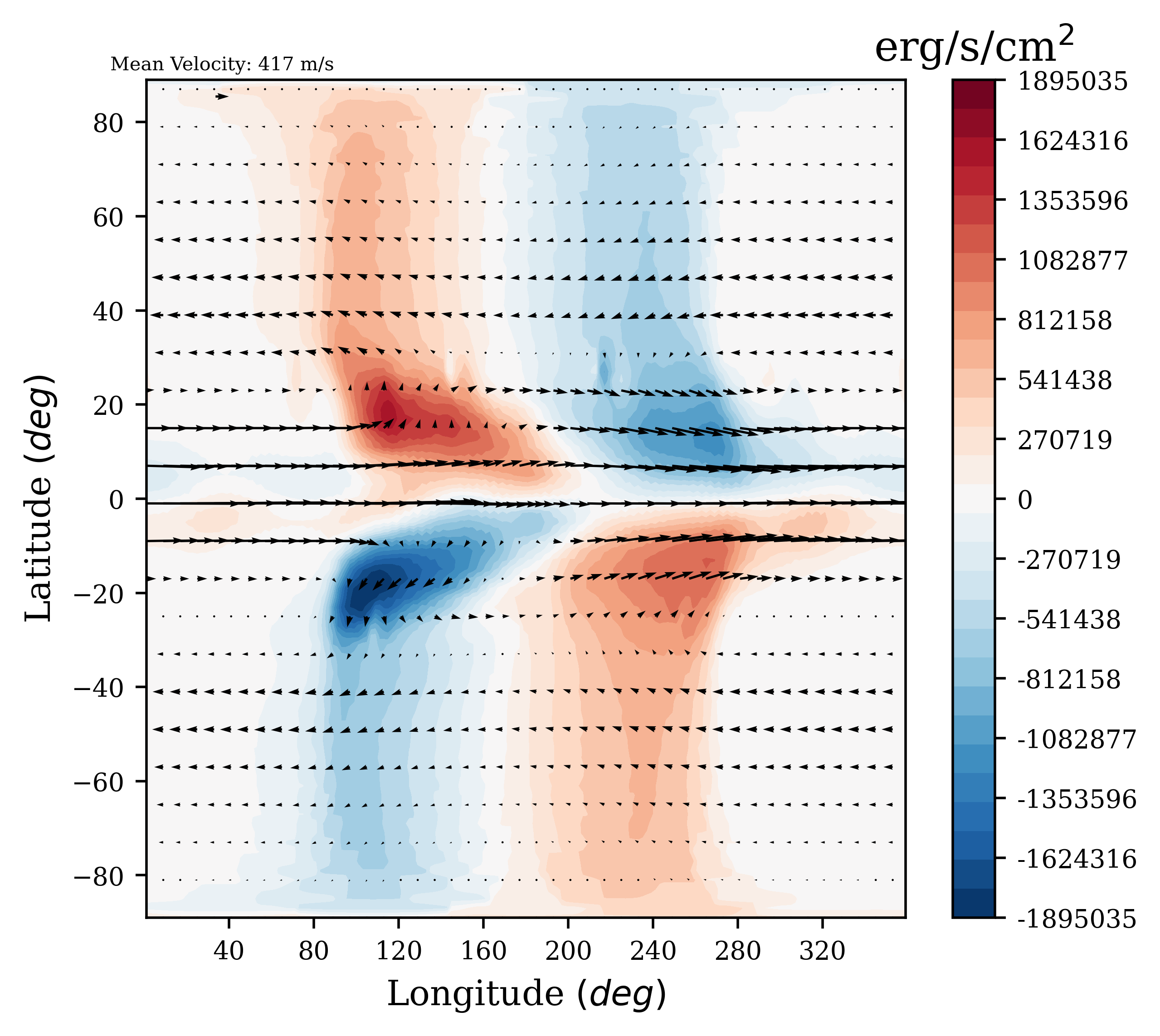}
\caption[]{$15\Omega_{0}$ - Meridional - 0.016 bar  \label{fig:Enthalpy_Wind_merid_15} }
\end{centering}
\end{subfigure}
\caption[Zonal, and Meridional enthalpy transport for three HD209458b-like models at different rotation rates]{Maps showing the Zonal, Meridional, and Vertical Enthalpy transport at select pressure levels for three exemplary HD209458b-like models with rotation rates of; $0.125\Omega_{0}$ - top, $\Omega_{0}$ - middle, and $15\Omega_{0}$ - bottom. Here, positive (red) fluxes represent eastward/polar/outwards flows for the zonal/meridional/vertical enthalpy flux maps respectively. Note: We include two maps of the zonal enthalpy transport at different pressure levels in order to emphasise how the zonal advection changes with height. \label{fig:Enthalpy_Wind} }
\end{centering}
\end{figure*} 
\begin{figure*}[tbp] %
\begin{centering}
\begin{subfigure}{0.3\textwidth}
\begin{centering}
\includegraphics[width=0.99\columnwidth]{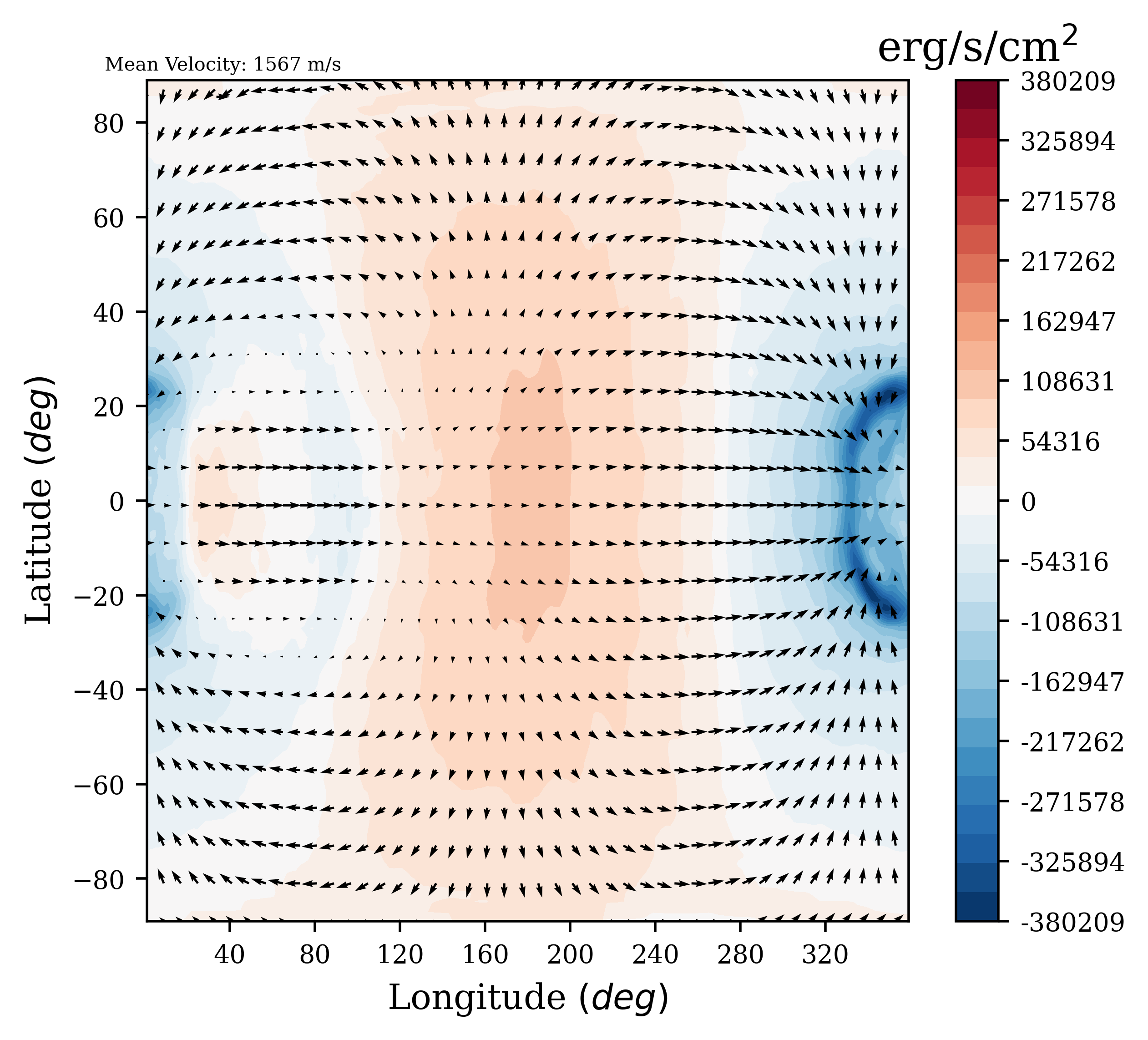}
\caption[]{$0.125\Omega_{0}$ - Vertical - 0.016 bar   \label{fig:Enthalpy_Wind_vert_0.125} }
\end{centering}
\end{subfigure}
\begin{subfigure}{0.3\textwidth}
\begin{centering}
\includegraphics[width=0.99\columnwidth]{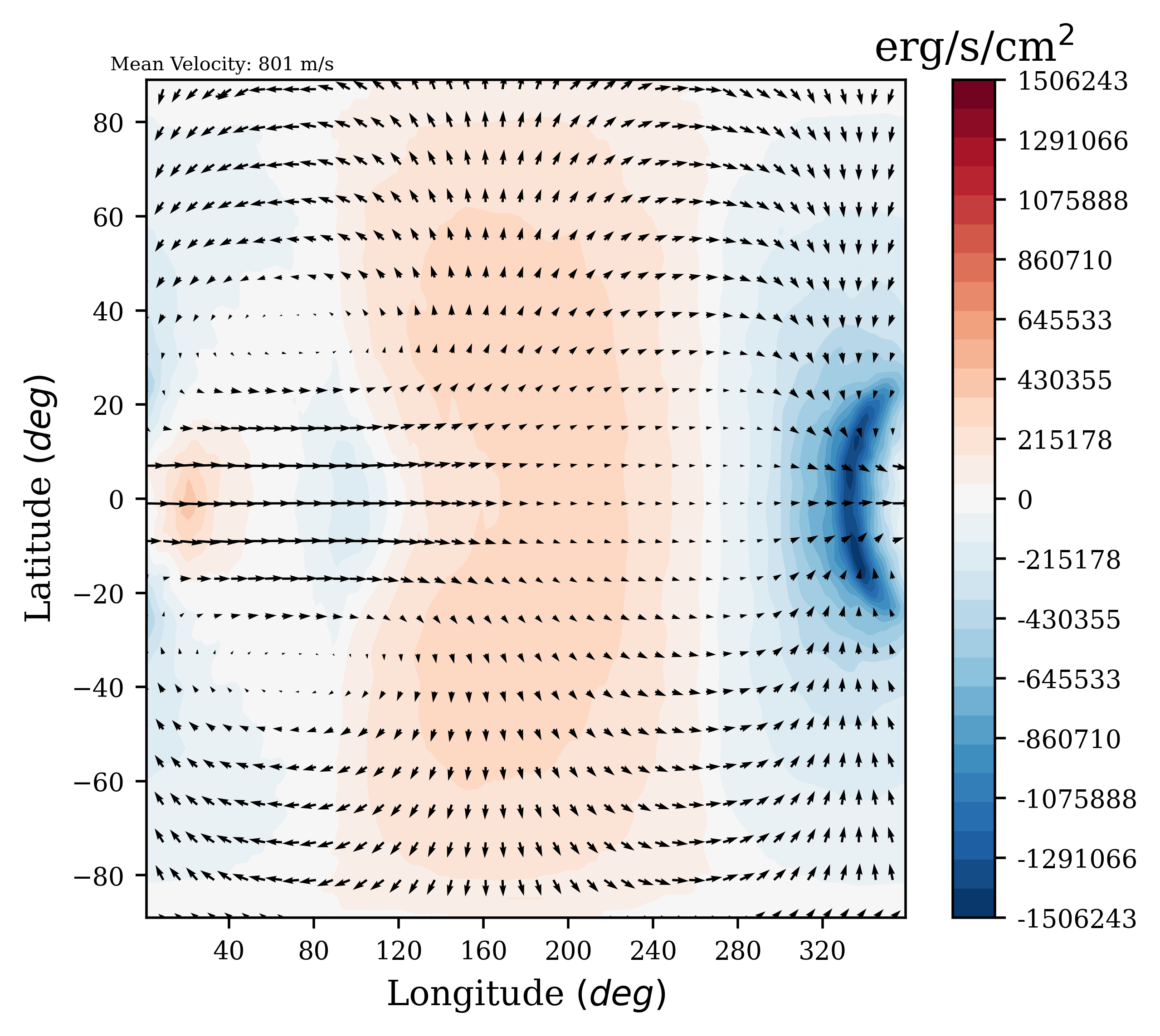}
\caption[]{$0.125\Omega_{0}$ - Vertical - 0.1 bar   \label{fig:Enthalpy_Wind_vert_0.125_0.1} }
\end{centering}
\end{subfigure}
\begin{subfigure}{0.3\textwidth}
\begin{centering}
\includegraphics[width=0.99\columnwidth]{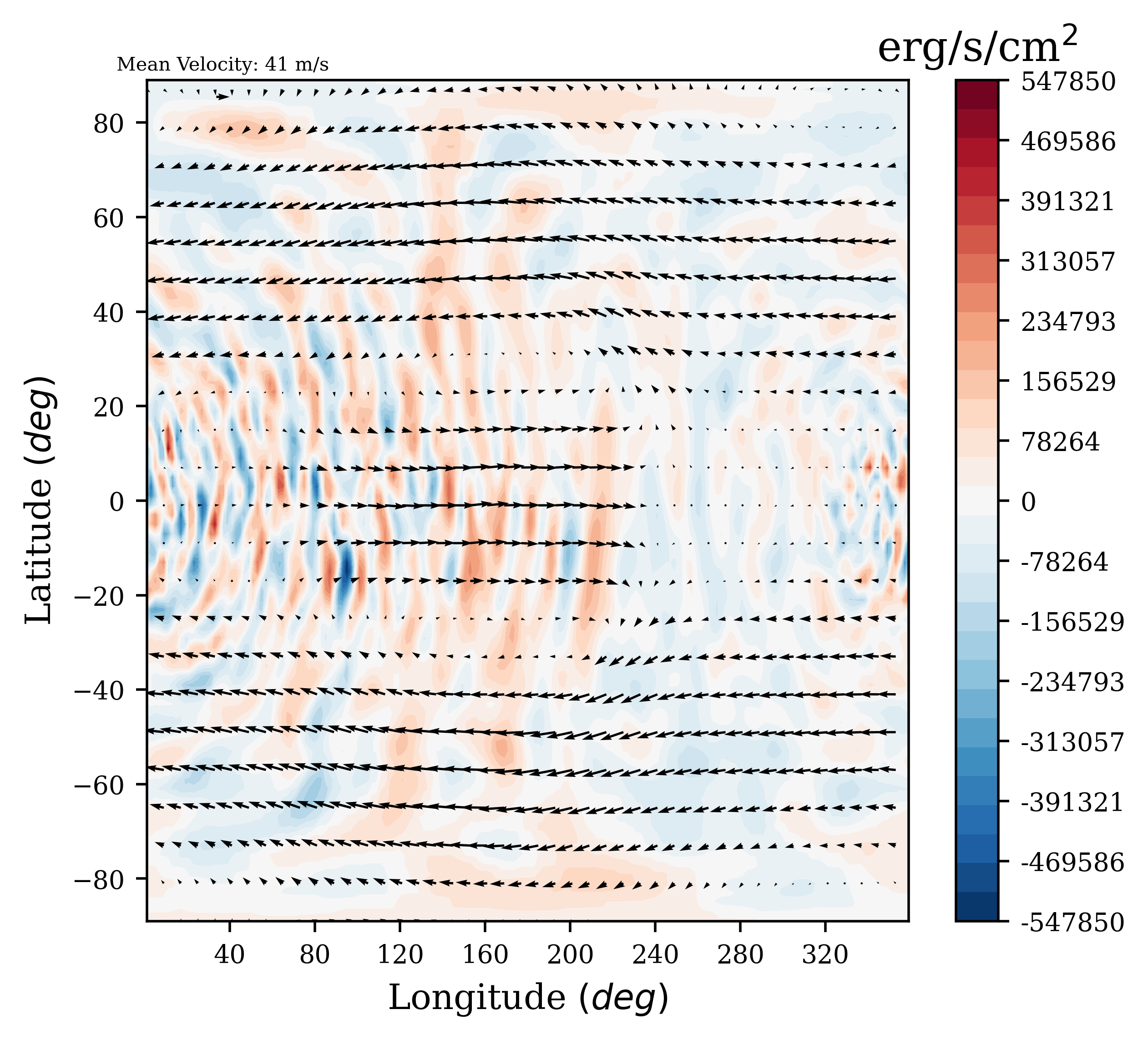}
\caption[]{$0.125\Omega_{0}$ - Vertical - 4 bar   \label{fig:Enthalpy_Wind_vert_0.125_4} }
\end{centering}
\end{subfigure}
\begin{subfigure}{0.3\textwidth}
\begin{centering}
\includegraphics[width=0.99\columnwidth]{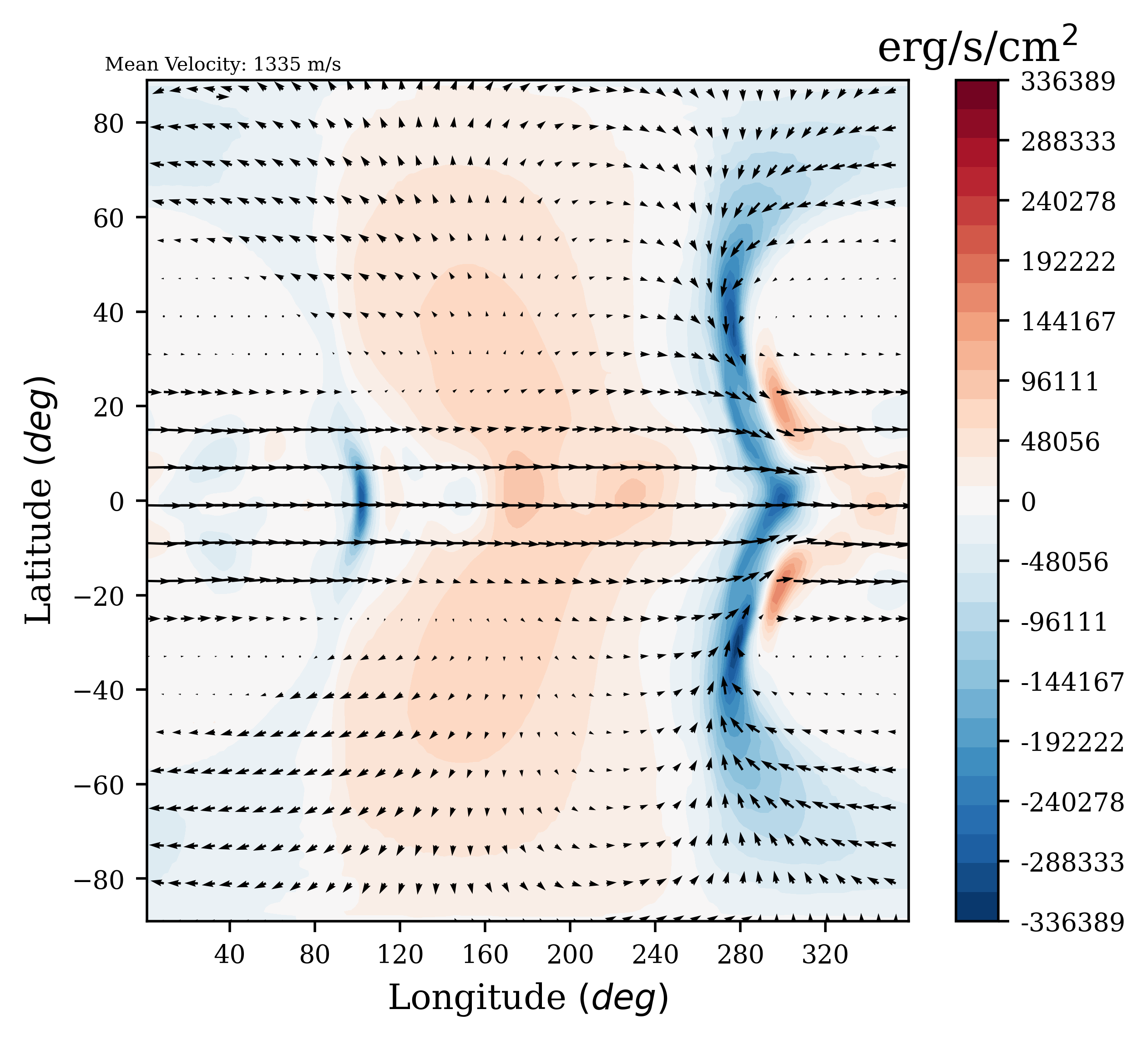}
\caption[]{$1\Omega_{0}$ - Vertical - 0.016 bar   \label{fig:Enthalpy_Wind_vert_1} }
\end{centering}
\end{subfigure}
\begin{subfigure}{0.3\textwidth}
\begin{centering}
\includegraphics[width=0.99\columnwidth]{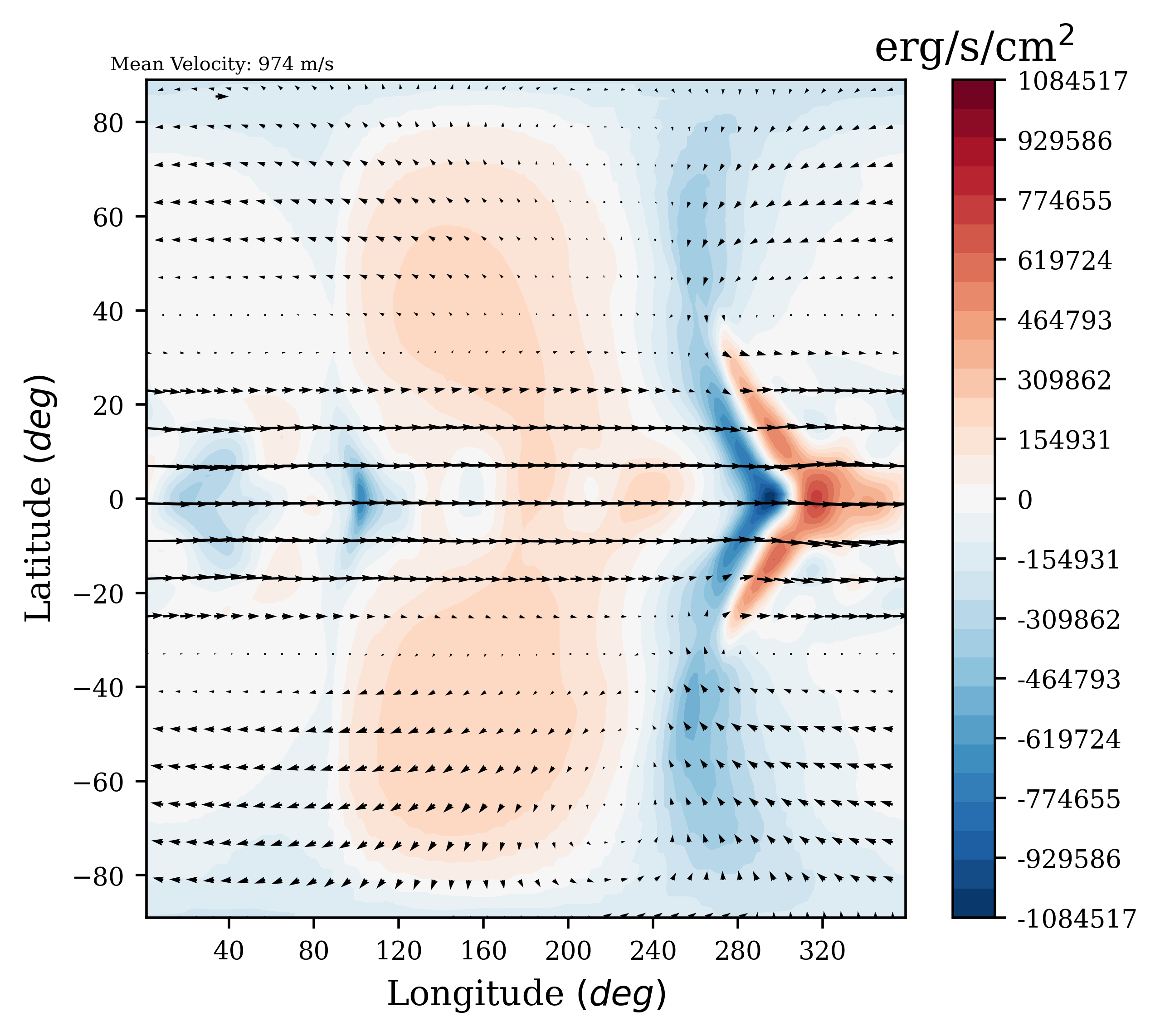}
\caption[]{$1\Omega_{0}$ - Vertical - 0.1 bar   \label{fig:Enthalpy_Wind_vert_1_0.1} }
\end{centering}
\end{subfigure}
\begin{subfigure}{0.3\textwidth}
\begin{centering}
\includegraphics[width=0.99\columnwidth]{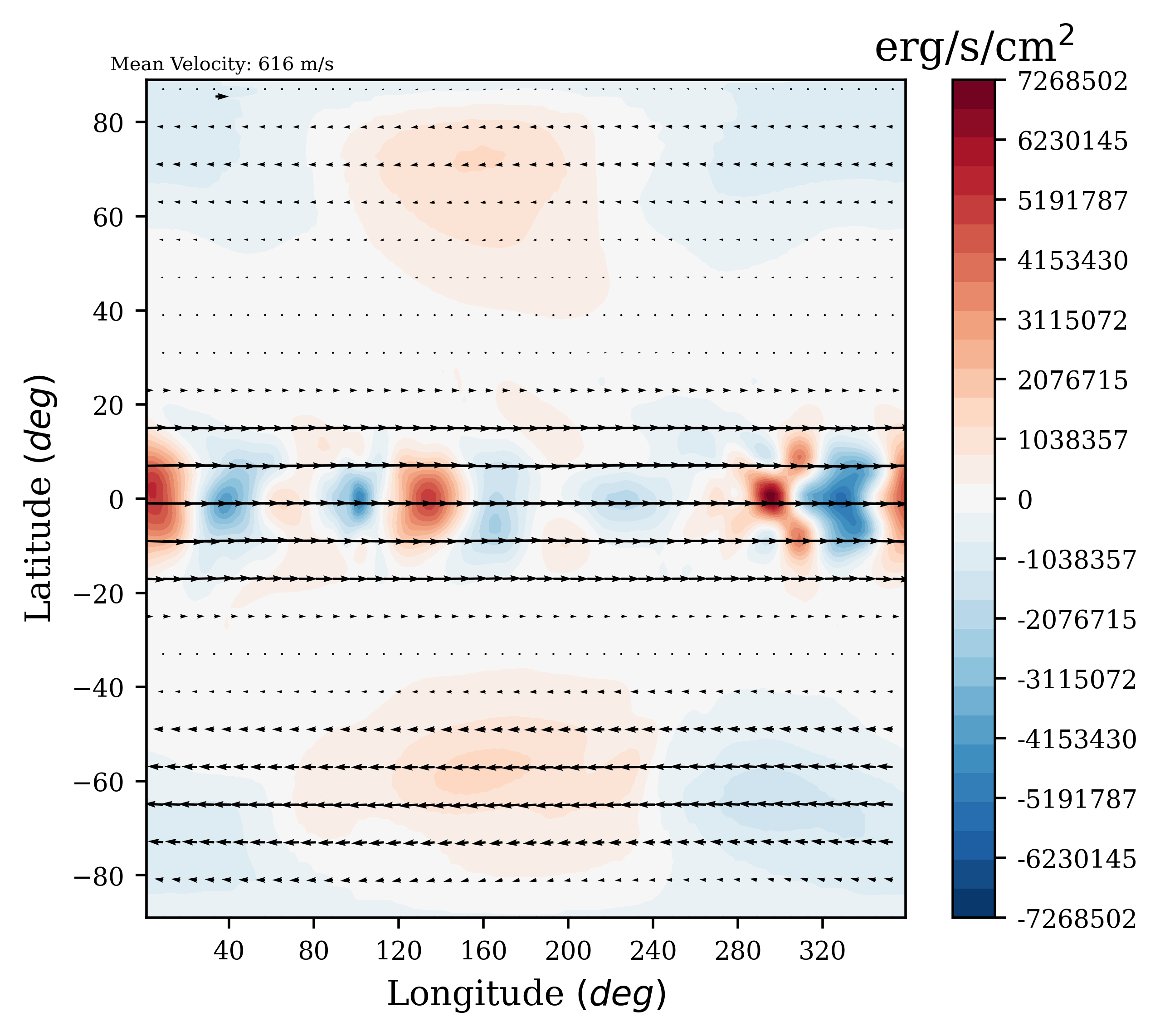}
\caption[]{$1\Omega_{0}$ - Vertical - 4 bar   \label{fig:Enthalpy_Wind_vert_1_4} }
\end{centering}
\end{subfigure}
\begin{subfigure}{0.3\textwidth}
\begin{centering}
\includegraphics[width=0.99\columnwidth]{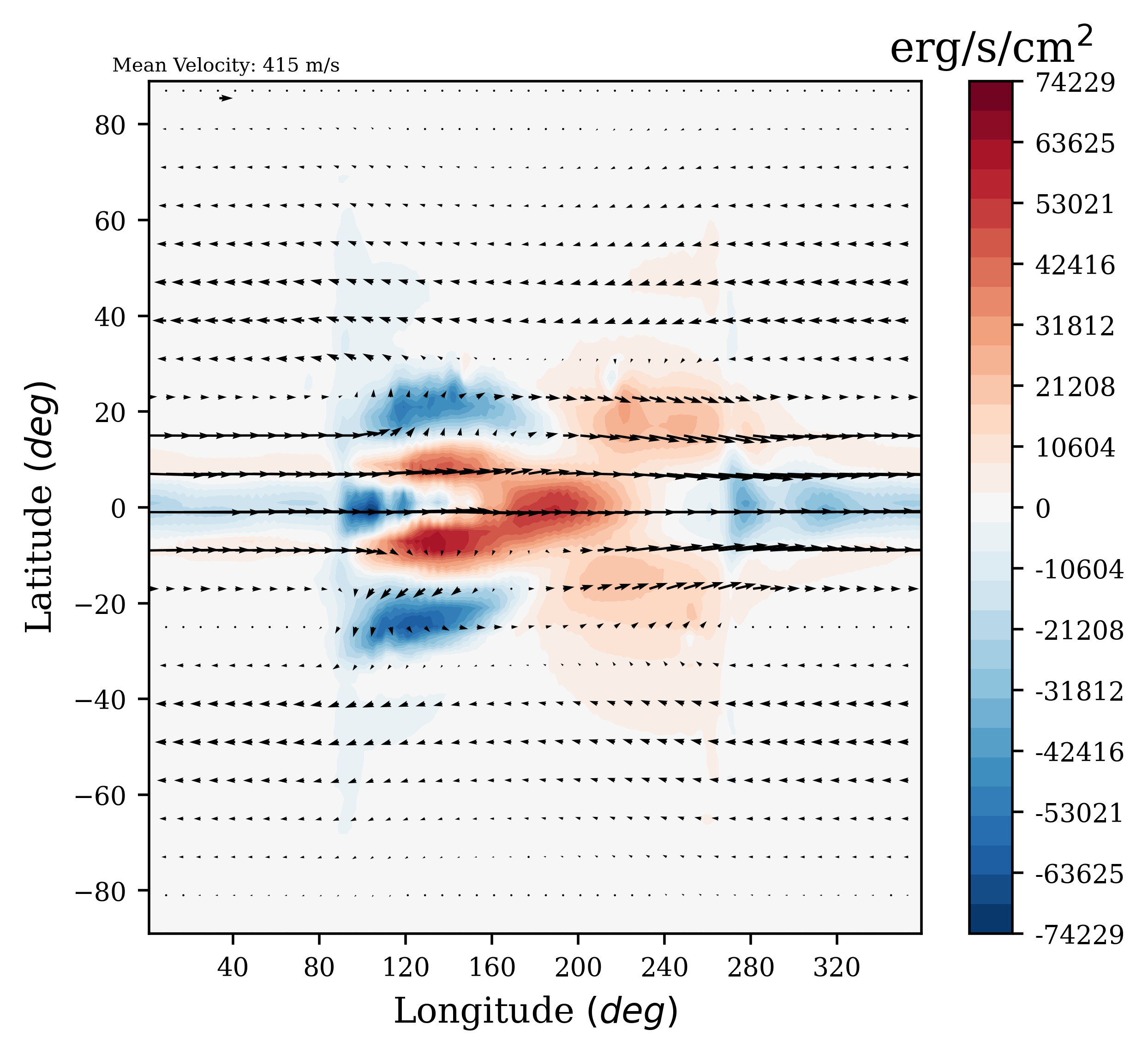}
\caption[]{$15\Omega_{0}$ - Vertical - 0.016 bar   \label{fig:Enthalpy_Wind_vert_15} }
\end{centering}
\end{subfigure}
\begin{subfigure}{0.3\textwidth}
\begin{centering}
\includegraphics[width=0.99\columnwidth]{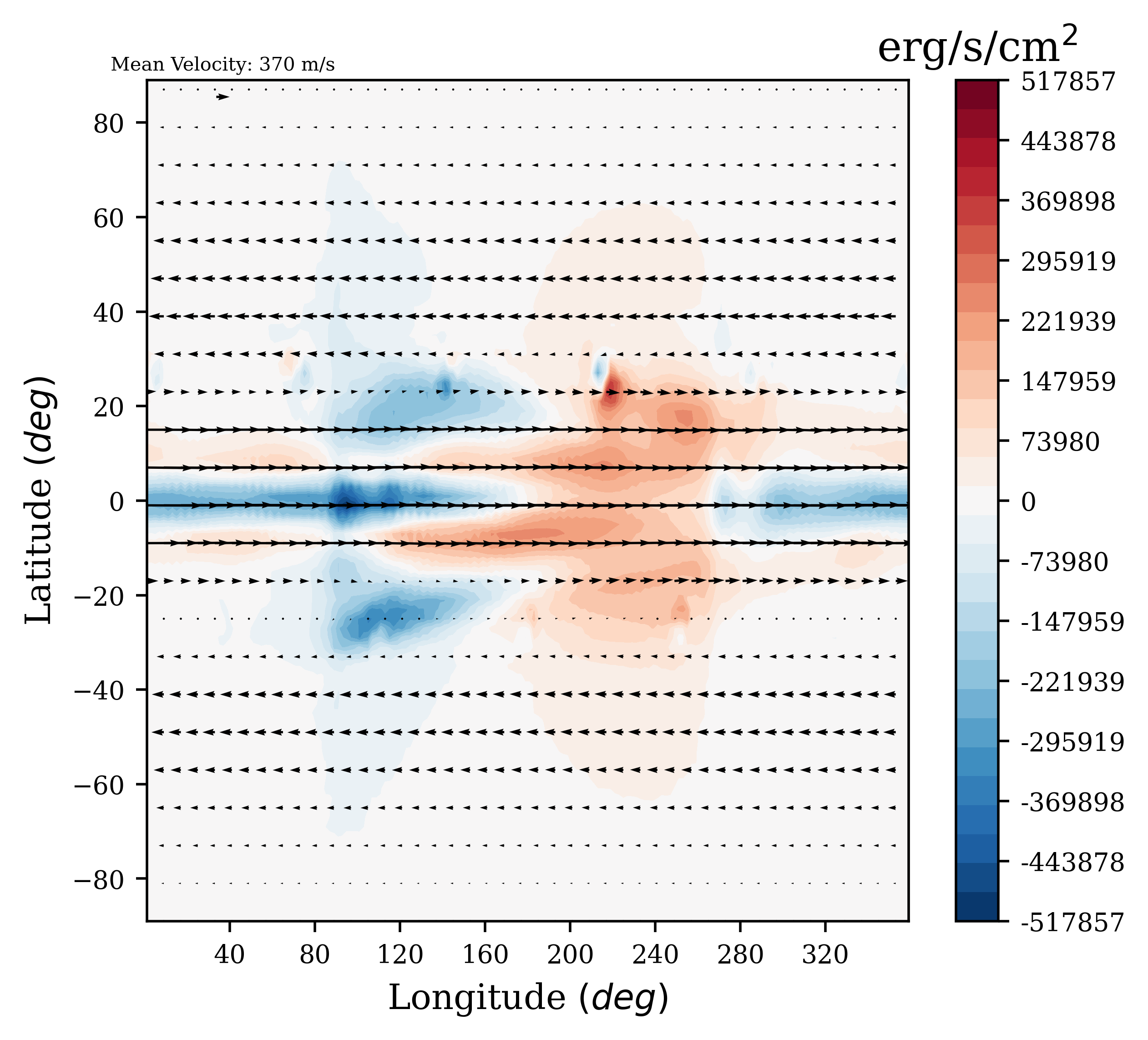}
\caption[]{$15\Omega_{0}$ - Vertical - 0.1 bar   \label{fig:Enthalpy_Wind_vert_15_0.1} }
\end{centering}
\end{subfigure}
\begin{subfigure}{0.3\textwidth}
\begin{centering}
\includegraphics[width=0.99\columnwidth]{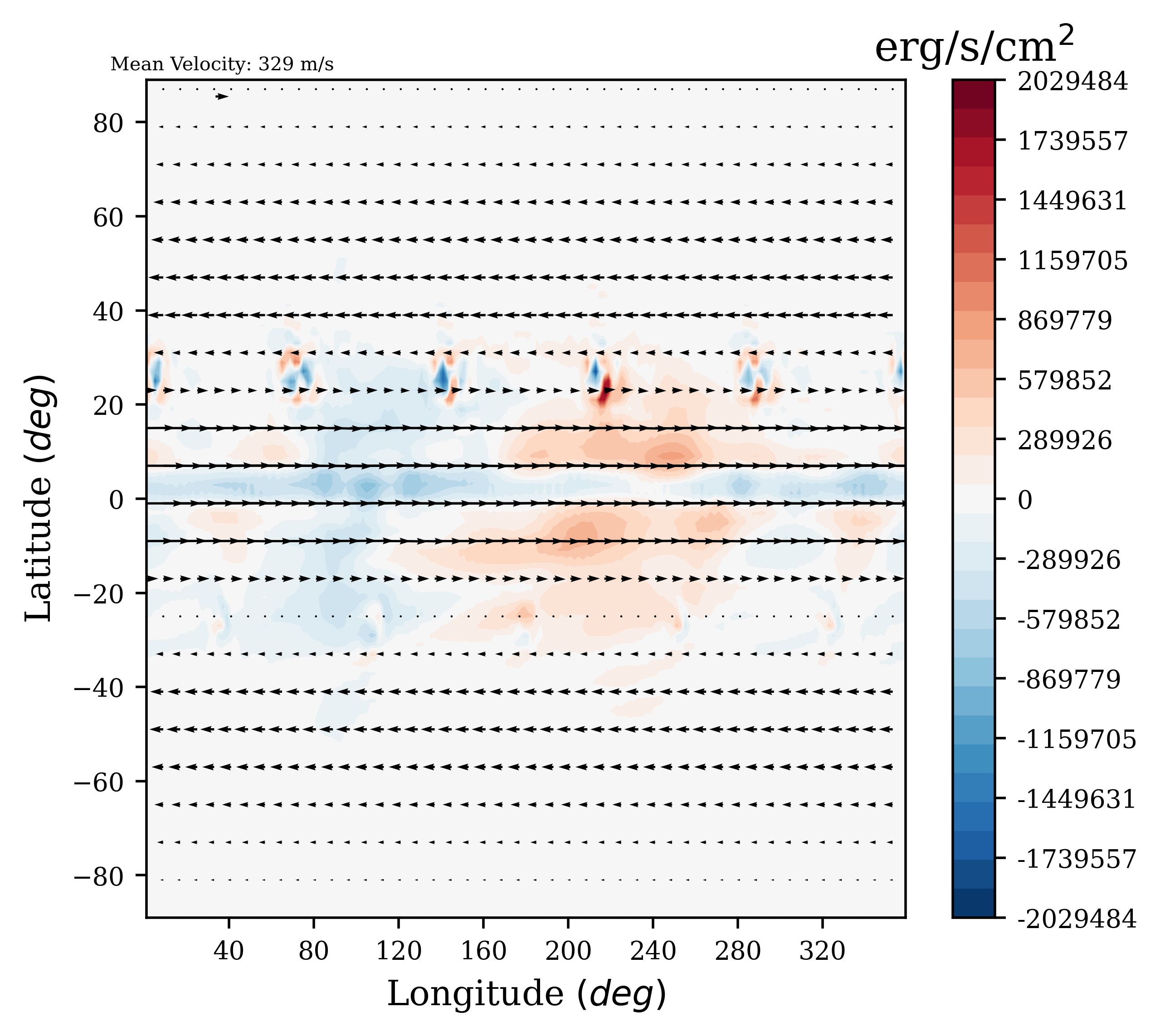}
\caption[]{$15\Omega_{0}$ - Vertical - 4 bar   \label{fig:Enthalpy_Wind_vert_15_4} }
\end{centering}
\end{subfigure}

\caption[Vertical enthalpy transport for three HD209458b-like models at different rotation rates]{Continued from previous page: Maps showing the Zonal, Meridional, and Vertical Enthalpy transport at select pressure levels for three exemplary HD209458b-like models with rotation rates of; $0.125\Omega_{0}$ - top, $\Omega_{0}$ - middle, and $15\Omega_{0}$ - bottom. Here, positive (red) fluxes represent eastward/polar/outwards flows for the zonal/meridional/vertical enthalpy flux maps respectively. Note: We include two maps of the zonal enthalpy transport at different pressure levels in order to emphasise how the zonal advection changes with height. \label{fig:Enthalpy_Wind_vertical} }
\end{centering}
\end{figure*} 

\begin{figure*}[tbp] %
\begin{centering}

\begin{subfigure}{0.3\textwidth}
\begin{centering}
\includegraphics[width=0.99\columnwidth]{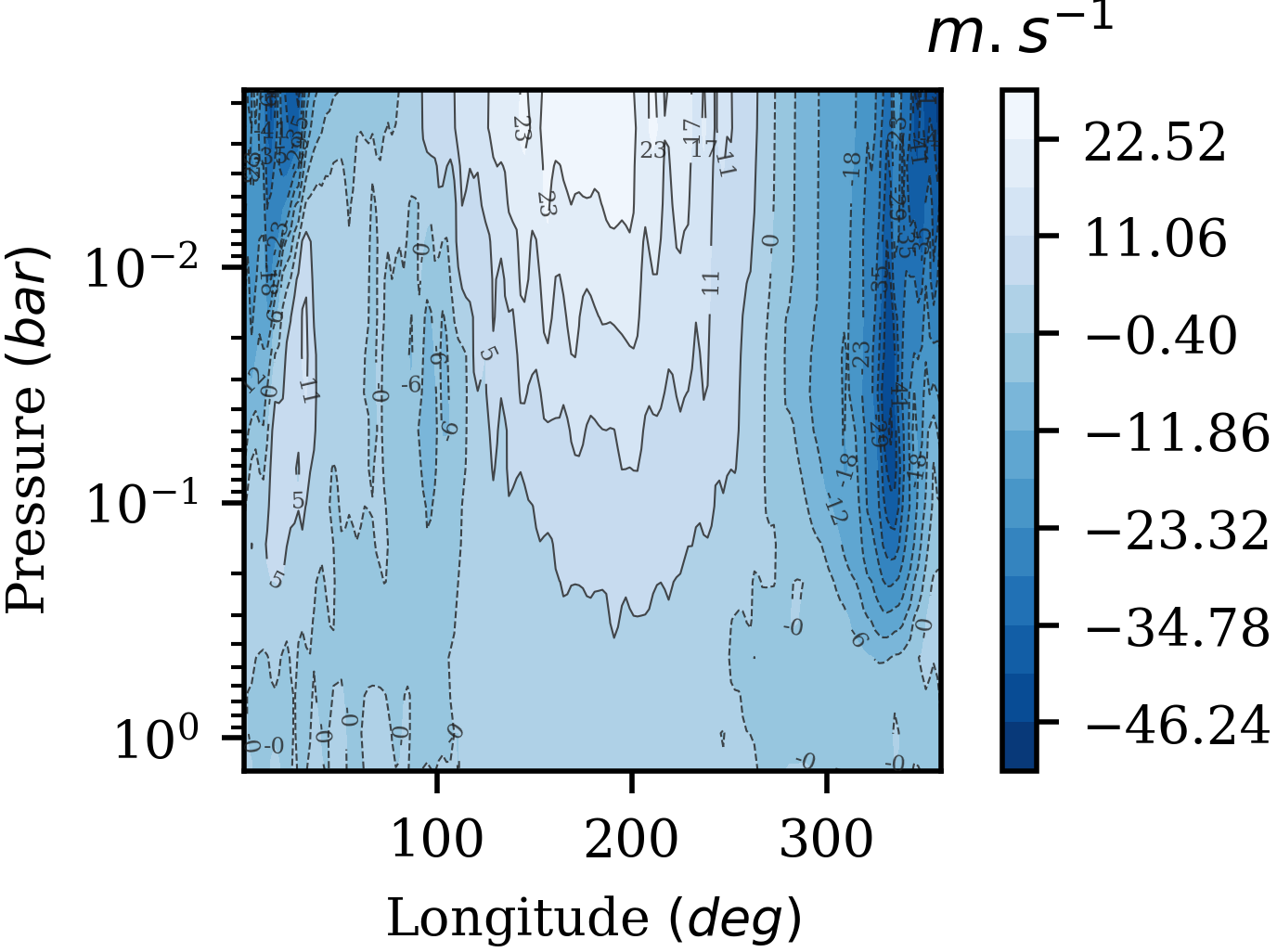}
\caption[]{$0.125\Omega_{0}$ - Vertical Wind  \label{fig:long_slices_W_0125} }
\end{centering}
\end{subfigure}
\begin{subfigure}{0.3\textwidth}
\begin{centering}
\includegraphics[width=0.99\columnwidth]{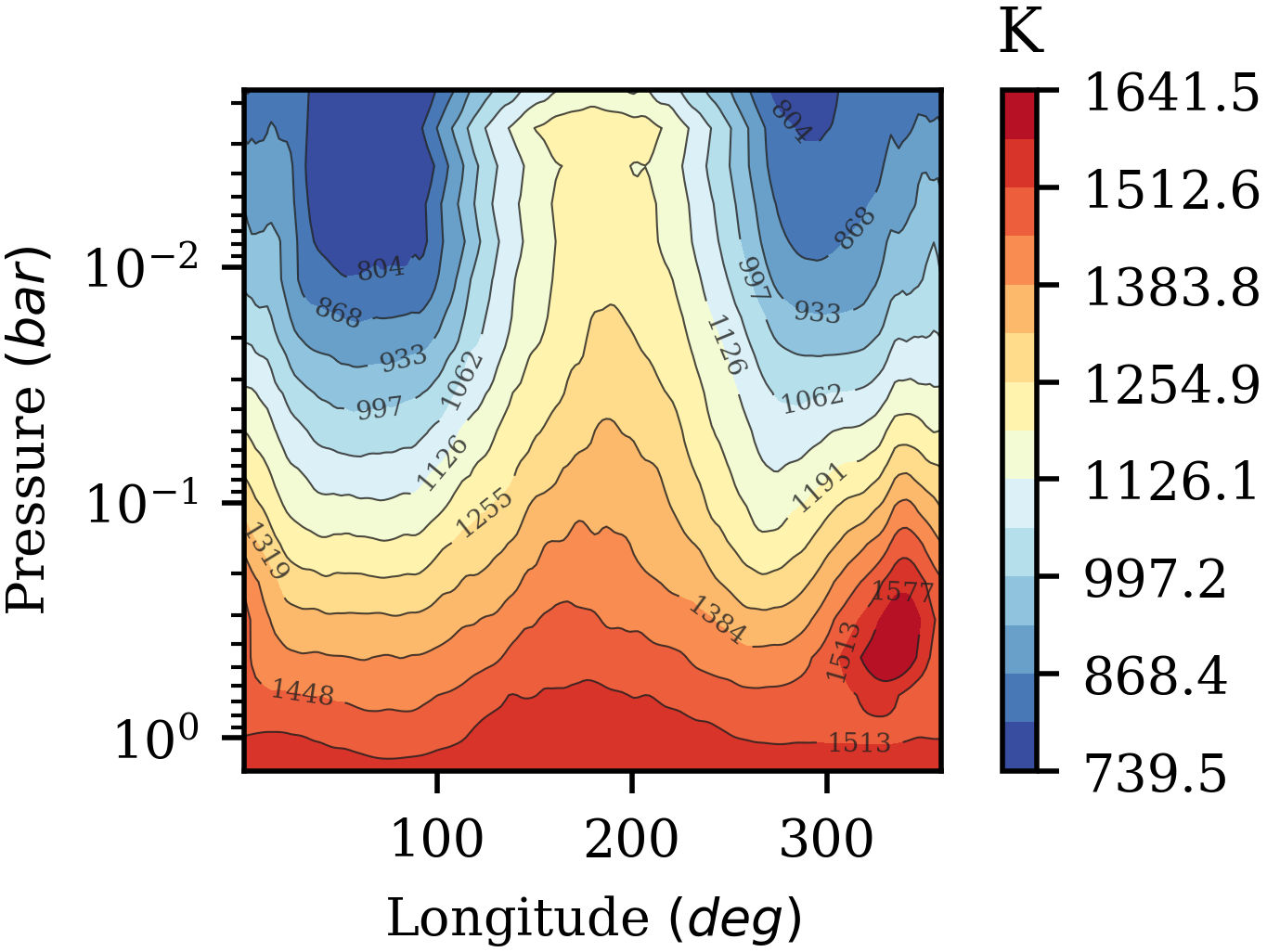}
\caption[]{$0.125\Omega_{0}$ - Temperature  \label{fig:long_slices_T_0125} }
\end{centering}
\end{subfigure}
\begin{subfigure}{0.31\textwidth}
\begin{centering}
\includegraphics[width=0.99\columnwidth]{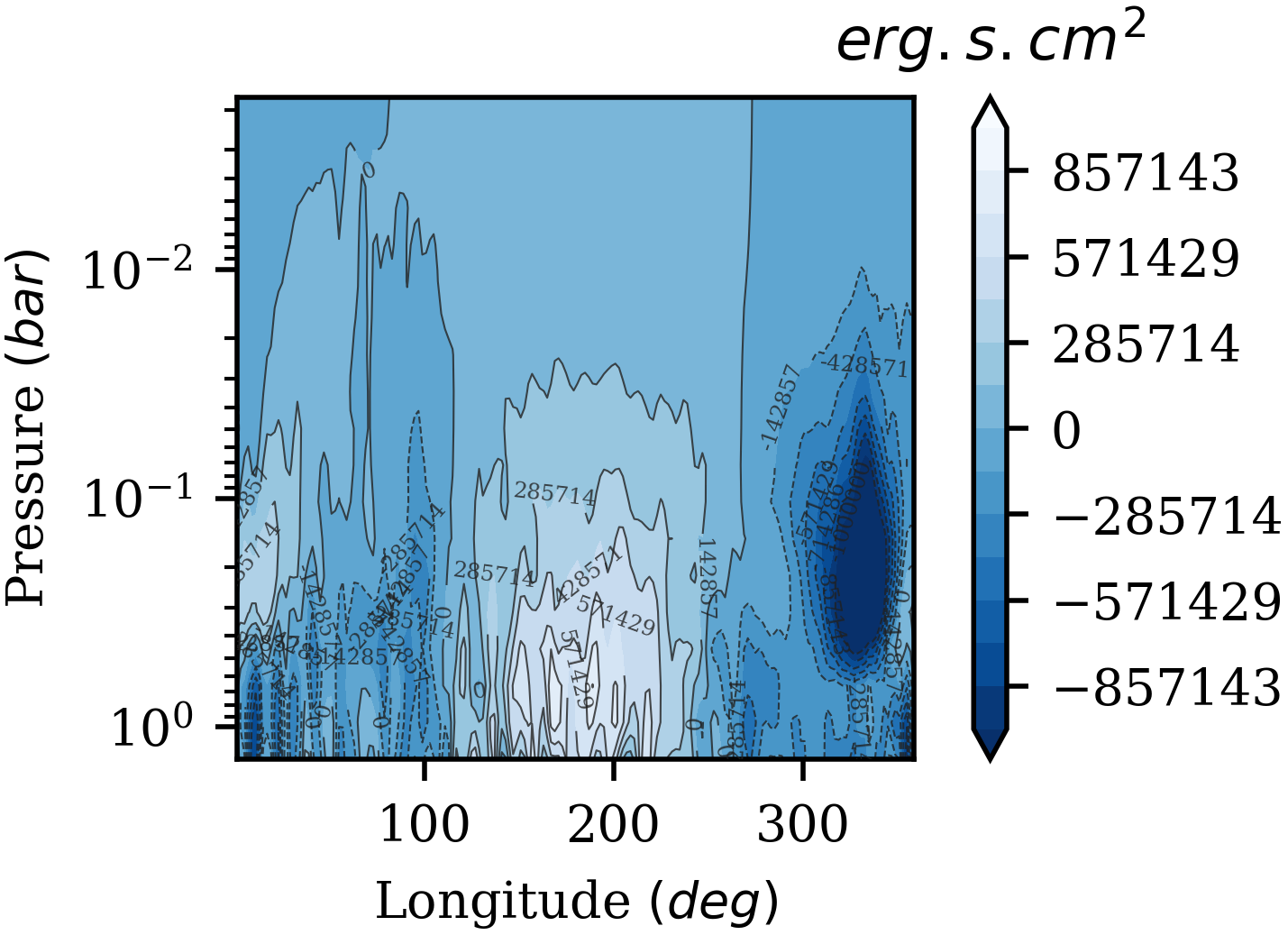}
\caption[]{$0.125\Omega_{0}$ - Vertical Enthalpy  \label{fig:long_slices_VE_0125} }
\end{centering}
\end{subfigure}

\begin{subfigure}{0.3\textwidth}
\begin{centering}
\includegraphics[width=0.99\columnwidth]{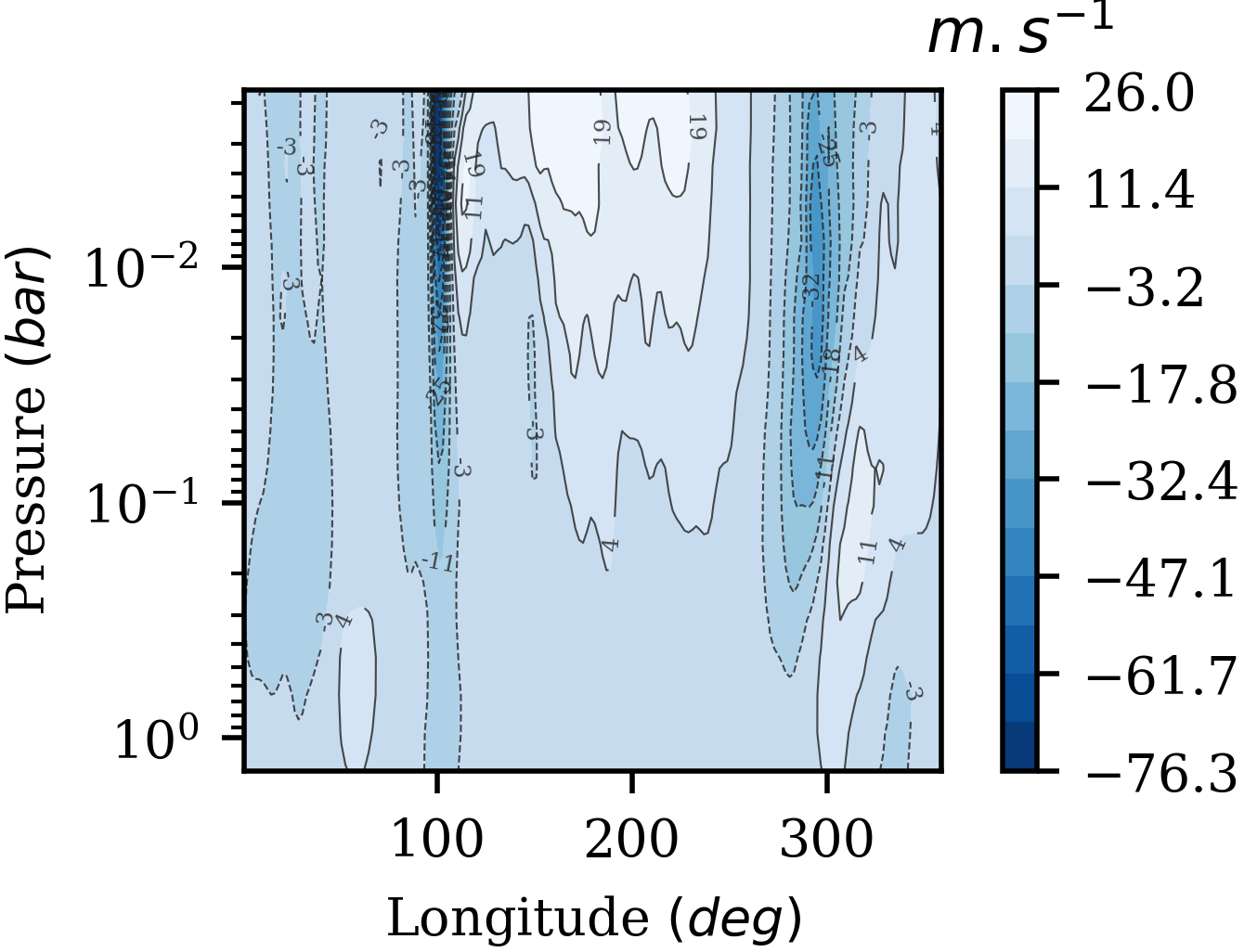}
\caption[]{$1\Omega_{0}$ - Vertical Wind  \label{fig:long_slices_W_1} }
\end{centering}
\end{subfigure}
\begin{subfigure}{0.3\textwidth}
\begin{centering}
\includegraphics[width=0.99\columnwidth]{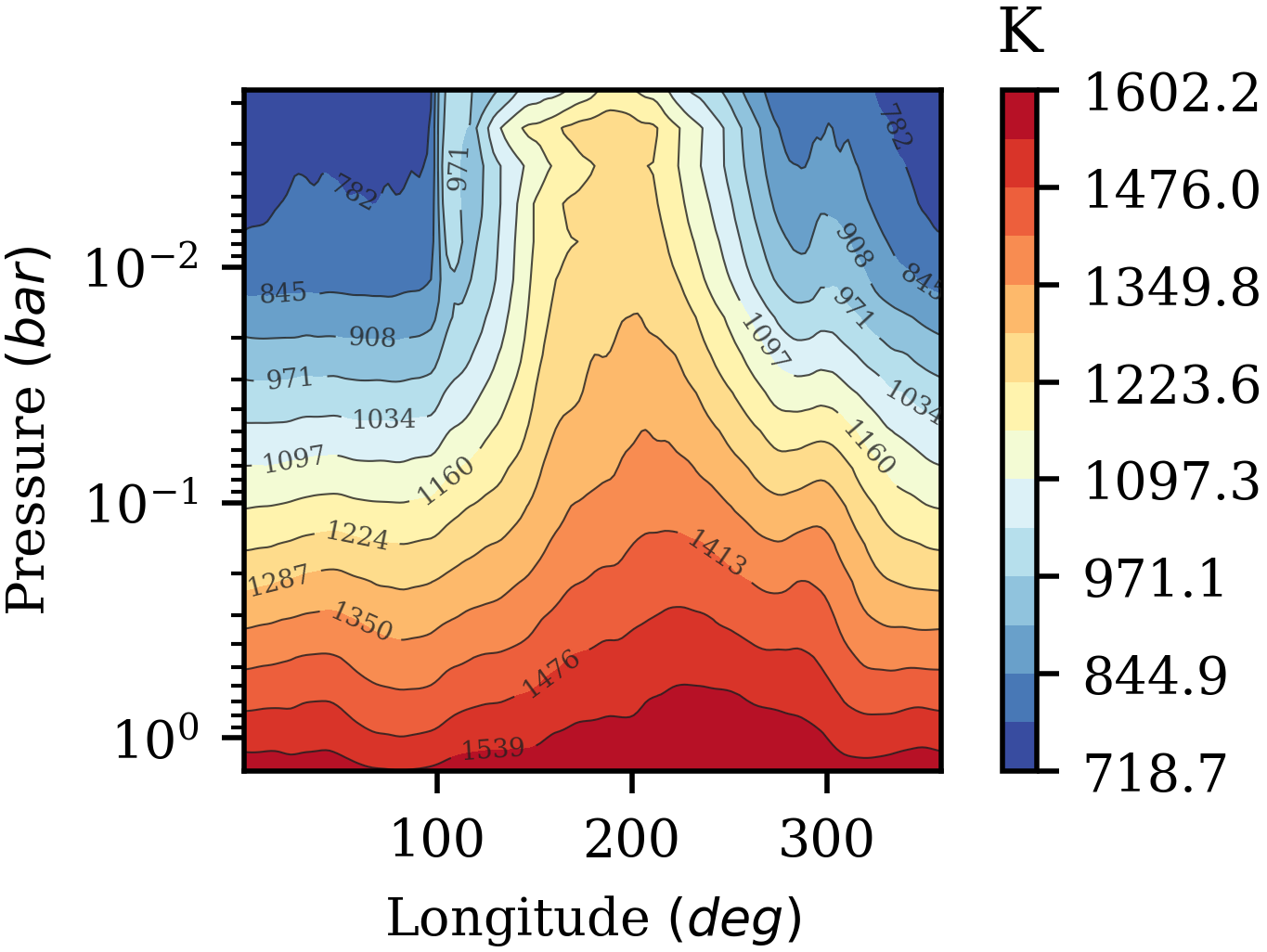}
\caption[]{$1\Omega_{0}$ - Temperature  \label{fig:long_slices_T_1} }
\end{centering}
\end{subfigure}
\begin{subfigure}{0.31\textwidth}
\begin{centering}
\includegraphics[width=0.99\columnwidth]{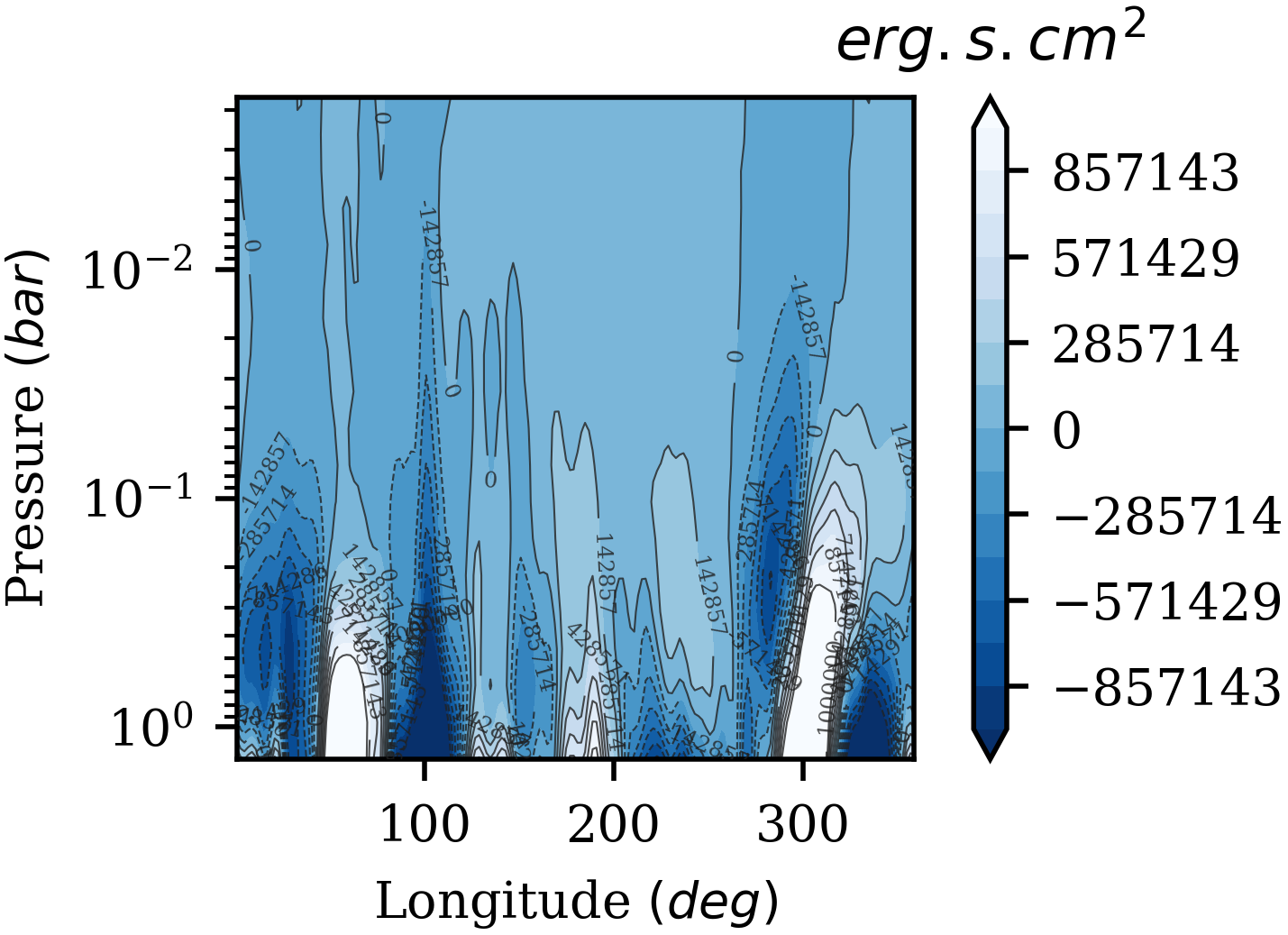}
\caption[]{$1\Omega_{0}$ - Vertical Enthalpy  \label{fig:long_slices_VE_1} }
\end{centering}
\end{subfigure}
\begin{subfigure}{0.3\textwidth}
\begin{centering}
\includegraphics[width=0.99\columnwidth]{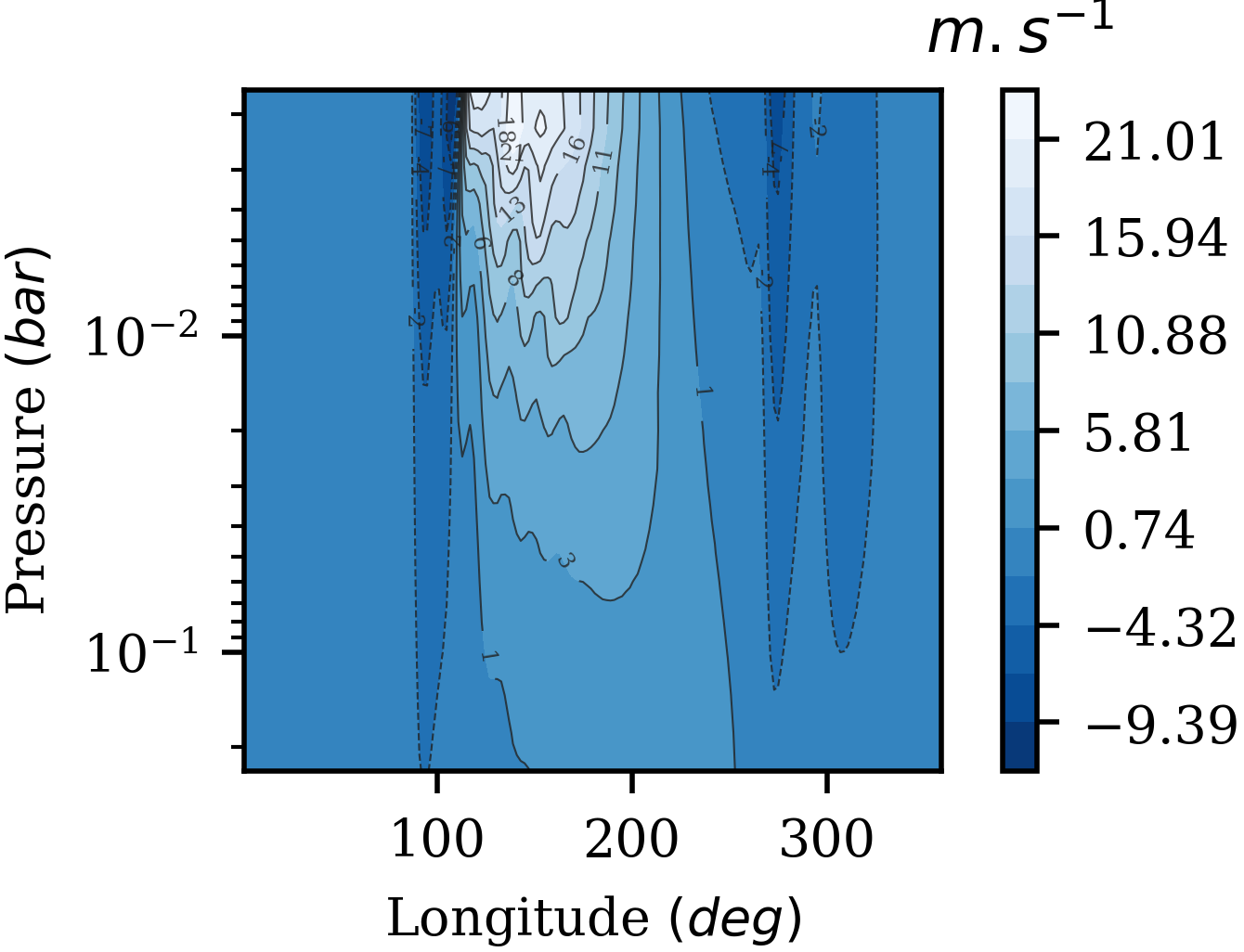}
\caption[]{$15\Omega_{0}$ - Vertical Wind  \label{fig:long_slices_W_15} }
\end{centering}
\end{subfigure}
\begin{subfigure}{0.3\textwidth}
\begin{centering}
\includegraphics[width=0.99\columnwidth]{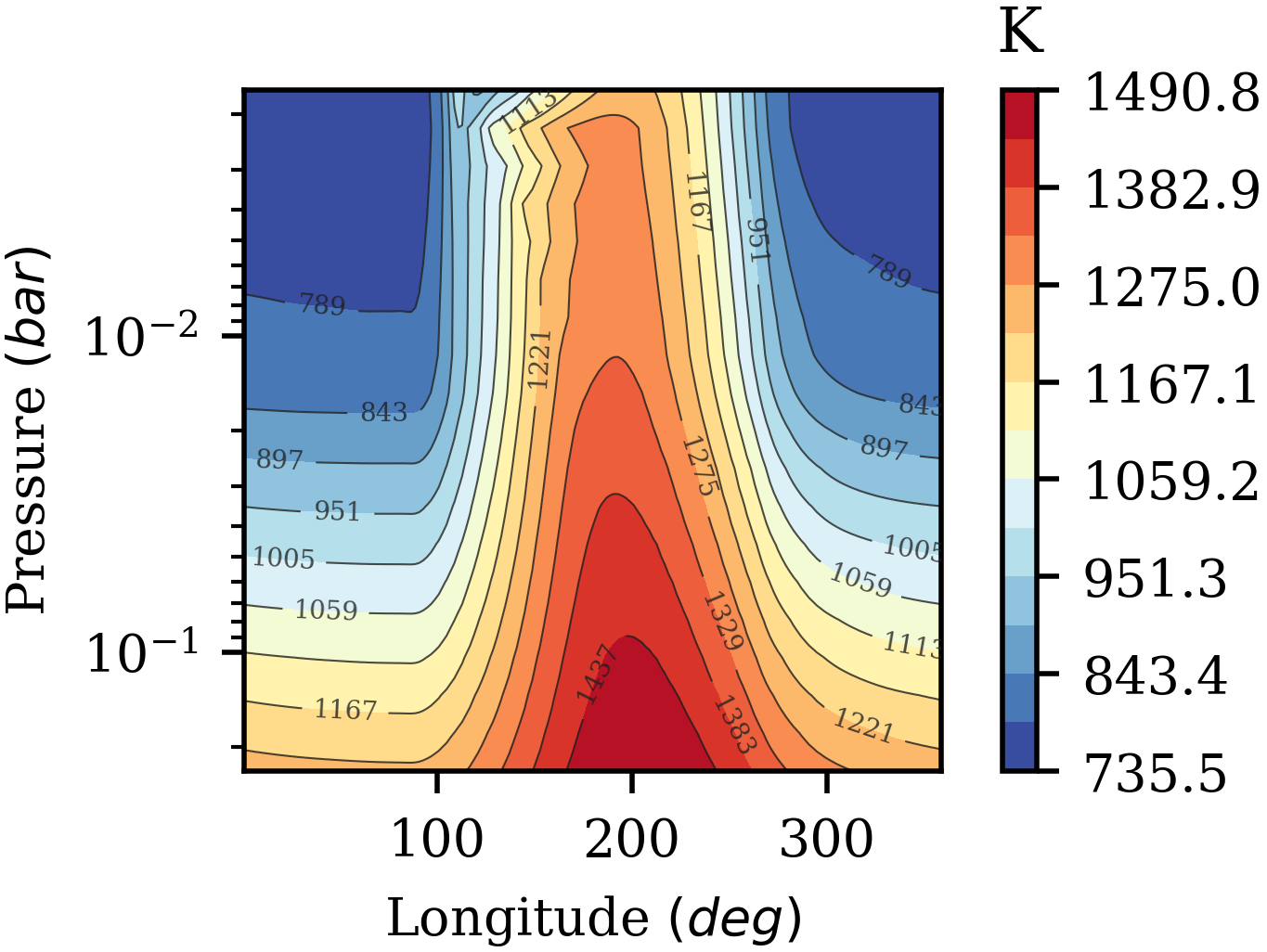}
\caption[]{$15\Omega_{0}$ - Temperature  \label{fig:long_slices_T_15} }
\end{centering}
\end{subfigure}
\begin{subfigure}{0.3\textwidth}
\begin{centering}
\includegraphics[width=0.99\columnwidth]{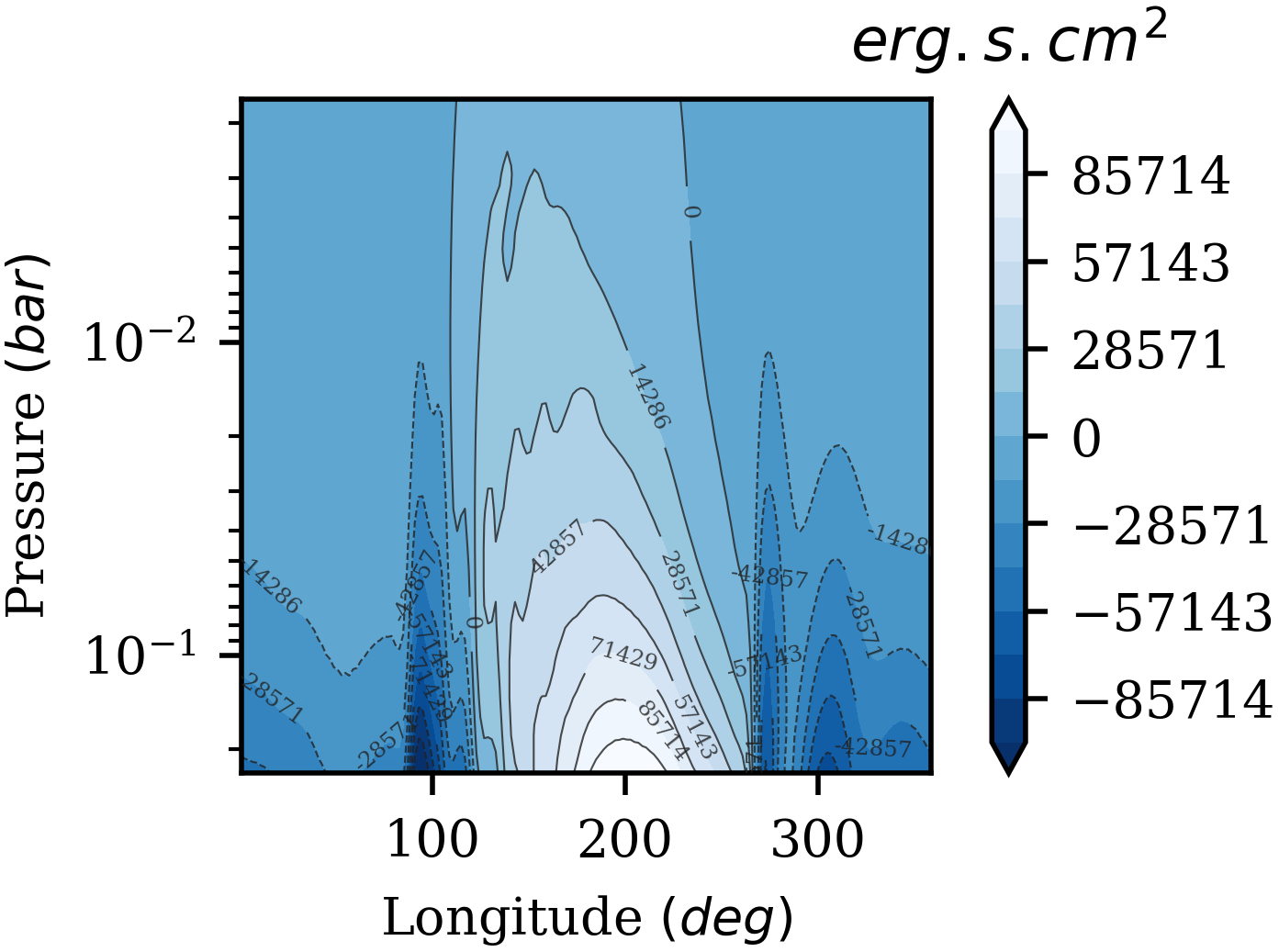}
\caption[]{$15\Omega_{0}$ - Vertical Enthalpy  \label{fig:long_slices_VE_15} }
\end{centering}
\end{subfigure}
\caption[Longitudinal slices at equator]{ Maps showing longitudinal slices of the Vertical Wind (left), Temperature (middle), and Vertical Enthalpy Flux (right), at the equator, for three exemplary HD209458b-like models with rotation rates of; $0.125\Omega_{0}$ - top, $\Omega_{0}$ - middle, and $15\Omega_{0}$ - bottom. Here we have confined our plots to pressures less than 0.26 bar in order to emphasise the dynamics and transport in the outer and mid atmosphere, particularly at pressure levels similar to those at which the night-side hot-spot forms. Note the difference in scale for the vertical enthalpy flux in our $15\Omega_{0}$ case, emphasising the weaker vertical transport, even at the equator, in the rapidly rotating regime. \label{fig:long_slices} }
\end{centering}
\end{figure*} }

\begin{figure*}[tbp] %
\begin{centering}
\begin{subfigure}{0.99\textwidth}
\begin{centering}
\includegraphics[width=0.67\columnwidth]{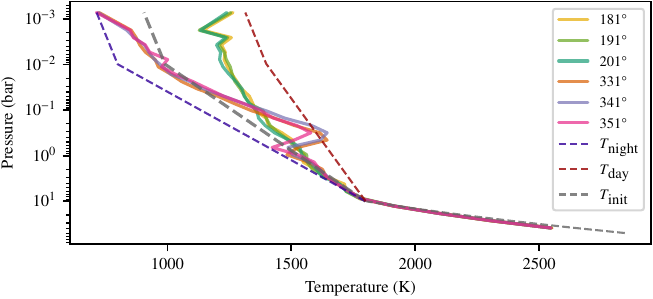}
\caption[]{$0.125\Omega_{0}$  \label{fig:Longitudinal_T_0125x} }
\end{centering}
\end{subfigure}
\begin{subfigure}{0.99\textwidth}
\begin{centering}
\includegraphics[width=0.67\columnwidth]{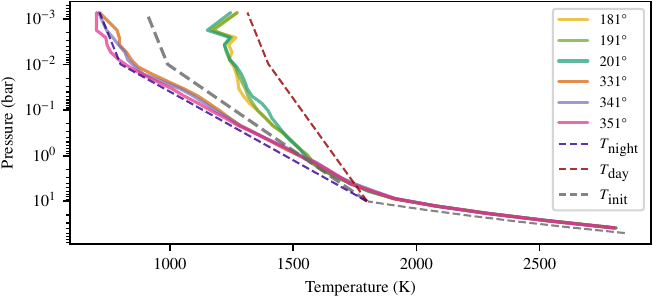}
\caption[]{$1\Omega_{0}$  \label{fig:Longitudinal_T_1x} }
\end{centering}
\end{subfigure}
\begin{subfigure}{0.99\textwidth}
\begin{centering}
\includegraphics[width=0.67\columnwidth]{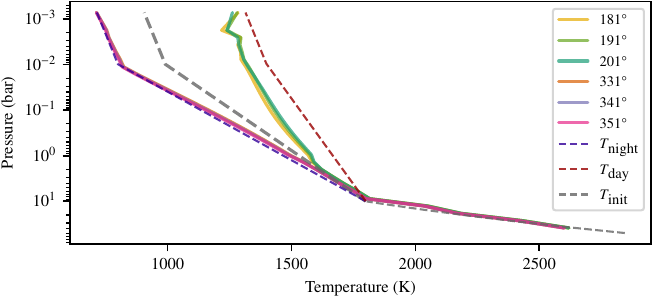}
\caption[]{$15\Omega_{0}$  \label{fig:Longitudinal_T_15x} }
\end{centering}
\end{subfigure}
\caption[Latitudinally averaged T-P profiles at six different longitudes and three different rotation rates]{Latitudinally averaged ({ over a latitudinal band spanning $\pm10^{\circ}$ of the equator}) T-P profiles for three HD209458b-like models at different rotation rates; $0.125\Omega_{0}$ - top, $1\Omega_{0}$ - middle, and $15\Omega_{0}$ bottom. Each plot includes profiles from 6 different longitudes, ranging from the sub-stellar point (whose equilibrium profile is shown in red) eastwards to the anti-stellar point (whose equilibrium profile is shown in blue) on the night-side. { Note that we have excluded the deepest regions of the atmosphere ($P>50\textrm{bar}$) from these plots so as to emphasise the changes that occur with longitude at lower pressures, whilst still showing how the deep atmospheric heating changes (compared to the initial adiabat, shown as a dark grey dash) with rotation rate. } \label{fig:Longitudinal_T} }
\end{centering}
\end{figure*} 

\begin{figure*}[tbp] %
\begin{centering}
\begin{subfigure}{0.3\textwidth}
\begin{centering}
\includegraphics[width=0.99\columnwidth]{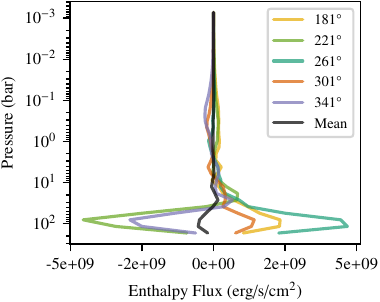}
\caption[]{$0.125\Omega_{0}$ - Longitudinal Variations  \label{fig:Enthalpy_Profiles_0125_slices} }
\end{centering}
\end{subfigure}
\begin{subfigure}{0.3\textwidth}
\begin{centering}
\includegraphics[width=0.99\columnwidth]{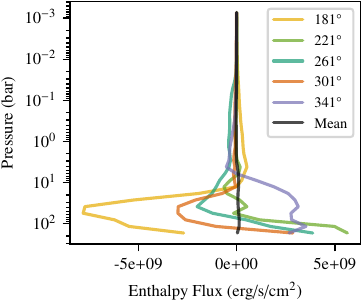}
\caption[]{$1\Omega_{0}$ - Longitudinal Variations   \label{fig:Enthalpy_Profiles_1_slices} }
\end{centering}
\end{subfigure}
\begin{subfigure}{0.3\textwidth}
\begin{centering}
\includegraphics[width=0.99\columnwidth]{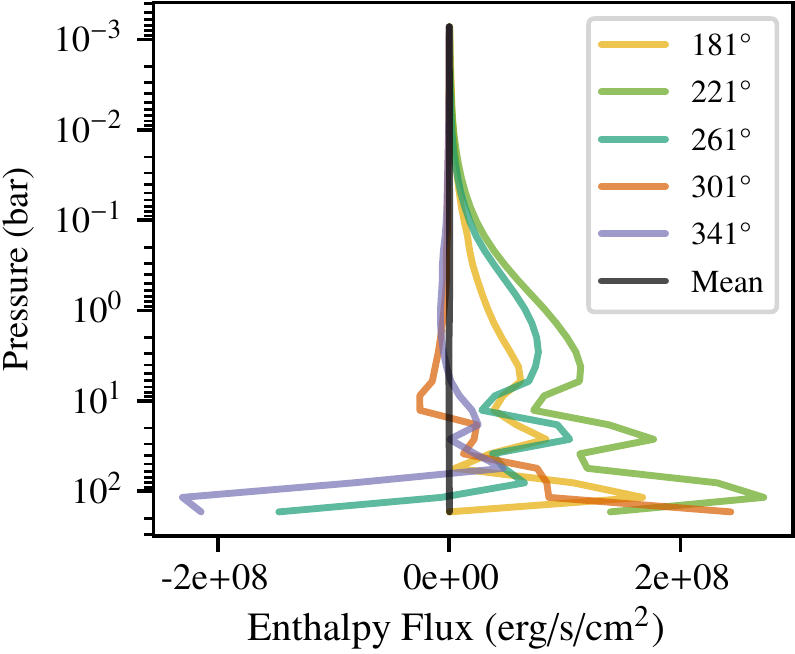}
\caption[]{$15\Omega_{0}$ - Longitudinal Variations   \label{fig:Enthalpy_Profiles_15_slices} }
\end{centering}
\end{subfigure}
\begin{subfigure}{0.3\textwidth}
\begin{centering}
\includegraphics[width=0.99\columnwidth]{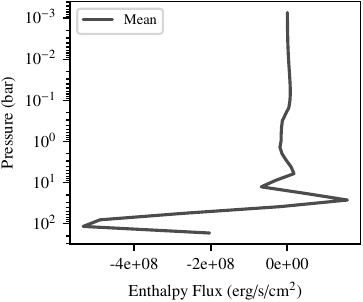}
\caption[]{$0.125\Omega_{0}$ - Global Mean  \label{fig:Enthalpy_Profiles_0125_mean} }
\end{centering}
\end{subfigure}
\begin{subfigure}{0.3\textwidth}
\begin{centering}
\includegraphics[width=0.99\columnwidth]{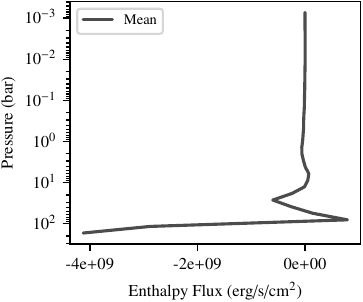}
\caption[]{$1\Omega_{0}$ - Global Mean   \label{fig:Enthalpy_Profiles_1_mean} }
\end{centering}
\end{subfigure}
\begin{subfigure}{0.3\textwidth}
\begin{centering}
\includegraphics[width=0.99\columnwidth]{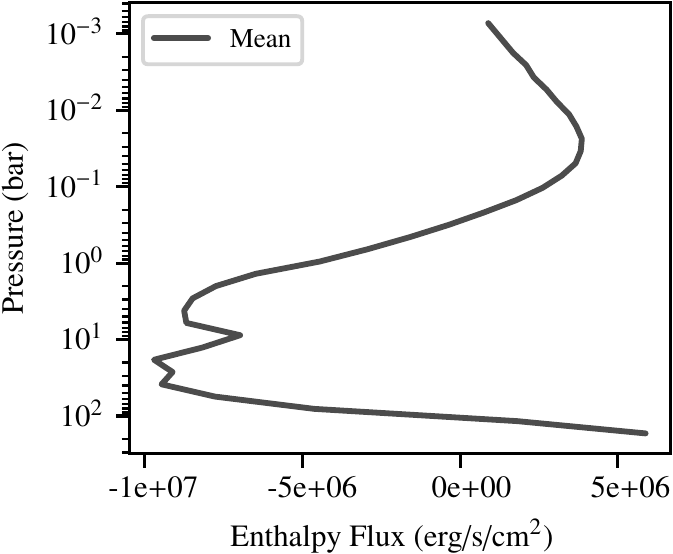}
\caption[]{$15\Omega_{0}$ - Global Mean   \label{fig:Enthalpy_Profiles_15_mean} }
\end{centering}
\end{subfigure}
\caption[Enthalpy transport profiles for three HD209458b-like models]{Longitudinal variations (top) and global means (bottom) of the vertical enthalpy flux $F_\mathcal{H}$ profiles for three exemplary HD209458b-like models; $0.125\Omega_{0}$ - left, $\Omega_{0}$ - middle, and $15\Omega_{0}$ - right. In the top row, we plot vertical enthalpy flux profiles at 5 different longitudes, ranging from the sub-stellar point to just west of the anti-stellar point, as well as the mean enthalpy flux profile. However, since the mean flux is significantly smaller than the local fluxes, we replot the mean profiles in the bottom row such as to better demonstrate the vertical variations in enthalpy transport.    \label{fig:Enthalpy_Profiles} }
\end{centering}
\end{figure*} 

\subsection{Enthalpy Transport} \label{sec:enthalpy_transport}

We expect differences in the horizontal, and hence vertical, wind dynamics to also manifest as significant differences in horizontal/vertical energy transport. 
In order to explore this, we consider the enthalpy flux:
\begin{equation}
F_{H}\left(r,\theta,\phi\right)= \rho c_{p} T \bm{U}\left(r,\theta,\phi\right)\,,
\end{equation}
where $\rho$ is the density, $T$ is the temperature, and $c_{p}$ is the specific heat. We plot the zonal, meridional, and vertical enthalpy flux for select pressure levels, chosen to emphasise differences in outer and deep atmosphere transport, for our three dynamical regimes in \autoref{fig:Enthalpy_Wind} and \autoref{fig:Enthalpy_Wind_vertical}.\\

Starting in the slowly rotating regime, shown in the top row of \autoref{fig:Enthalpy_Wind} and \autoref{fig:Enthalpy_Wind_vertical}, we find that the zonal and meridional transport in the outer atmosphere (\autoref{fig:Enthalpy_Wind_zonal_0016_0.125}/c) is dominated by the divergent day-night wind, with eastward/westward enthalpy transport east/west of the sub-stellar point, and polewards/equatorwards meridional transport of the day/night side. Taken together, these flows suggest that enthalpy is transported from the irradiated day-side to slightly west of the anti-stellar point on the night-side, i.e. exactly where the night-side hot-spot eventually forms (\autoref{fig:Wind_Temp_0.125_0016}). \\
{ In order to better demonstrate the vertical advection of potential temperature/enthalpy in the outer atmosphere, and in particular the role that vertical winds play in forming the night-side hot-spot in the slowly rotating regime, we also plot longitudinal slices of the vertical wind, temperature, and vertical enthalpy flux in \autoref{fig:long_slices}. For example, we can see how vertical transport by a strong night-side downflow, and the associated adiabatic heating, leads to the formation of a hot-spot at around 0.2 to 0.5 bar in the slowly rotating regime. We can also see slight evidence for heating at pressures above the hot-spot as energy is deposited by the descending fluid parcels. Overall, we find a global overturning circulation, with a significant upwelling on the day-side, which is aligned with the upwelling, and the aforementioned downwelling on the night side. This is {typical} of atmospheres whose dynamics are shaped by divergent flows. Note that we find a similar overturning circulation in the classical hot Jupiter and rapidly rotating regimes, although here the advective heat transport is either deep (in the classical regime) or weak (in the rapidly rotating regime). This is in agreement with the differences in deep heating rates found between the three rotational regimes. \\}
The impact that the night-side hot-spot has on enthalpy transport becomes even more apparent at higher pressures. For example, in the mid-atmosphere, the zonal enthalpy transport (\autoref{fig:Enthalpy_Wind_zonal_02_0.125}) is now no longer controlled just by the divergent day-night flows (although divergence driven dynamics remain dominant at high latitudes). Instead, we find significant, equatorial, enthalpy transport eastwards from the night-side towards the day-side/sub-stellar-point, driven by the `reversed' day/night temperature difference. Note that this night to day transport essentially acts to close the overturning-like circulation described above, which has important implications for the deep atmosphere.  Specifically, it means that the deep atmosphere is relatively quiescent and not as strongly linked to the outer atmosphere, leading to { weak vertical enthalpy transport at high pressures (e.g. \autoref{fig:Enthalpy_Wind_vert_0.125_4}, where we find a vertical enthalpy flux an order of magnitude smaller than the flux found in the classical hot Jupiter regime; \autoref{fig:Enthalpy_Wind_vert_1_4})} and hence the suppression of advectively driven radius inflation (\autoref{fig:Longitudinal_T_0125x}). 
Evidence for this can be seen in both \autoref{fig:Enthalpy_Profiles_0125_slices} and \autoref{fig:Enthalpy_Profiles_0125_mean}, which show the vertical enthalpy transport against pressure for various longitudes/the global mean respectively: here we find a significant vertical enthalpy flux in the mid-atmosphere, around the same pressure as the night-side hot-spot, as well as a net (global mean) downward enthalpy flux in the deep atmosphere. { Note that, despite the similarities in enthalpy flux magnitudes in local regions (compare \autoref{fig:Enthalpy_Profiles_0125_slices} and \autoref{fig:Enthalpy_Profiles_1_slices}), we find that overall (i.e. when taking the global mean/calculating the net flux), the downward enthalpy advection is almost an order of magnitude lower here than in the classical hot Jupiter regime (compare \autoref{fig:Enthalpy_Profiles_0125_mean} and \autoref{fig:Enthalpy_Profiles_1_mean}). This can be linked to the different vertical wind structures that the different horizontal wind profiles drive, see for example the much narrower downflow in \autoref{fig:Enthalpy_Wind_vert_0.125} as opposed to \autoref{fig:Enthalpy_Wind_vert_1}.}\\
Finally, we note that the results of this enthalpy transport are clearly visible in the T-P profiles of the mid/outer atmosphere, as shown in \autoref{fig:Longitudinal_T_0125x}. For example, on the night-side between $\sim0.05\rightarrow\sim2\si{\bar}$ we find clear evidence of the night-side hot-spot, including the significant cold-trap associated with the thermal inversion. We also find a reduced day-night temperature difference, suggesting that divergent day-night energy transport is highly efficient in the outer atmosphere, and that the day-night contrast of hot Jupiters might be sensitive to rotation rate. { This is in agreement with the analytic shallow-water models of \citet{2013ApJ...776..134P}, which showed that the day/night temperature contrast should increase with rotation rate (i.e. as the efficient overturning circulation is suppressed by rotational effects). } \\

Moving onto the classical hot Jupiter regime, which we plot on the middle row of \autoref{fig:Enthalpy_Wind}, \autoref{fig:Enthalpy_Wind_vertical}, { and \autoref{fig:long_slices},} we find that the primary driver of enthalpy transport is now the rotational component of the wind. For example, at low pressures, we find that both the zonal and meridional enthalpy transport (\autoref{fig:Enthalpy_Wind_zonal_0016_1}/g) traces the eddy component of the wind (\autoref{fig:Helmholtz_1_eddy}). { As a result we find a significant asymmetry between westwards and eastwards day-night enthalpy transport: westwards of the sub-stellar point we find broad poleward transport which crosses the entire night-side, whereas eastwards of the sub-stellar point we find a much more longitudinally constrained profile linked with the off-equator wings that from in the outer and mid atmosphere (e.g. \autoref{fig:Wind_Temp_1_0016}). In turn, these off-equator wings are associated with a strong downward vertical enthalpy transport (\autoref{fig:Enthalpy_Wind_vert_1}) that is somewhat balanced by a broad upwelling near the sub-stellar point and to the east of the off-equator wings (\autoref{fig:Enthalpy_Wind_vert_1_0.1}). Together forming a global circulation reminiscent of the global `overturning' circulation.} \\
However unlike in {the slowly rotating regime}, we find that this global `overturning' circulation extends deep into the atmosphere, where radiative forcing is weak, thus connecting the outer and deep atmospheres, and allowing for deep heating { (\autoref{fig:Longitudinal_T_1x} and \autoref{fig:long_slices_VE_1})}.
As for the reason why this circulation penetrates deeper into the atmosphere, this is because of the strong, deep zonal jet which drives strong mixing near the equator, matching what we find in the meridional circulation streamfunction (\autoref{fig:Streamfunction_1}). Evidence for this deeper zonal jet can be seen in \autoref{fig:Enthalpy_Wind_zonal_02_1}, which reveals strong eastwards advection near the equator at all longitudes, paired with weak off-equator transport linked to the equatorward pumping of eastwards angular momentum from mid-latitudes. \\
Overall, these strong zonal/equatorial dynamics drive significant vertical enthalpy transport, as shown in { \autoref{fig:Enthalpy_Wind_vert_1_4},  \autoref{fig:long_slices_VE_1}}, and \autoref{fig:Enthalpy_Profiles_1_slices}/e, where we find the highest net/global-mean downwards vertical enthalpy of our three regimes. A strong enthalpy flux which drives significant deep heating (see \autoref{fig:Longitudinal_T_1x} and the dashed line in the deep atmosphere which shows the initial adiabat), suggesting that the planet is inflated relative to both more slowly rotating hot Jupiters and 1D atmospheric models \citep{2019A&A...632A.114S}. \\

Finally we come to the rapidly rotating regime, which we plot on the bottom row of \autoref{fig:Enthalpy_Wind}, {  \autoref{fig:Enthalpy_Wind_vertical}, and \autoref{fig:long_slices}}. Here we find that advective heat transport has become increasingly confined to low latitudes, with the equatorial jet acting as the main source of horizontal transport at almost all pressures. For example, in the outer atmosphere, \autoref{fig:Enthalpy_Wind_zonal_0016_15}/k, we find horizontal enthalpy transport which is either driven by the jet (zonally) or by the very highly tilted and latitudinally constrained $m=1$ standing wave pattern (meridional). The jet transports enthalpy eastwards from the sub-substellar point, across the night-side, leading to the band of equatorial heating/mixing seen in \autoref{fig:Wind_Temp_15_02}, whilst the tilted standing wave primarily drives transport on the day-side (poleward/westward advection west of the sub-stellar point and equatorward/eastward advection to the east) and leads to the warped/compressed thermal butterfly also seen in \autoref{fig:Wind_Temp_15_02}. \\
{ The influence of rotation on the enthalpy transport is also evident in the vertical enthalpy flow which is highly confined to low-latitude/equatorial regions. For example, as shown in \autoref{fig:Enthalpy_Wind_vert_15} both the days-side upwelling and night-side downflow are confined to be within $\sim20^{\circ}$ of the equator,} which explains the lack of high-latitude downflows found at low Rossby numbers (\autoref{fig:Rossby_vs_W_mean}). Further, due to the day-side confinement of the off-equator circulations, we find that the night-side downflows tend to be closer to the terminators/day-side than in the more slowly rotating regimes, although this changes as we move deeper into the atmosphere and the downflows become more tightly associated with a zonally-uniform zonal jet, as seen in { \autoref{fig:Enthalpy_Wind_vert_15_0.1}/i} and \autoref{fig:Enthalpy_Wind_zonal_02_15}. { Evidence for this confinement of the downflows to the day-side can be seen in the left-hand column of \autoref{fig:long_slices}. }  \\ 
Overall, the Coriolis driven confinement of the horizontal and vertical enthalpy fluxes has a suppressing effect on the driving of deep atmospheric heating. That is to say that we find that, as shown in \autoref{fig:Enthalpy_Profiles_15_slices} downward enthalpy advection is significantly weaker than both of the other regimes considered here, suggesting that deep heating, and hence radius inflation, should be strongly suppressed. And the global-mean deep T-P profile confirms that this is indeed the case. However. if we confine our investigation of the deep atmosphere to equatorial regions, as we do in { \autoref{fig:long_slices_VE_15} and} \autoref{fig:Longitudinal_T_15x}, we do find slight hints of deep heating. This occurs because the local vertical enthalpy advection is orders of magnitude stronger near the equator than elsewhere (see { \autoref{fig:Enthalpy_Wind_vert_15_0.1}/i}), and the latitudinal confinement for the off-equator circulations mean that latitudinal mixing is weak. With time, this mixing might lead to global heating of the deep atmosphere, albeit to a much lesser extent than in the classical regime, however a quick test with even weak radiative forcing back towards the initial adiabat (designed to reproduce the slight radiative cooling of the hot deep atmosphere) is more than enough to prevent this heating. As such, we feel confident in arguing that models in the rapid rotation regime should show little to no radius inflation, in agreement with the SDSS1411b results of \citet{2021A&A...656A.128S}. { Note that the slight northward shift of the downward enthalpy flux seen in { \autoref{fig:Enthalpy_Wind_vert_15_0.1}/i} is driven by the slight asymmetry in the zonal winds, as seen in \autoref{fig:Zonal_Wind_15}.} \\

\subsection{The Rossby Deformation Radius and Rhines Length Scales} \label{sec:length_scales}
\begin{figure}[tbp] %
\begin{centering}
\includegraphics[width=0.75\columnwidth]{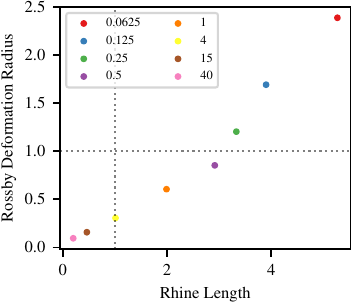}
\caption[Synchronous Rossby Deformation and Rhines Length]{ A comparison of the maximum normalised Rossby deformation radius ($R_{rossby}$) and Rhines length ($R_{rhine}$) for eight of our synchronously rotating HD209458b-like atmospheric models which span the three rotation regimes discussed here. This figure can be compared with Figure 4 of \citet{2018ApJ...852...67H}.  \label{fig:rossby_rhines_comp} }
\end{centering}
\end{figure}

{ We finish by briefly comparing our three rotational regimes to those proposed by \citet{2018ApJ...852...67H} for terrestrial exoplanets. \citet{2018ApJ...852...67H} use the non-dimensional Rossby Deformation Radius, 
\begin{equation}
  R_{rossby} = \frac{NH}{2\Omega R_{p}}
\end{equation}
where H is the atmospheric scale height and N is the Brunt–Väisälä Frequency, and the Rhine Length, 
\begin{equation}
  R_{rhine} =\frac{1}{R_{p}}\sqrt{\frac{U_{rms}R_{p}}{2\Omega}}, 
\end{equation}

to define three dynamical regimes. {The slow rotation regime when both the normalised Rossby deformation radius ($R_{rossby}$) and Rhine length ($R_{rhine}$) are $>1$, the Rhine rotation regime when $R_{rossby}>1$ and $R_{rhine}<1$, and the rapidly rotating regime when both $R_{rossby}<1$ and $R_{rhine}<1$. They also propose a forth regime when $R_{rossby}<1$ and $R_{rhine}>1$, although none of their models fall into this regime.} 
Each of these regimes are characterised by different dynamics, including global overturning circulation in the slowly rotating regime, significant equatorial super-rotation in the rapidly rotating regime, and a mix of dynamics in the Rhines regime. These regimes are broadly in line with the regimes we discuss here with a a global overturning circulation which dominates for slow rotators ($\Omega<0.5\Omega_{0}$) and a zonal-jet driven by the standing Rossby and Kelvin waves which dominates for rapid rotators ($\Omega\gtrsim5\Omega_{0}$). 
{However at intermediate rotation rates, our `classical' hot Jupiter regime, instead of our models falling into the Rhines rotator regime, we instead fall into their theoretical fourth regime which, like the Rhines rotator regime, is characterised by a combination of a global overturning circulation and a standing-wave driven jet.}
This alignment with the regimes of \citet{2018ApJ...852...67H} can be seen in \autoref{fig:rossby_rhines_comp}, where we plot the maximum values of the normalised Rossby deformation radius and Rhines length for eight of our synchronous models which span the three rotation regimes. \\}
{ It is important to note that both the Rossby deformation radius and the Rhines length become very short in the rapidly rotating regime. In fact, depending upon the location in which the Rhines length is calculated, for our most rapidly rotating model ($\Omega=40\Omega_{0}$) the Rhines length can approach the resolution of our model: if we consider the global root mean square velocity ($U_{rms}$), we find that $R_{rhine}=2.5^\circ$, which is essentially identical to the horizontal extent of a grid cell in our models. However this value is so low in part because much of the atmosphere is quiescent in the rapidly rotating regime due to the suppression of off-equator motions by Coriolis forces. As such, if we instead constrain our calculation of $U_{rms}$ to winds within 10 degrees of the equator, we find that $R_{rhine}=6.0^\circ$, which we can resolve, if only just.\\
Consequently, given the interesting dynamics found here, we suggest that future studies investigate the rapidly rotating regime at a significantly higher resolution in order to fully capture all scales of the atmospheric dynamics at all latitudes.}

\section{Non-Synchronous Rotation} \label{sec:NS}
Many of the dynamics found in the synchronously rotating regime (\autoref{sec:synchronous}) rely on the presence of an $m=1$ standing wave (\autoref{fig:Helmholtz}), which develops thanks to the strong, persistent, day-night forcing associated tidally-locked irradiation \citep{2011ApJ...738...71S}. However for warm Jupiters with longer orbital radii (i.e. $a=0.2\si{\astronomicalunit}$ for a Sun-like star, like HD209458A, even closer for an M-Dwarf), the synchronisation time-scale starts to approach the system/planetary age. \\
Thus there is a strong possibility that next-generation exoplanet hunting missions will start to detect warm/hot Jupiters whose rotation period is not perfectly synchronised with their orbital period. This might have profound implications for their atmospheric dynamics. As such, we next explore how both small and large differences between the orbital and rotation period (i.e. non-synchronicity) affect the dynamics and resulting heat transport of such objects. Here our models fall into two regimes: the first is models in which the difference between the orbital ($\Omega_{\mathrm{orb}}$) and rotation ($\Omega_{\mathrm{rot}}$) angular rotation rates is a fraction of the overall rotation rate (e.g. $\Omega_{\mathrm{rot}}=0.9\Omega_{\mathrm{orb}}=0.9\Omega_{0}$ - more generally when the difference is less than $\sim15\%$ of the rotation rate - here that is the rotation rate of HD209458b), and the effect of non-synchronicity is relatively minor. And the second regime is when the difference is significant (e.g. $\Omega_{\mathrm{rot}}=4\Omega_{\mathrm{orb}}=4\Omega_{0}$), on the order of or greater than either the orbital or angular rotation rate, leading to significant disruption of the driving mechanisms discussed in \autoref{sec:synchronous}. { This work builds upon the non-synchronous models of, for example, \citet{Showman_2009,2014arXiv1411.4731S} by more thoroughly exploring the non-synchronous parameter space and investigating the effects of non-synchronous rotation on the vertical energy transport.}

\begin{figure*}[tbp] %
\begin{centering}
\begin{subfigure}{0.48\textwidth}
\begin{centering}
\includegraphics[width=0.8\columnwidth]{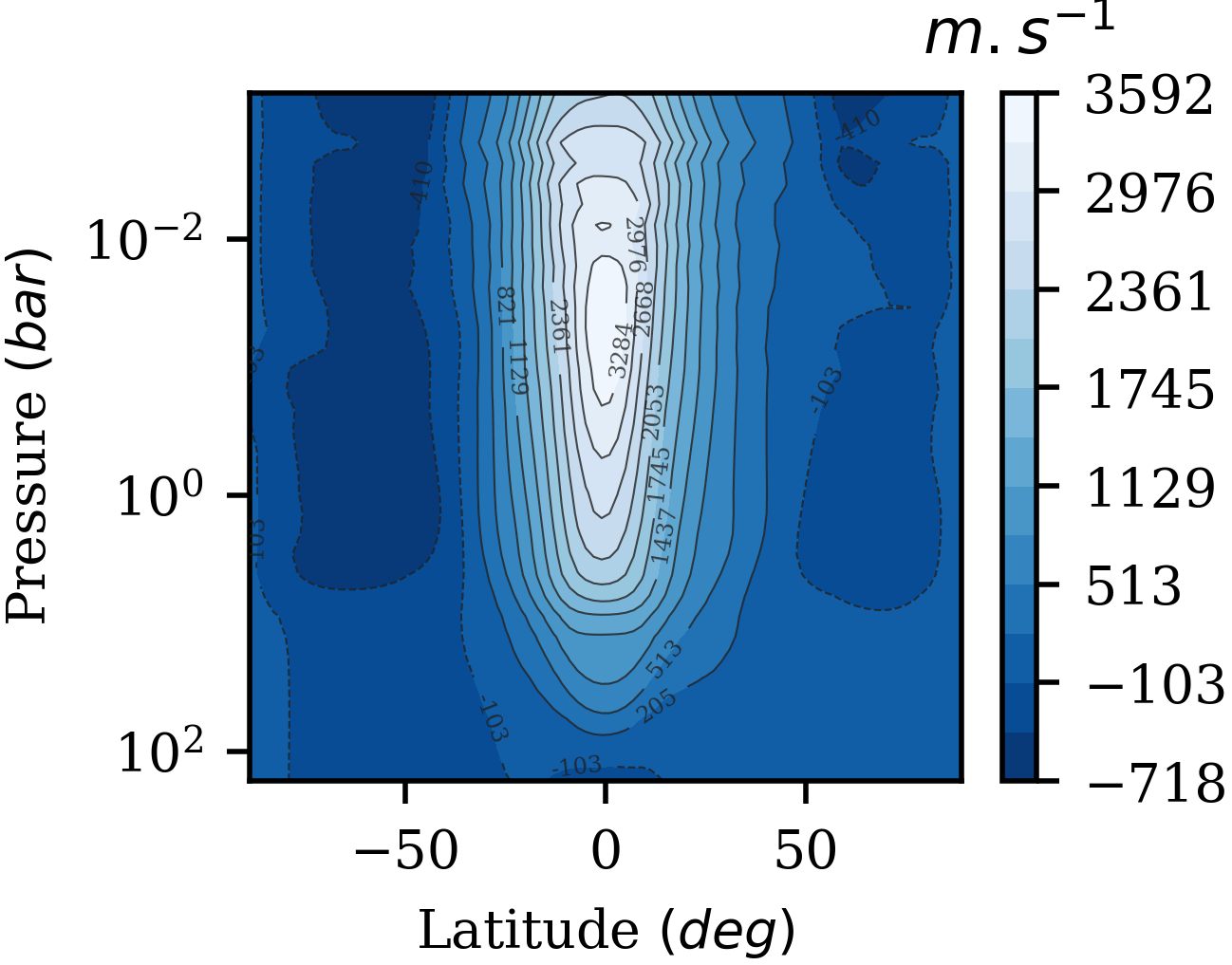}
\caption[]{$\Omega_{\mathrm{rot}}=0.9\Omega_{0}$ - Non-Synchronous  \label{fig:Zonal_wind_NS_09} }
\end{centering}
\end{subfigure}
\begin{subfigure}{0.48\textwidth}
\begin{centering}
\includegraphics[width=0.8\columnwidth]{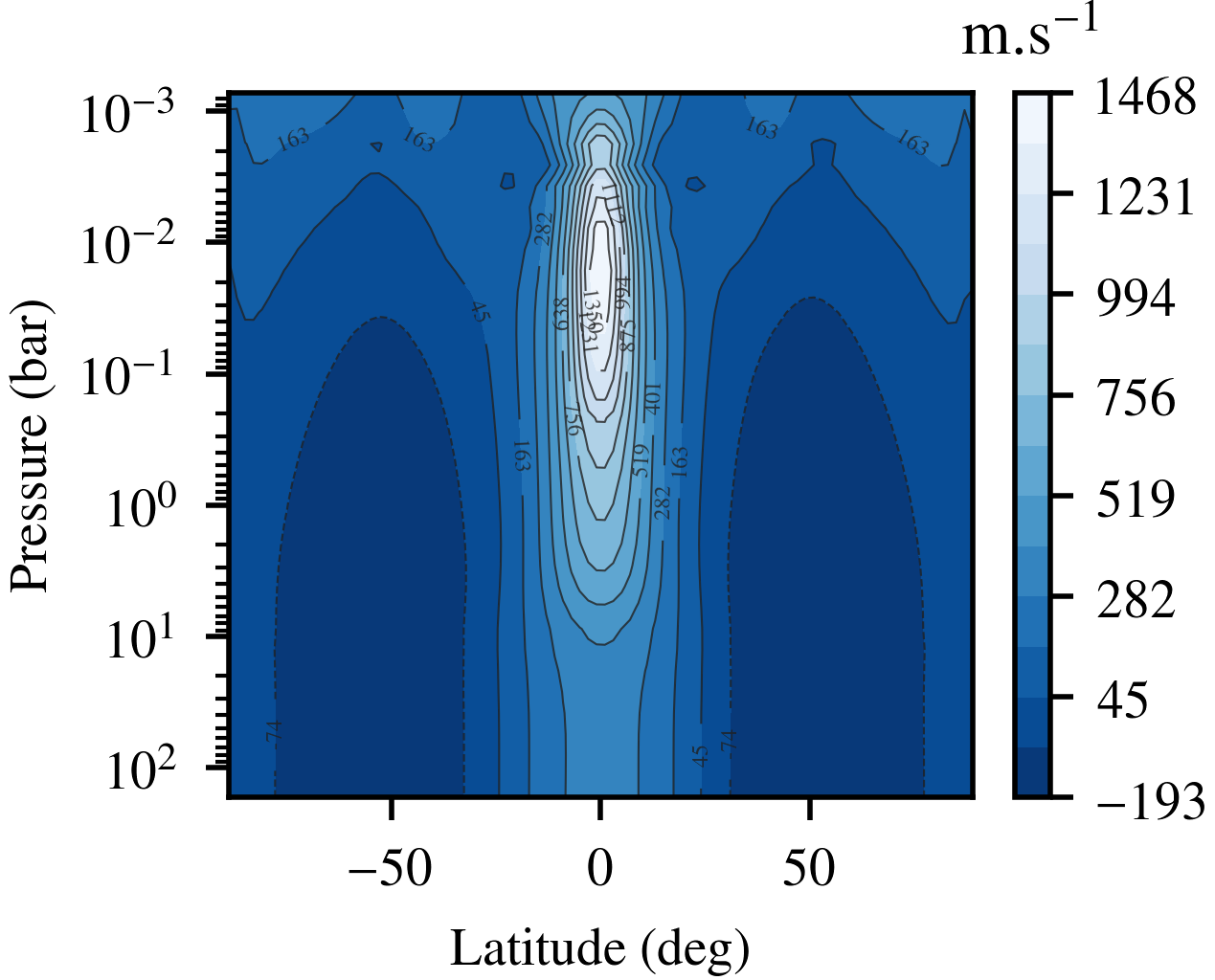}
\caption[]{$\Omega_{\mathrm{rot}}=4.0\Omega_{0}$ - Non-Synchronous   \label{fig:Zonal_wind_NS_4} }
\end{centering}
\end{subfigure}
\begin{subfigure}{0.48\textwidth}
\begin{centering}
\includegraphics[width=0.8\columnwidth]{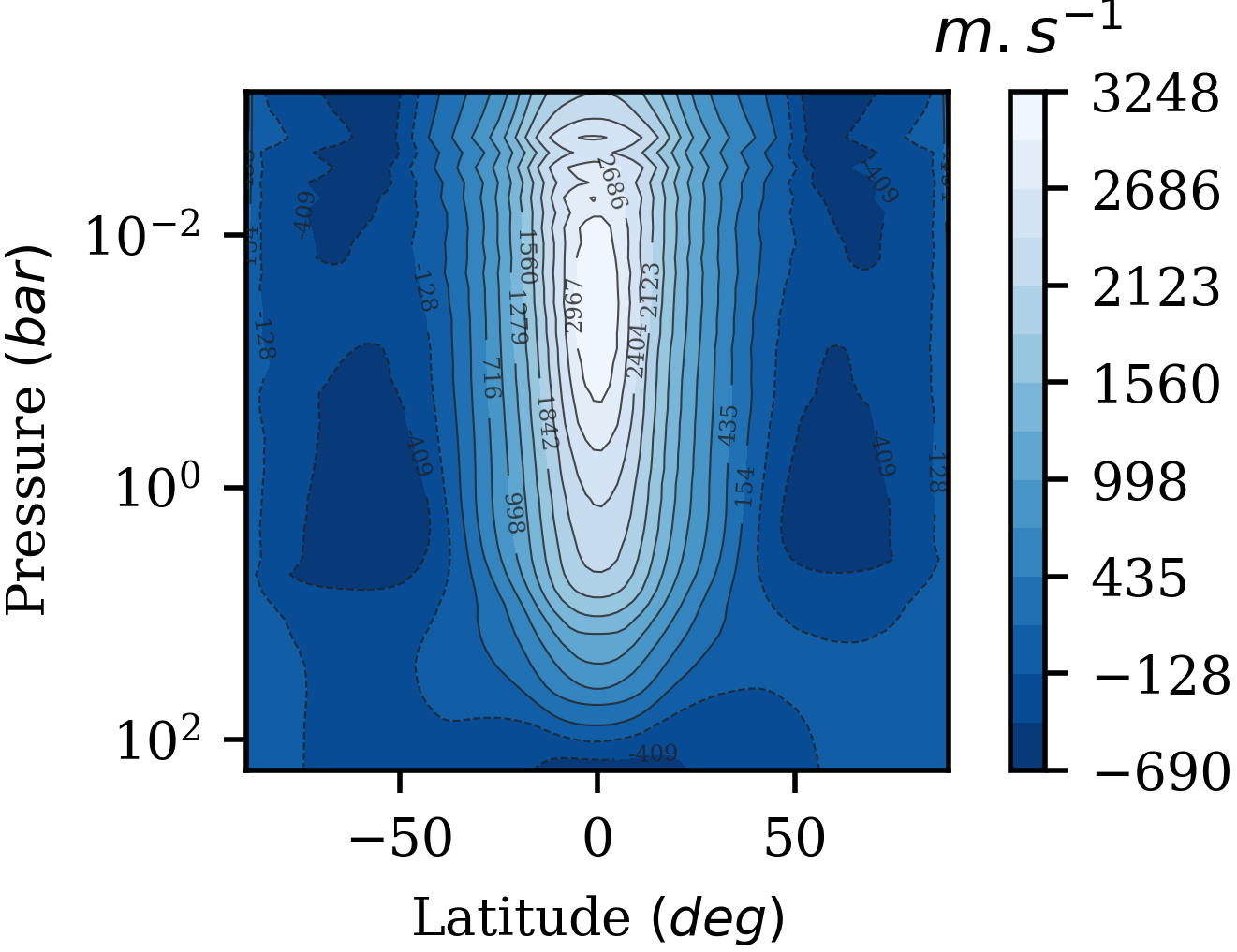}
\caption[]{$\Omega=0.9\Omega_{0}$ - Synchronous  \label{fig:Zonal_wind_SYNC_09} }
\end{centering}
\end{subfigure}
\begin{subfigure}{0.48\textwidth}
\begin{centering}
\includegraphics[width=0.8\columnwidth]{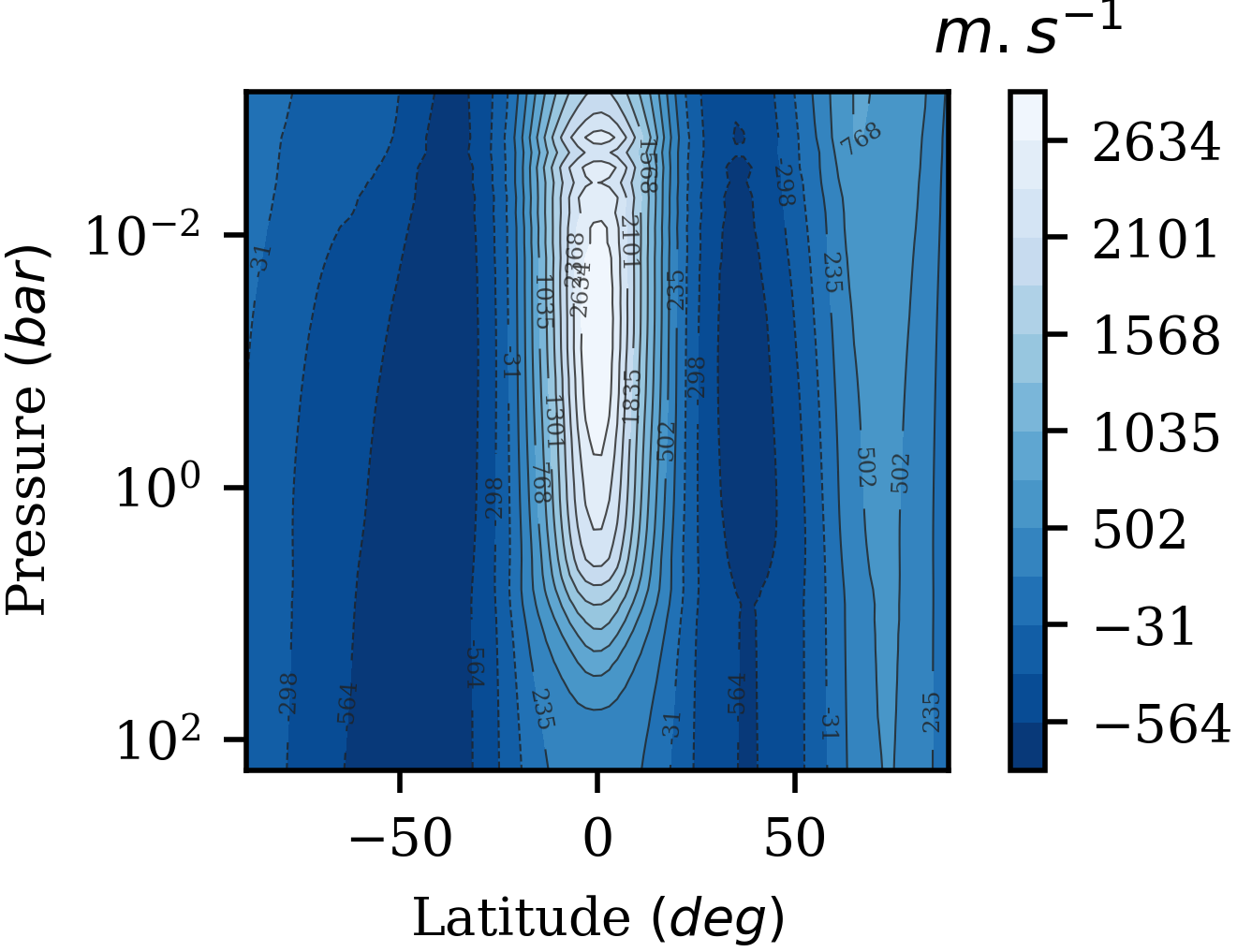}
\caption[]{$\Omega=4.0\Omega_{0}$ - Synchronous   \label{fig:Zonal_wind_SYNC_4} }
\end{centering}
\end{subfigure}
\caption[Zonal wind profiles for two synchronous and two non-synchronous HD209458b-like models ]{ Zonally and temporally, { over five Earth-years of simulation time,} averaged zonal wind profiles for four HD209458b-like models, two of which are non-synchronously rotating (top) and two of which are synchronously rotating (bottom), with rotation rates chosen to fall into our weak-non-synchronicity { (left - $\Omega_{\mathrm{rot}}=0.9\Omega_{\mathrm{orb}}=0.9\Omega_{0}$ in the non-synchronous model and $\Omega_{\mathrm{rot}}=\Omega_{\mathrm{orb}}=0.9\Omega_{0}$ in its synchronous equivalent) and strong-non-synchronicity (right - $\Omega_{\mathrm{rot}}=4\Omega_{\mathrm{orb}}=4.0\Omega_{0}$ in the non-synchronous model and $\Omega_{\mathrm{rot}}=\Omega_{\mathrm{orb}}=4.0\Omega_{0}$ in its synchronous equivalent)} regimes. \label{fig:Zonal_wind_NS_Sync} }
\end{centering}
\end{figure*} 
\begin{figure}[tbp] %
\begin{centering}
\includegraphics[width=0.75\columnwidth]{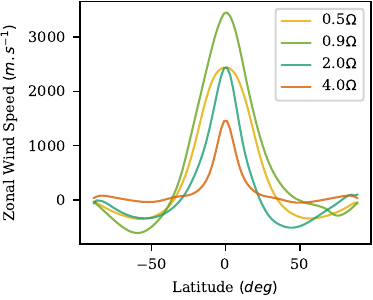}
\caption[Non-Synchornous zonal wind slices for 4 HD209458b-like models]{ Longitudinally averaged zonal wind profiles in the outer atmospheres (P $\simeq0.016\textrm{bar}$) of four non-synchronous HD209458b-like models, with rotation rates ($\Omega_{\mathrm{rot}}$)  chosen to fall into our weak-non-synchronicity ($\Omega_{\mathrm{rot}}=0.9\Omega_{0}$ - green, and $\Omega_{\mathrm{rot}}=2.0\Omega_{0}$ - teal) and strong-non-synchronicity ($\Omega_{\mathrm{rot}}=0.5\Omega_{0}$ - yellow, and $\Omega_{\mathrm{rot}}=4.0\Omega_{0}$ - orange) regimes. \label{fig:OA_ZW_Non_Sync} }
\end{centering}
\end{figure} 
\subsection{Zonal-Mean Zonal-Winds} \label{sec:NS_zonal}
We start by exploring how the zonal-mean zonal-wind changes as we increase the difference between the rotational and orbital angular rotation rates (i.e. as we increase the non-synchronicity). \\
To that end we plot, on the top row of \autoref{fig:Zonal_wind_NS_Sync} the zonal-mean zonal-wind for models with both a small difference between the orbital and rotational angular rotation rates ($\Omega_{\mathrm{rot}}=0.9\Omega_{\mathrm{orb}}=0.9\Omega_{0}$ - left - weakly non-synchronous) and a large difference ($\Omega_{\mathrm{rot}}=4\Omega_{\mathrm{orb}}=4.0\Omega_{0}$ - right - highly non-synchronous). Further, to aid in our comparisons, we also plot models with the same rotation rates, but with their orbital periods adjusted such that the rotation is once again synchronous ($\Omega_{\mathrm{rot}}=\Omega_{\mathrm{orb}}=0.9\Omega_{0}$ on the left and $\Omega_{\mathrm{rot}}=\Omega_{\mathrm{orb}}=0.9\Omega_{0}$ on the right).\\
We start in the weakly non-synchronous regime, where we find a zonal-mean zonal-wind (\autoref{fig:Zonal_wind_NS_09}) that is very similar to its synchronously-rotating counterpart (\autoref{fig:Zonal_wind_SYNC_09}), with differences on the order of that found when changing the averaging period of the temporal mean. This suggests that, when the difference between the rotational and orbital periods is small, the day-night temperature forcing that develops due to the strong stellar irradiation is sufficient to drive a long-lived standing Rossby and Kevin wave pattern leading to the development of a super-rotating jet equivalent to that found in the synchronous regime. { This is consistent with the super-synchronous (1.5 times; see Figure 14) models of \citet{Showman_2009}}.  \\
However as we increase the difference between the orbital and rotational angular rotation rates, we find that the differences between non-synchronous (\autoref{fig:Zonal_wind_NS_4}) and synchronous (\autoref{fig:Zonal_wind_SYNC_4}) models with the same planetary rotation rate starts to grow. In this highly non-synchronous regime, we find that: a) both eastwards (equatorial jet) and westwards (mid-latitude) winds are slower than in the synchronous regime and b) the weak eastwards equatorial jet that does develop is notably shallower than its synchronous counterpart, which in turn implies that vertical enthalpy advection/heating will be weakened/suppressed (see \autoref{sec:NS_meridional} and \autoref{sec:NS_enthalpy}). These changes suggest that the time-dependent shift in the outer atmospheric irradiation (caused by the difference between the rotational and orbital period) hampers the formation of the standing wave pattern which drives both on and off-equator dynamics. We investigate the exact nature of these changes in \autoref{sec:NS_horizontal_wind_temp}. { This shift is similar to that found by \citet{Showman_2009}, a slower, shallower, equatorial jet driven by a (then) theorised weakening in standing wave driven dynamics. Furthermore, like \citet{Showman_2009}, we also find that a high latitude and low pressure eastward flows form when the angular rotation rate significantly exceeds the orbital rotations rate. Although here our flows are slower, likely as a result of differences in the radiative treatment.} \\

We can better understand how a difference between the rotational and orbital periods can affect the shape/speed of the zonal jet by looking at the latitudinal variations in the zonal-mean outer atmosphere wind, as shown in \autoref{fig:OA_ZW_Non_Sync}. Comparing this with \autoref{fig:Latitudinal_Wind}, we can see that, when the level of non-synchronicity is small (e.g. $\Omega_{\mathrm{rot}}=0.9\Omega_{\mathrm{orb}}$), the jet closely resembles that found in the synchronous regime. However as the difference between the rotational and orbital periods grows, things start to change, with both the peak jet speed and jet width decreasing. Eventually we reach the point ($\Omega_{\mathrm{rot}}=4.0\Omega_{\mathrm{orb}}$) that off-equator counterflows have almost dropped to zero, which reinforces our conclusion that it is a lack of off-equator standing-waves which leads to the lack of a strong super-rotating jet in highly non-synchronous atmospheres. \\
Note that the influence of rotation can also be seen in \autoref{fig:OA_ZW_Non_Sync}, with models in which we double/half the rotation rate with respect to the orbital period showing similar peak jet speeds and off-equator flows, but different jet widths: the jet appears wider in the more slowly rotating model due to the increased influence of divergent winds on the dynamics, as was seen in the synchronous regime (\autoref{sec:helmholtz_wind}). A similar story holds true for the cases discussed here. For example, our model with $\Omega_{rot}=1.1\Omega_{\mathrm{orb}}$ is very similar to the case with $\Omega_{\mathrm{rot}}=0.9\Omega_{\mathrm{orb}}$, just with a slightly narrower jet. This is because, in both cases, the level of non-synchronicity is low so the main driver of differences is the relative influence of rotation. \\ 

\begin{figure*}[tbp] %
\begin{centering}
\begin{subfigure}{0.48\textwidth}
\begin{centering}
\includegraphics[width=0.8\columnwidth]{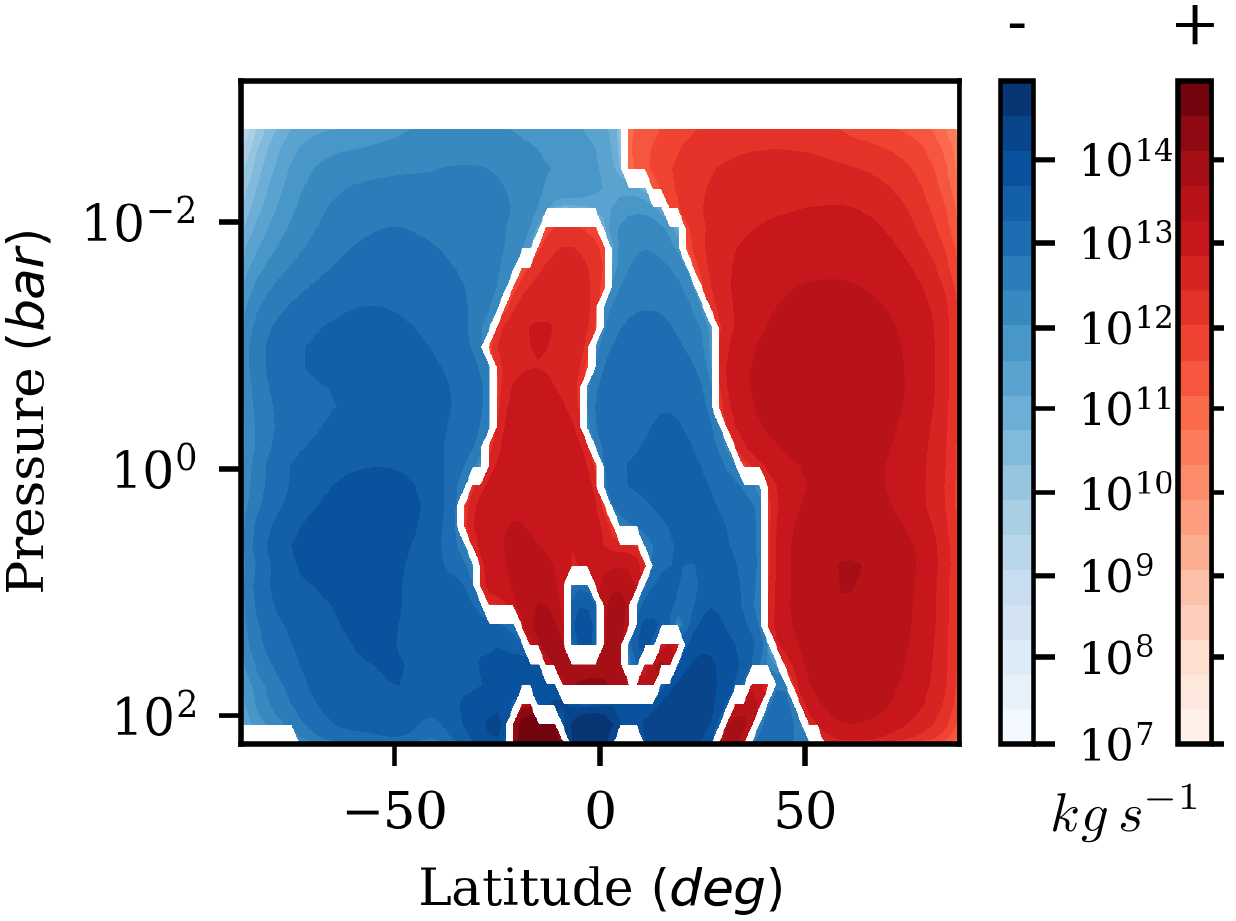}
\caption[]{$\Omega_{\mathrm{rot}}=0.9\Omega_{0}$ - Non-Synchronous  \label{fig:SF_NS_09} }
\end{centering}
\end{subfigure}
\begin{subfigure}{0.48\textwidth}
\begin{centering}
\includegraphics[width=0.8\columnwidth]{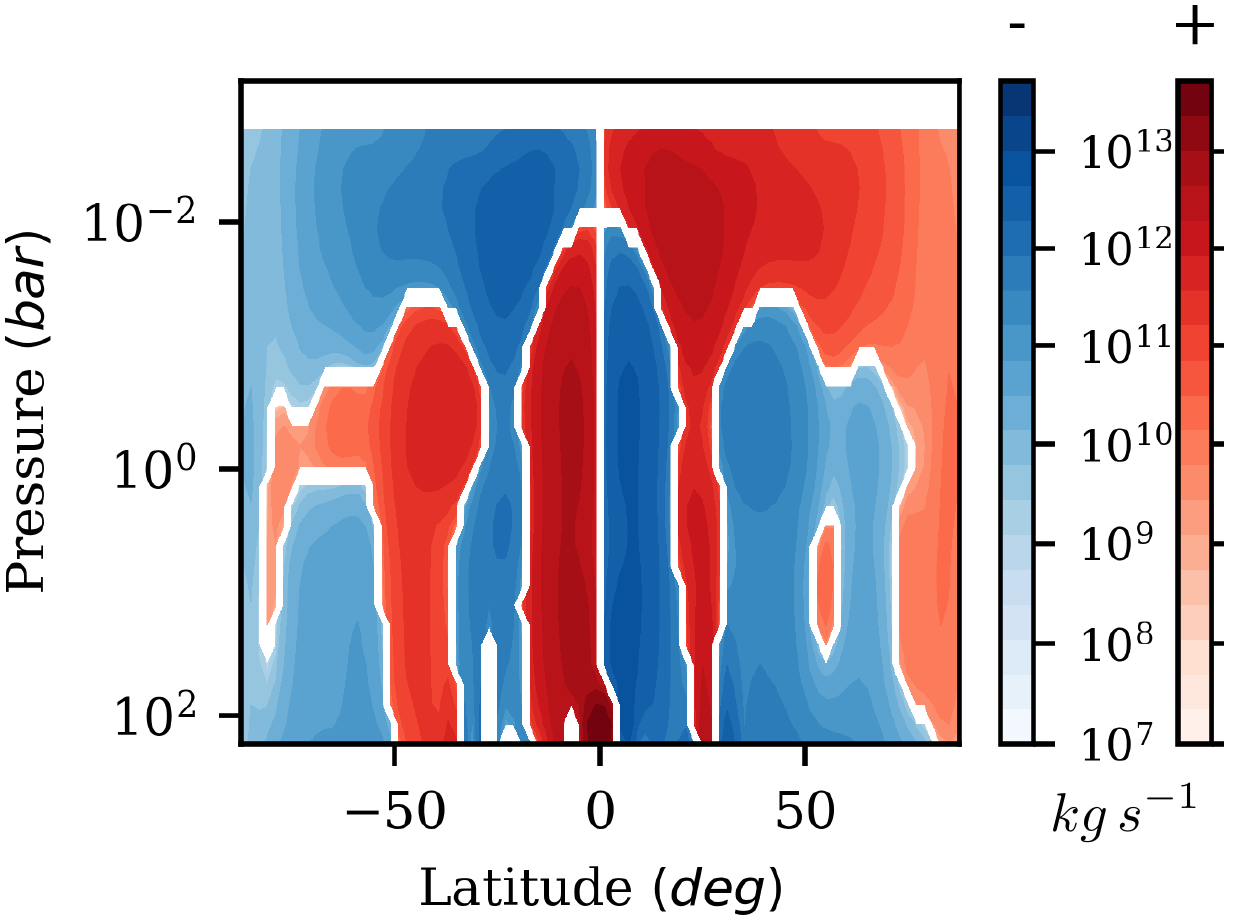}
\caption[]{$\Omega_{\mathrm{rot}}=4.0\Omega_{0}$ - Non-Synchronous   \label{fig:SF_NS_4} }
\end{centering}
\end{subfigure}
\begin{subfigure}{0.48\textwidth}
\begin{centering}
\includegraphics[width=0.8\columnwidth]{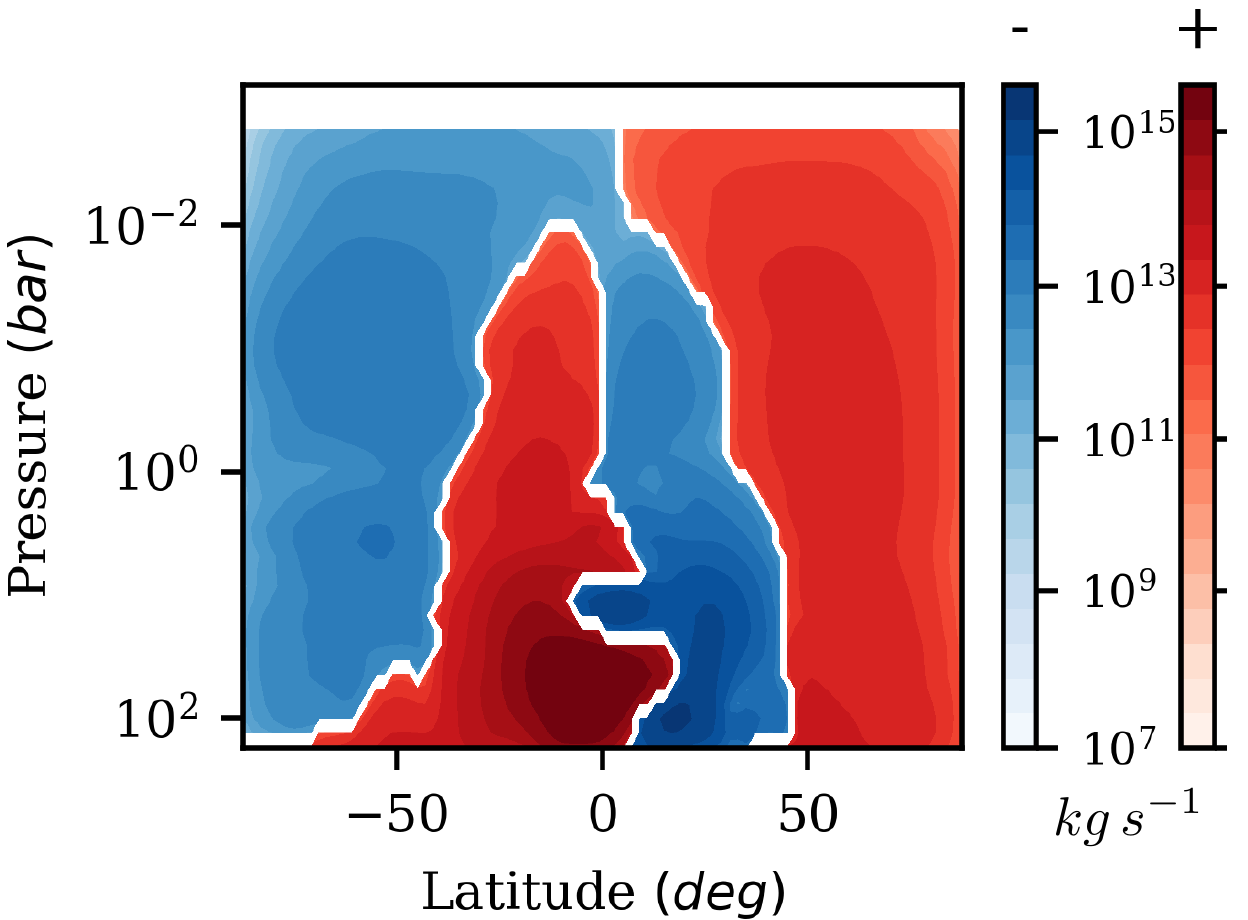}
\caption[]{$\Omega=0.9\Omega_{0}$ - Synchronous  \label{fig:SF_SYNC_09} }
\end{centering}
\end{subfigure}
\begin{subfigure}{0.48\textwidth}
\begin{centering}
\includegraphics[width=0.8\columnwidth]{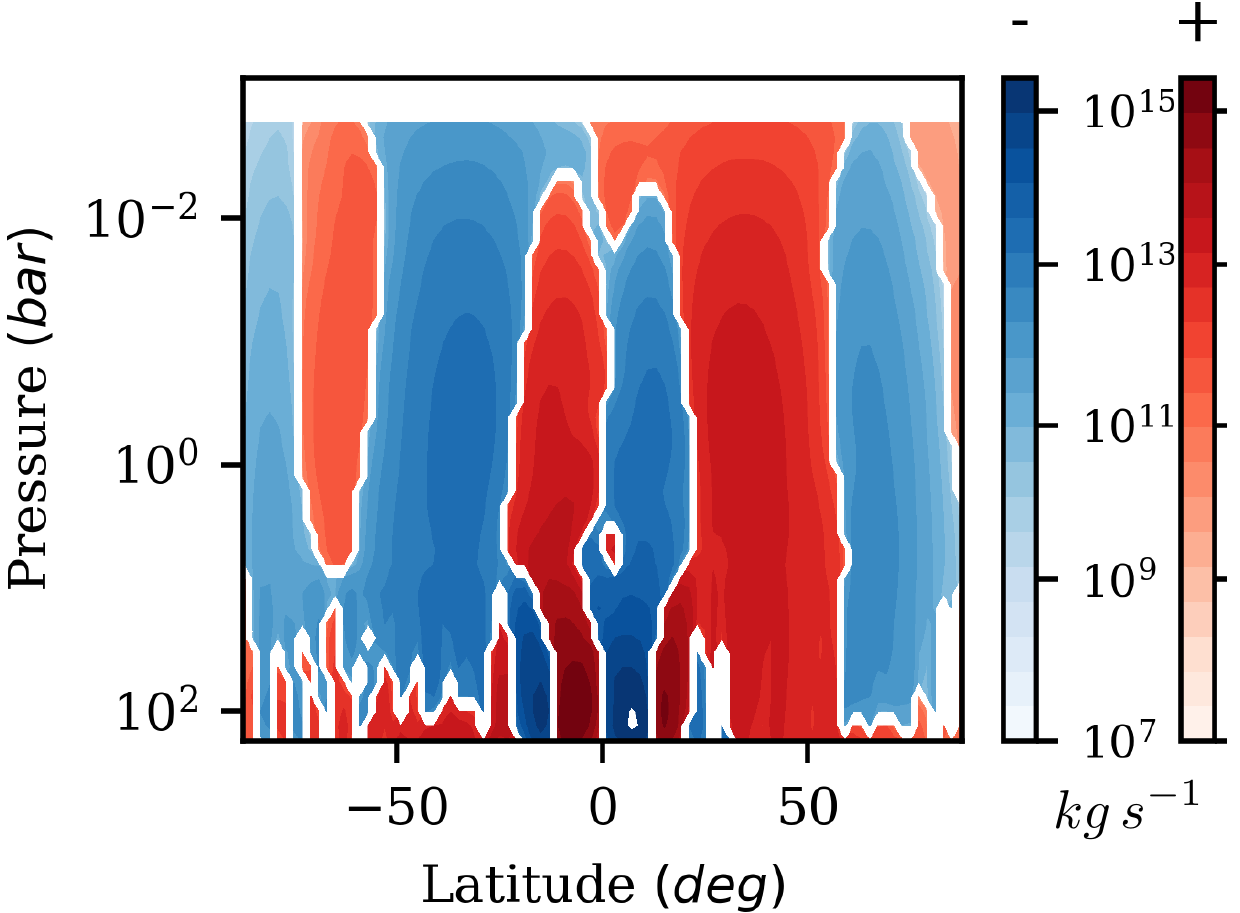}
\caption[]{$\Omega=4.0\Omega_{0}$ - Synchronous   \label{fig:SF_SYNC_4} }
\end{centering}
\end{subfigure}
\caption[Meridional mass streamfunctions for two synchronous and two non-synchronous HD209458b-like models ]{ Zonally and temporally, { over five Earth-years of simulation time,} averaged meridional circulation streamfunctions for four HD209458b-like models: two non-synchronous models (top), one in the weak-non-synchronicity regime (left) and one on the strong-non-synchronicity regime (right), paired with two synchronous models at the same rotation rates (bottom). \label{fig:SF_NS_Sync} }
\end{centering}
\end{figure*} 
\subsection{Meridional Circulation} \label{sec:NS_meridional}
We next explore how the changes induced by a difference between the rotational and orbital angular rotation rates impacts the meridional mass stream-function, as shown in \autoref{fig:SF_NS_Sync}. \\

As was the case for the zonal-mean zonal-wind, when the difference between the rotational and orbital periods is small (i.e. in the weakly non-synchronous regime), we find a circulation profile that is broadly equivalent to that found in for a synchronous model (\autoref{fig:Streamfunction}), just a little weaker. In turn, this suggests that vertical enthalpy advection, and hence deep heating, should remain significant. We explore if this is the case below (\autoref{sec:NS_enthalpy}). \\
However, as the difference between the rotational and orbital periods grows, the above no longer holds true: Whilst we do find general agreement, visually, with our equivalently rotating synchronous model, including the development of an additional circulation cells in each hemisphere as the influence of rotation increases, we also find that the strength of these circulations is significantly reduced, reaching a peak difference of almost 2 orders of magnitude. Furthermore, when we look at snapshots instead of the temporal average, we find that the circulation is highly time dependent, particularly in the deep atmosphere. Taken together, these two factors suggest that the vertical advection of enthalpy is likely to be significantly suppressed as the difference between the rotational and orbital periods grows. \\

\begin{figure*}[tbp] %
\begin{centering}
\begin{subfigure}{0.32\textwidth}
\begin{centering}
\includegraphics[width=0.95\columnwidth]{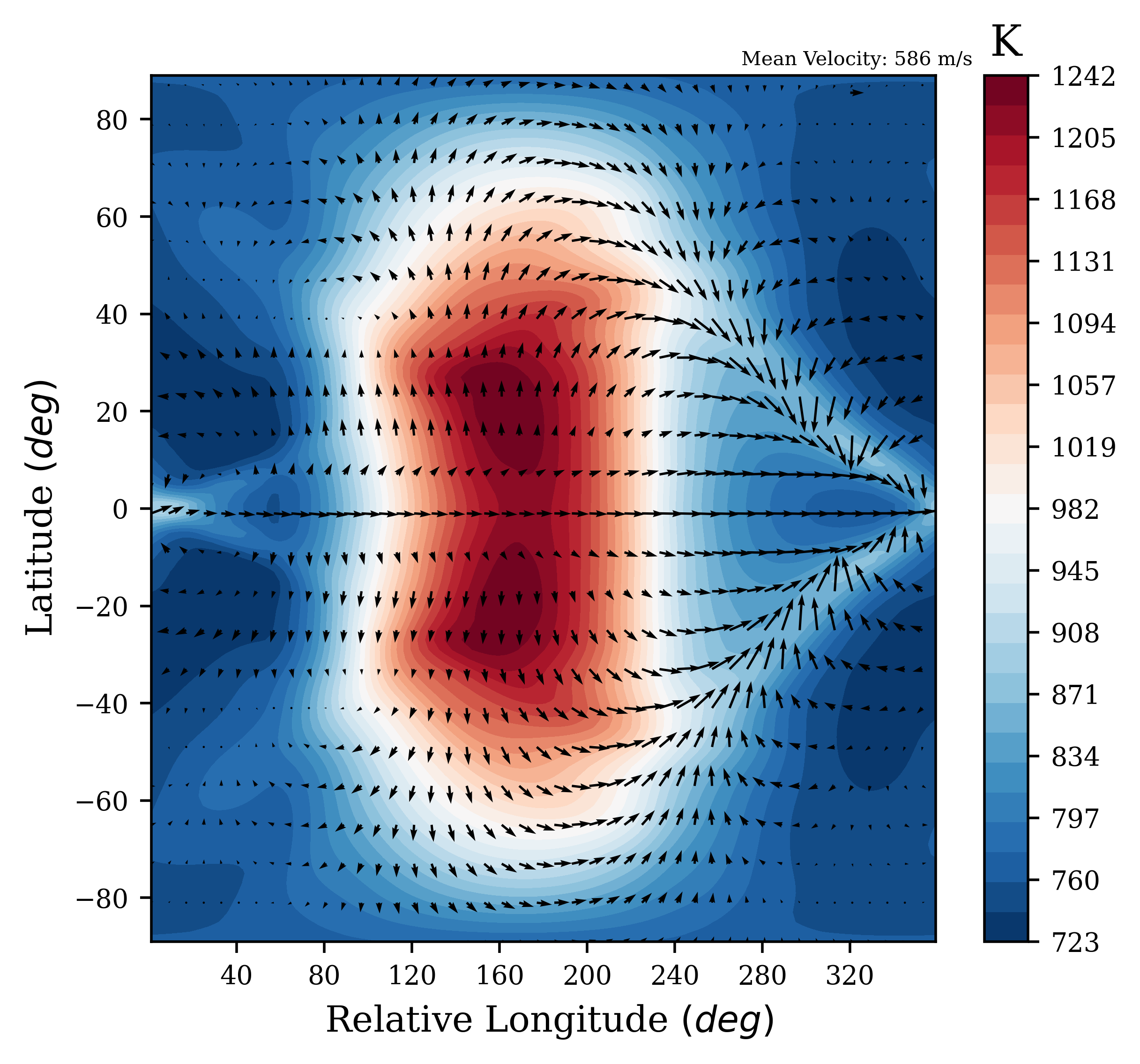}
\caption[]{0.0026 bar   \label{fig:WT_Maps_4_00026} }
\end{centering}
\end{subfigure}
\begin{subfigure}{0.32\textwidth}
\begin{centering}
\includegraphics[width=0.95\columnwidth]{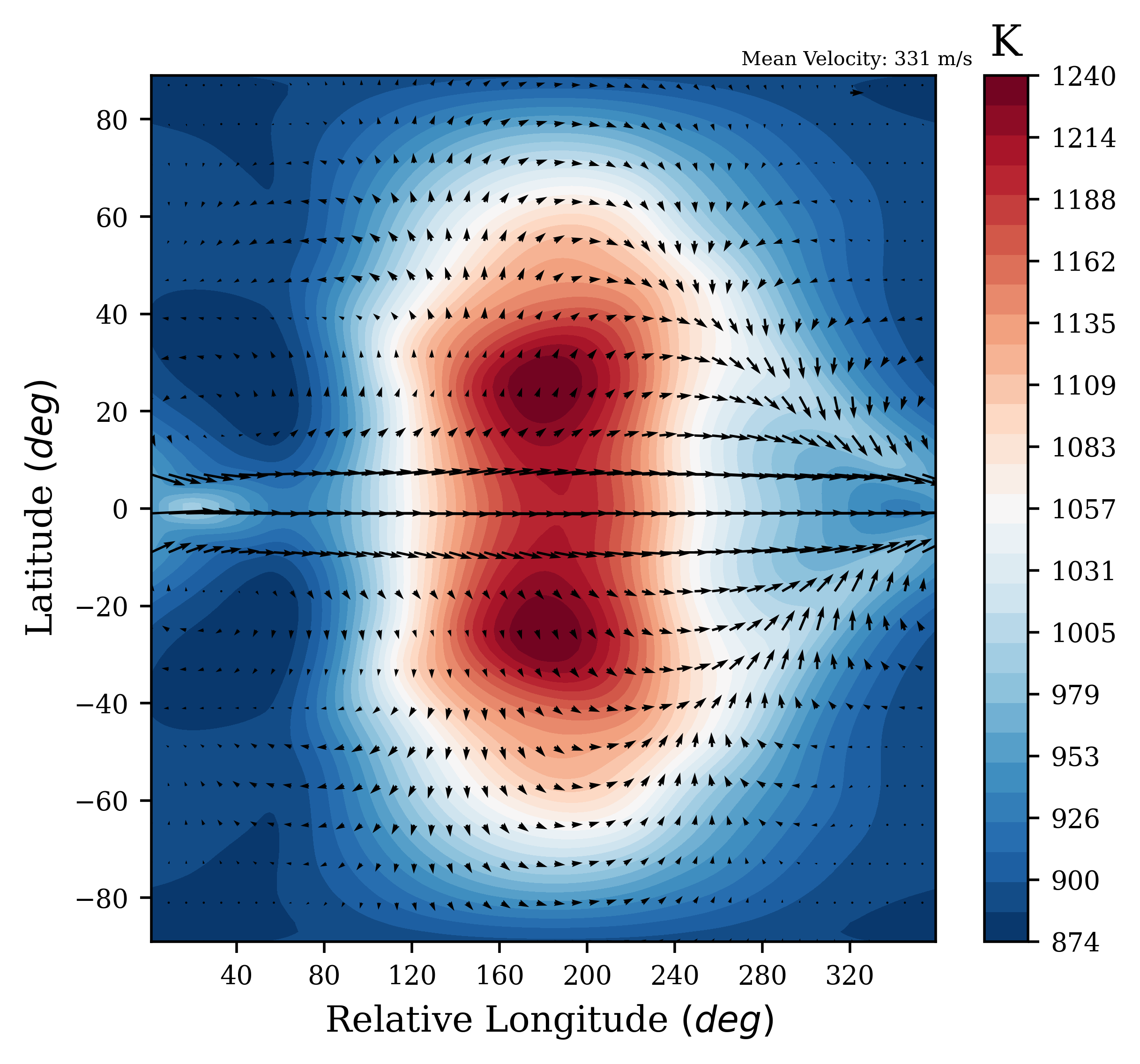}
\caption[]{0.016 bar   \label{fig:WT_Maps_4_0016} }
\end{centering}
\end{subfigure}
\begin{subfigure}{0.32\textwidth}
\begin{centering}
\includegraphics[width=0.95\columnwidth]{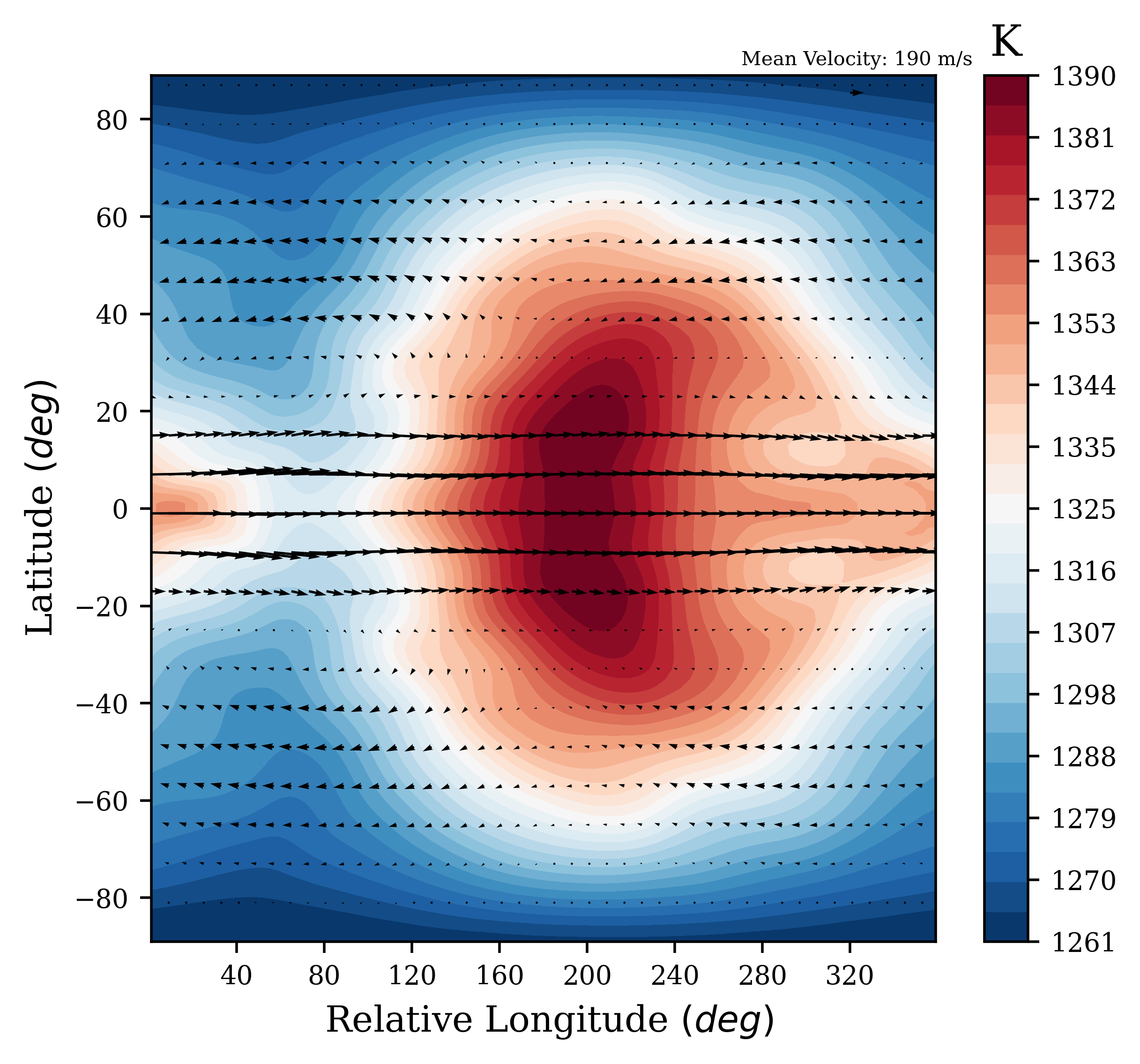}
\caption[]{0.2 bar   \label{fig:WT_Maps_4_02} }
\end{centering}
\end{subfigure}
\caption[Wind-Temp maps at three different pressures for a NS HD209458b-like model]{ Temporally averaged, { over five Earth-years of simulation time,} zonal wind (arrows) and temperature { maps} at { three} different pressures ({ 0.0026 bar - left, 0.016 bar - centre, and 0.2 bar - right})  for a non-synchronous HD209458b-like model in which the difference between the rotational and orbital periods is large ($\Omega_{\mathrm{rot}}=4\Omega_{\mathrm{orb}}=4\Omega_{0}$). Here we have forgone showing the weakly non-synchronous regime as we have already established its similarity to its synchronous equivalent.
Note that the x-axis of these plots is relative-longitude, indicating that we have taken the temporal mean of the wind/temperature in the frame of reference of the irradiation.  \label{fig:WT_Maps_NS} }
\end{centering}
\end{figure*}

\subsection{Horizontal-Wind and Temperature Maps} \label{sec:NS_horizontal_wind_temp}
In order to better understand how differences between the rotational and orbital periods impact the dynamics, we next explore how the horizontal wind/temperatures profiles in the outer atmosphere (where the influence of the time-dependent radiative forcing profile is likely to be largest) change with increasing non-synchronicity. \\

Briefly, when the difference between the orbital and rotational periods is small, we find a wind/temperature structure that is almost identical to that found in the classical hot Jupiter regime. This includes a strong equatorial jet which is driven by off-equator $m=1$ standing wave circulations, as well as the classical thermal butterfly. 
Note that we find slightly more asymmetry in the flows here than in the synchronous case, asymmetry which we attribute to the time-dependent radiative forcing profile at least slightly disrupting the $m=1$ standing wave structure. Other evidence for the time-dependent radiative forcing can be found in the temperature profile of the non-synchronous mean night-side, which { is} a hint hotter here than in an equivalent synchronous model, likely as a result of remnant thermal energy (that is still dissipating) from when that physical region of the atmosphere was irradiated. \\

On the other hand, when the difference between the rotational and orbital periods is large, other than in the fully radiatively controlled regions of the very outer atmosphere, significant changes to the wind and temperature structure become evident, changes which we can be seen in \autoref{fig:WT_Maps_NS}. \\
{ At low pressure, e.g. $0.0026\si{\bar}$, we find that the `mean' (where the mean is taken in the frame of reference of the irradiation) hot-spot lies slightly to the west, approximately $15^{\circ}$, of the sub-stellar point. Such a westwards shift, if paired with opacity sources such as clouds which lead to phase curves probing low-pressure regions of the atmospheres, may explain the significant westwards phase curves observed for CoRoT-2b \citep{2018NatAs...2..220D}. {Note that a westward shift in the planetary-scale wave pattern at similarly low pressures was proposed by \citet{2022ApJ...941..171L} as an explanation for westward hot-spot offsets.} } 
Moving deeper, at $0.016\si{\bar}$, we find strong thermal `wings' that extend from mid-latitudes on the `mean' day-side to near the equator on the `mean' night-side, `wings' which trace the relatively weak off-equator standing wave circulation the develops here. These `wings' only increase in strength as we move deeper into the atmosphere ($0.2\si{\bar}$ - \autoref{fig:WT_Maps_4_02}), where we find that the non-synchronous mean day-side temperature map has becoming increasingly smeared eastwards of the (moving) sub-stellar point. Whilst this description make the temperature map sound a little like the classical thermal butterfly, differences in advective transport mean that this is not the case. \\
The formation of the classical thermal butterfly is driven by the eastwards advection of heat at the equator and the westwards advection of heat at higher latitudes. However, as seen in \autoref{sec:NS_zonal} and \autoref{fig:WT_Maps_4_02}, the horizontal winds are much weaker in a highly non-synchronous atmosphere than its synchronous equivalent. For example, the eddy wind in our $\Omega_{\mathrm{rot}}=4\Omega_{\mathrm{orb}}=4\Omega_{0}$ model is a third of that found in the synchronous $\Omega_{\mathrm{rot}}=\Omega_{\mathrm{orb}}=4\Omega_{0}$ model ($\left|u_{e}\right|=63\si{\meter\per\second}$ vs $\left|u_{e}\right|=179\si{\meter\per\second}$), whilst the zonal component is less than a quarter of its synchronous counterpart ($\left|u_{z}\right|=146\si{\meter\per\second}$ vs $\left|u_{z}\right|=645\si{\meter\per\second}$).
Instead, analysis of snapshots of the temperature/wind profiles reveal that the `mean' dynamics found in \autoref{fig:WT_Maps_NS} are the result of the sub-stellar point sweeping across the planets `surface' at a rate that depends upon { the difference between the rotational and orbital angular rotation rates}. This leads to the smeared { and shifted} thermal structure seen in \autoref{fig:WT_Maps_4_02}, which occurs due to residual night-side temperature/heat from when that region of the atmosphere was irradiated (i.e. the same effect which lead to slight night-side heating in the weakly non-synchronous regime) { and the interaction between zonal winds and the shifting insolation}. Essentially, we find that increasing the difference between the orbital and rotational periods decreases the day-night temperature difference by acting as a form of `advection' from the day-side to the night-side. { This is similar to the decrease in phase curve amplitude found in the rapidly rotating non-synchronous models of \citet{Showman_2009}.}  \\
Note that the dynamics are slightly different when the rotational period is less than the orbital period. This occurs for two reasons, the first is that the direction of the hot-spot shift is reversed westwards rather than eastwards and the second in the decreasing influence of rotation at low $\Omega_{\mathrm{rot}}$ resulting in divergent rather than rotational dynamics. Note that the aforementioned smearing of the temperature map by the shifting sub-stellar point, as well as the above factors, means that we also no longer find any evidence for the formation of the night-side hot-spot that was a defining feature of our slow-rotation-regime models (\autoref{sec:horizontal_wind_temp}). \\
Overall we find that, regardless of its sign, when the difference between the rotational and orbital angular rotation rates is large the underlying dynamics which lead to the features found in the synchronous regime are unable to develop.

\begin{figure}[tbp] %
\begin{centering}
\begin{subfigure}{0.99\columnwidth}
\begin{centering}
\includegraphics[width=0.76\columnwidth]{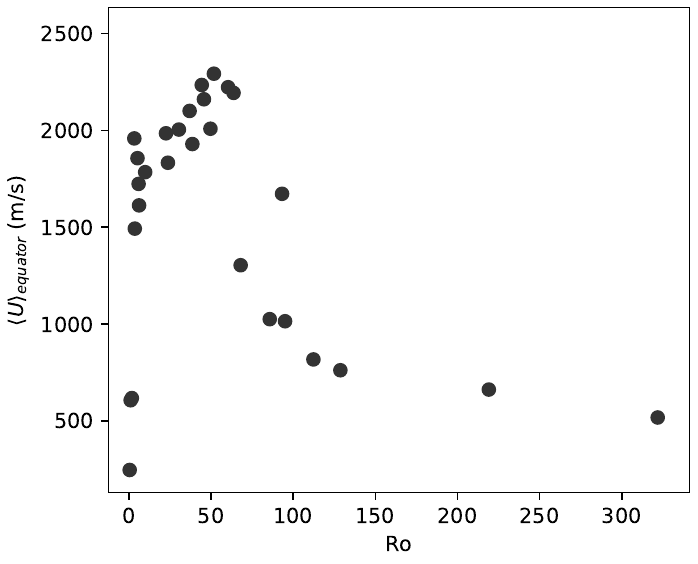}
\caption[]{$R_{0}$ vs. $\left<U\right>_{equator}$  \label{fig:Rossby_vs_U_mean} }
\end{centering}
\end{subfigure}
\begin{subfigure}{0.99\columnwidth}
\begin{centering}
\includegraphics[width=0.76\columnwidth]{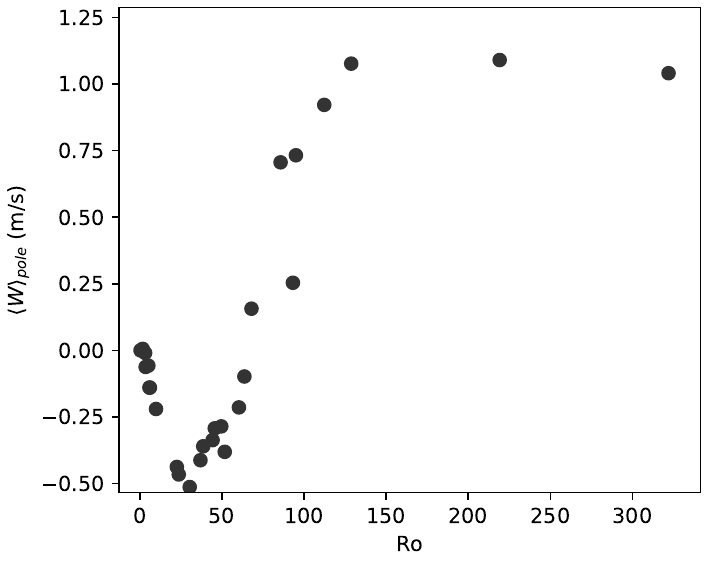}
\caption[]{$R_{0}$ vs. $\left<W\right>_{pole}$   \label{fig:Rossby_vs_W_mean} }
\end{centering}
\end{subfigure}
\caption[Zonal and vertical wind vs Rossby number]{A pair of plots showing how the mean zonal wind at the equator - top - and mean vertical wind near the northern pole - bottom - vary with global mean Rossby number for 29 HD209458b-like models spanning the rotation range $0.125\Omega_{0}\rightarrow15\Omega_{0}$. \label{fig:Rossby_vs} }
\end{centering}
\end{figure}
\begin{figure}[tbp] %
\begin{centering}
\includegraphics[width=0.75\columnwidth]{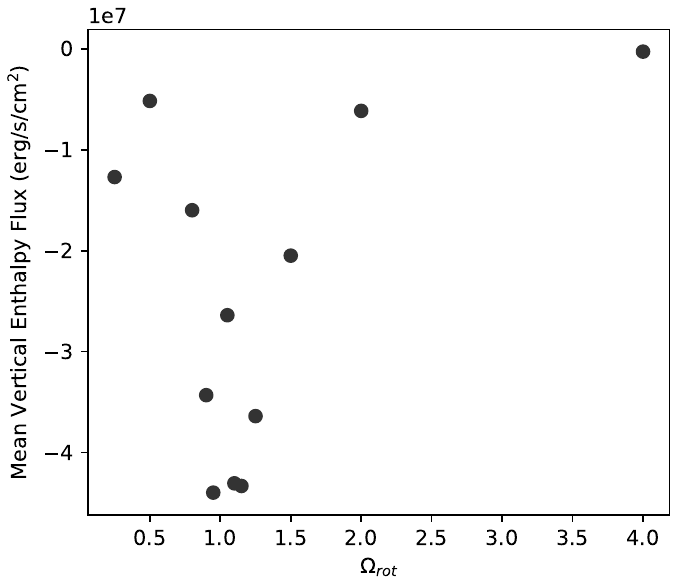}
\caption[Global mean vertical enthalpy flux]{Global mean vertical enthalpy flux for 13 non-synchronous HD209458b-like models spanning the rotation range $\Omega_{\mathrm{rot}}=0.5\Omega_{\mathrm{orb}}\rightarrow4\Omega_{\mathrm{orb}}$, where $\Omega_{\mathrm{orb}}=\Omega_{0}$ is held constant. Note that negative fluxes indicate net downward enthalpy transport. { Note that there is a slight scatter in the mean vertical enthalpy flux around $\Omega_{\mathrm{rot}}=\Omega_{\mathrm{orb}}=\Omega_{0}$ can be linked with the averaging window and the current state on long-timescale oscillations in the atmosphere. The main takeaway from this figure should be that the vertical enthalpy flux rapidly drops off when the different between the orbital and rotational periods increases. }   \label{fig:Global_Enthalpy_vs_Rot_NS} }
\end{centering}
\end{figure} 
\subsection{Vertical Enthalpy Transport} \label{sec:NS_enthalpy}
We finish our analysis with a brief investigation into how the above changes to the flows and circulations due to a difference between the rotational and orbital periods impacts the global mean vertical enthalpy flux, as shown in \autoref{fig:Global_Enthalpy_vs_Rot_NS}. \\
When the difference between the rotational and orbital periods is small, we find global mean vertical enthalpy fluxes that are at most 25$\%$ weaker than that found in an equivalent synchronous model. { However as the difference between the rotational and orbital periods grows, and we enter the highly non-synchronous regime in which the horizontal dynamics have significantly slowed relative to a synchronous equivalent, we find that the vertical enthalpy transport has become increasingly suppressed, eventually dropping by over two orders of magnitude when $\Omega_{\mathrm{rot}}=4\Omega_{\mathrm{orb}}=4\Omega_{0}$. \\}
{ This suggests that only weakly non-synchronous atmospheres should exhibit any significant deep heating/radius inflation. And indeed an analysis of the deep Temperature-Pressure profiles reveals this to be the case, with our weakly non-synchronous models showing signs of warming from their initial adiabat whilst the deep atmospheres of our highly non-synchronous models are highly sensitive to even the weakest of deep radiative forcing, suggesting that the link between the deep and outer atmosphere is nebulous at best in these models. Note that, as reflected in the weaker vertical enthalpy fluxes, we do find that the deep adiabats of our weakly non-synchronous atmospheres is slightly cooler than a synchronous equivalent. }

\section{Discussion and Conclusion} \label{sec:conclusion}
In this work, we have investigated how the planetary rotation rate affects the atmospheric dynamics, including the vertical energy transport, of an HD209458b-like hot Jupiter (with $\Omega_{0}=2.1\times10^{5}\,\mathrm{s^{-1}}$ based upon the model of \citealt{2019A&A...632A.114S}) for both synchronous ($\Omega_{\mathrm{rot}}=0.125\Omega_{0}\rightarrow40\Omega_{0}$ - 28 models) and non-synchronous ($\Omega_{\mathrm{rot}}=0.5\Omega_{0}\rightarrow4\Omega_{0}$ - 13 models) rotation. \\
{ This work builds upon previous studies of the influence of orbtial radii on atmospheric dynamics, such as \citet{Showman_2009,2014arXiv1411.4731S}, by isolating the effects of rotation from the change in insolation that occurs when modifying the orbital radii.}
Our analysis revealed that, for the rotation rate ranges considered, our models fall into one of a number of distinct regimes. \\

For synchronous rotation we find three distinct dynamical regimes: slow rotation when $\Omega_{\mathrm{rot}}<0.5\Omega_{0}$, the classical hot Jupiter regime when $\sim0.5\Omega_{0}<\Omega_{\mathrm{rot}}<\sim5\Omega_{0}$ (an analysis of which can be found in the main text), and the rapidly rotating regime when  $\Omega_{\mathrm{rot}}\gg5\Omega_{0}$. { Note that, as discussed in \autoref{sec:length_scales}, these regimes are broadly in line with the rotation regimes for terrestrial exoplanet atmosphere dynamics discussed by \citet{2018ApJ...852...67H}, albeit with differences in dynamics due to differences between the insolation of hot Jupiters and habitable zonal terrestrial exoplanets. They are also different from the rotation regimes discussed by \citet{2014arXiv1411.4731S} in that we do not consider the weakly irradiated regime that they find at larger orbital radii and our rapidly rotating regime maintains a strong zonal jet due to the combination of significant, synchronous, day-side irradiation (i.e. a high day-night temperature contrast) and the suppression of off-equator dynamics by the coriolis effect in our models (we consider much shorter orbital periods than \citealt{2014arXiv1411.4731S}).} \\

We find that the slowly rotating regime is characterised by a slower zonal-mean zonal-jet (\autoref{fig:Zonal_Wind_0125}) which does not extend deep into the atmosphere ($P>0.5\si{\bar}$ is quiescent). The underlying cause of this slow wind can be understood via the Helmholtz wind decomposition, which reveals that the strongest component of the horizontal wind, in the outer atmosphere, is the divergent component associated with a global overturning circulation as heat rises on the day-side, diverges from the sub-stellar point, before converging on the night-side, just to the west of the sub-stellar point. Note that the exact location at which the winds converge is dependent upon the rotation rate, with the convergence point shifting westwards towards the eastern terminator as the influence of rotation grows. This divergent wind, { and the associated global overturning circulation} acts as the main driver of { vertical} enthalpy transport { and deep adiabatic heating}, leading to the formation of a rather unexpected feature: a night-side hot-spot (\autoref{fig:Wind_Temp_0.125_02}, \autoref{fig:Longitudinal_T_0125x}, { and \autoref{fig:long_slices_T_0125}}). This night-side hot-spot appears to be a robust feature of the slowly rotating regime and plays a role in shaping the mid-atmosphere dynamics. For example, around $0.2\si{\bar}$, we find that the night-side hot spot is hotter than any point on the day-side, an effect which reshapes the winds, driving a strong night-day flow. \\
As a result we find that the { aforementioned} global overturning circulation, associated with the divergent day-side wind and a broad sub-stellar point upflow, is closed in the mid-atmosphere, helping to explain the quiescent deep atmosphere. This effect can be seen in the meridional circulation profile, where we find an equatorial downflow that is confined to low pressures, and horizontal mixing at the bottom of the jet, where the night-side hot-spot forms. Finally this lack of a strong equatorial downflow that connects the outer atmosphere with the deep, $P>10\,\si{\bar}$, atmosphere means that the net deep heating is also weak (\autoref{fig:Enthalpy_Profiles_0125_slices} and \autoref{fig:Enthalpy_Profiles_0125_mean}), leading to a deep atmosphere that is not significantly heated (\autoref{fig:Longitudinal_T_0125x}), thus suggesting that slowly rotating warm/hot Jupiters should exhibit little to no radius inflation. \\
It is important to note that this wind structure and the associated night-side hot spot may also have implications beyond suppressing radius inflation. For example, linked with the night-side hot-spot is a thermal inversion and cold-trap (\autoref{fig:Longitudinal_T_0125x}), an effect which may have important implications for observations of these planets and in particular their atmospheric chemistry: these observations are starting to become possible thanks to next-generation telescopes, such as JWST, that allow us to characterise hot/warm gaseous exoplanets on longer orbits around hotter stars. \\
{ As for where we might observe such a planet, including a the possible presence of a night-side hot-spot, the most likely scenario is in a longer orbit around a K/M dwarf. Here the synchronisation time-scale is no longer an issue due to the expected lifetimes of cool stars (greater than the current age of the universe), although the reduced insolation may effect the strength of the night-side heating. }

\begin{figure}[tbp] %
\begin{centering}
\includegraphics[width=0.99\columnwidth]{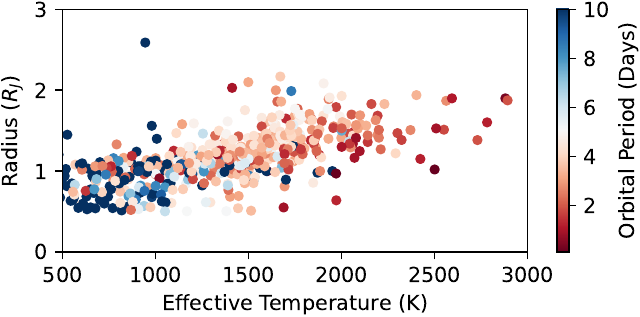}
\caption[Radius vs effective temperature vs orbital period]{ A plot showing the relationship between the radius and effective temperature (i.e. insolation) of Jupiter-mass planets, and how the rotation rate (represented by the color of each point) modulates inflation. This plot was made using data from \url{exoplanet.eu}.  \label{fig:rad_rot} }
\end{centering}
\end{figure}

On the other hand, we find that the rapidly rotating regime is characterised by a narrow, but deep, equatorial jet braced by easterly counterflows (\autoref{fig:Zonal_Wind_15}). This is a result of the increasing influence that the Coriolis effect plays on the atmospheric dynamics, suppressing high-latitude off-equator flows. Specifically we find that, as the rotation rate rises, the standing $m=1$ Rossby and Kelvin waves which drive the archetypal super-rotating jet become increasingly tilted and confined to low latitudes, leading not only to the narrower jet seen here as the circulation cells become latitudinally compressed, but also a weakening of the overall equatorial winds as the angular momentum pumping (from mid-latitudes to the equator) mechanism becomes less efficient. Note that this compression only grows as we increase the rotation rate, leading to noticeably slow jets at very high $\Omega_{\mathrm{rot}}$ (\autoref{fig:Rossby_vs_U_mean}). 
This latitudinal compression of the zonal-winds and off-equator flows also impacts the advection of enthalpy away from the sub-stellar point. For example, we find that the thermal butterfly (\autoref{fig:Wind_Temp_15_00026}/h/l) has become somewhat confined to equatorial regions, with eastwards advection at the equator, westwards/polewards advection at low latitudes and almost no advection at higher latitudes where the Coriolis forces are strongest (and winds are weakest). { Note that this slight westwards advection at low latitudes might explain the weak westwards phase-curve offsets found in observations by, for example, \citet{2022AJ....163..256M}. } 
\\ 
Vertically, the latitudinal compression and resulting banding of the zonal-mean zonal-wind leads to a meridional circulation profile (\autoref{fig:Streamfunction_15}) which contains multiple circulation cells in each hemisphere, although these circulations cells do still combine at the equator to drive a weakened downflow at all pressures, connecting the outer and deep atmospheres. Note that, much like the zonal wind, the meridional circulation is significantly weakened at high latitudes thanks to the Coriolis effect, hence most vertical mixing is also confined to low-latitudes (i.e. the equatorial region).  This has important implications for the vertical enthalpy transport and hence heating of the deep atmosphere, both of which we find to be globally weak (\autoref{fig:Enthalpy_Wind_vert_15},  \autoref{fig:Enthalpy_Profiles_15_mean}, and \autoref{fig:Longitudinal_T_15x}). Note however that this is just the global value. If we instead consider the equatorial mean (where vertical mixing is strongest), we find a net downward enthalpy flux comparable to that found in the classical hot Jupiter regime, and which leads to slight warming of the equatorial regions deep T-P profile (and only the equatorial regions due to the quiescence of the off-equator deep atmospheric winds).
As such, we might expect such objects to be slightly puffed up near the equator, although such an effect is likely to remain minor and may be entangled with other effects that rotation and irradiation has on the atmospheres shape. \\
This may help to explain the radius inflation, or lack-thereof, observed for ultra-hot Jupiters: planets with an orbital period of less than around 20 hours tend to exhibit little to no radius inflation e.g. Wasp-18b - \citealt{2019A&A...625A.136A} - or Wasp-43b - \citealt{2011A&A...535L...7H}), whereas planets with orbital periods of $>22$ hours start to show signs of inflation (e.g. Wasp-103b - \citealt{2014A&A...562L...3G} - or KELT-16b - \citealt{Oberst_2017}), similar to the transition seen with rotation rate here. This is also in agreement with the results of \citet{2021A&A...656A.128S}, who found similar dynamics and a similar lack of radius inflation for SDSS1411b. Note that these examples suggest a rather sharp transition, however when we consider the larger sample of known Jupiter-sized exoplanets, the transition is less distinct. { As shown in \autoref{fig:rad_rot}, which plots the relationship between planetary Radius, Effective Temperature, and Orbital Periods, {as expected} there appears to be a general relationship between increased effective temperature (i.e. insolation) and planetary radius \citep{2011ApJ...729L...7L,Demory_2011}. However, there is also a trend of decreasing radius with decreased orbital periods, albeit with some exceptions/outliers, suggesting that rotation is {may play a role in} modulating radius inflation. }  \\

{ The confinement of dynamics to low latitudes also comes with a corresponding decrease in typical dynamical length-scales, such as the Rossby deformation radius or the Rhine length. In our most rapidly rotating models, these length-scales can approach the resolution of our grid, suggesting that our models might be struggling to fully capture some of the dynamics. As such, we suggest that a future study investigate the most rapidly rotating hot Jupiter regime with a significantly increased horizontal resolution in order to fully resolve the dynamics at play.}

Thanks to JWST, as well as other next-generation ground and space based telescopes, there is likely to be in influx of long orbital period hot/warm Jupiters in the near future. Not only would these objects likely fall into the slowly rotating regime, it is also increasingly likely that at least some will not be perfectly synchronously rotating. This can occur because the synchronisation timescale increases with orbital radii, approaching the system age for a hot/warm Jupiter orbiting a sun-like star at $0.2\si{\astronomicalunit}$. There are also a number of scenarios which might lead to a non-synchronous but highly irradiated hot Jupiter being observed. For example, we know that hot Jupiters cannot have formed so close to their host star and hence must have undergone planetary migration \citep{1996Natur.380..606L,2019A&A...628A..42H}. If we where to observe a hot Jupiter that had recently, or is still undergoing, migration, then it might not yet have had time to synchronise. { Another possibility is that we may find a hot/warm Jupiter that in stable 3:2 orbital resonance instead of full 1:1 spin synchronisation. This would lead to dynamics that are similar to our slightly non-synchronous models, with the solar insolation `slowly' sweeping across the planetary atmosphere.} \\
As a result, and since many of the dynamics observed in hot Jupiter atmospheres are directly or indirectly linked to the strong day/night thermal forcing (from the day/night divergent flows in the slowly rotating regime to the thermally driven, $m=1$ Kelvin and Rossby standing waves which are responsible for the equatorial jet when rotational effects dominate), we next investigated how varying levels of non-synchronicity, as characterised by a difference between the varying planetary rotation rate and a fixed orbital period affects atmospheric dynamics. Here we found two additional regimes, depending upon the difference between the planetary rotational and orbital periods: when the difference is small, on the order of a few percent ($<15\%$) of the smaller period, we find dynamics that are qualitatively similar to the synchronous regime, just a little weaker. On the other hand when the difference is large, we find significant differences in the dynamics, especially in the driving of the super-rotating jet. { This is in agreement with the results of \citet{Showman_2009}.}\\

Briefly, when the difference between the rotational and orbital periods is small, we find dynamics, including vertical enthalpy transport, that are structurally similar to those found in the synchronous regime, but just a little bit weaker. The biggest difference comes about when we look at the horizontal temperature map in the outer atmosphere, where we find that the `sweeping' of the stellar irradiation essentially acts as a form of day-night transport, with residual heat leading to a warmer night-side and hence a reduced day-night temperature difference. \\
The above holds true when we also vary $\Omega_{\mathrm{orb}}$. As such, we expect { a} small difference between the rotational and orbital periods to have little effect on observable features, such as the phase curve (hot-spot location) or radius. Furthermore, since the synchronisation timescale rapidly decreases with decreasing orbital radii it is highly likely that the majority of warm/hot Jupiters detected in the near future will be either synchronously rotating or only weakly non-synchronous.  Whilst this is good in one sense, i.e. we can understand the dynamics of such objects quite easily, it also means that, with current tools, there is a distinct possibility that some weakly non-synchronous objects (which are potentially young, depending upon the local synchronisation time-scale), might be erroneously labelled as tidally-locked. \\

But what about a scenario in which a planet has recently migrated onto a shorter orbit and not yet synchronised its rotation with its host star, leading to a large difference between the rotational and orbital angular rotation rates? Here our models suggest that the dynamics should be distinct from those found in the synchronous regime. For example, we find a zonal-mean zonal-jet (\autoref{fig:Zonal_wind_NS_4}) which is both slower and shallower than its synchronous counterpart. This has a knock-on effect on the vertical dynamics, weakening both the meridional circulation and vertical enthalpy flux, the latter of which rapidly decreases as the difference between the rotational and orbital periods grows (\autoref{fig:Global_Enthalpy_vs_Rot_NS}). The underlying cause of these changes can be best understood by exploring the horizontal wind/temperature maps in the outer atmosphere (\autoref{fig:WT_Maps_NS}). 
One immediate observation is that the thermal butterfly that typically forms in the mid-atmosphere of non-slowly-rotating hot Jupiters is not present. Instead, we find that the temperature map has become longitudinally smeared, with the strength of the smearing depending upon $\Delta\Omega$: i.e. as the difference between the rotational and orbital periods increases, the time since the night-side was last irradiated gets smaller, leading to increased residual heating and a hot-spot that appears to trail the irradiation. Not only does this have the effect of reducing the strength of the day/night forcing, it also prevents the formation of the standing-wave structures which drive much of the dynamics of hot Jupiters, including the super-rotating jet and hence meridional circulation and vertical enthalpy flux. This can also shift the transition between the rotational-dominated and divergence-dominated dynamical regimes to faster planetary rotation rates, which will lead to additional changes in the strength of the vertical advection, since divergence-dominated winds are less efficient at vertical transport. { Note that the `smearing' of the temperature profile found in the highly non-synchronous regime is similar to non-synchronous models of \citet{2014arXiv1411.4731S} - see their Figure 4.} \\
Overall, this suggests that hot/warm Jupiters with large differences between their rotational and orbital periods should be observationally distinct from a synchronously rotating planet at the same orbital radius. This includes both a reduced phase curve amplitude as well as a reduced hot-spot offset, the latter of which could be either to the east or the west of the sub-stellar point depending upon the sign of the difference between the rotational and orbital periods { as well as the pressure-level probed. Evidence for a significant ($15^{\circ}$) west-ward hot-spot offset was found at low-pressures (e.g. $P=0.0026$ bar)
in our highly non-synchronous model (with $\Omega_\mathrm{rot}=4\Omega_\mathrm{orb}=4\Omega_0$). Such a shift, if paired with atmospheric opacity features, for example clouds, that lead to phase curves observations probing lower pressure regions of the atmosphere, may explain the westwards phase curve offset and broad phase-curve minimum of CoRoT-2b \citep{2018NatAs...2..220D}.}\\
We might also expect to see a reduced level of radius inflation for such objects thanks to the reduced vertical heat flux, although this conclusion is more tenuous since newly formed Jupiter-like objects which have recently undergone planetary migration are likely still dissipating excess energy, associated with for example gravitational collapse, from the formation process, meaning that their envelopes are likely puffed up \citep{10.1093/mnras/stw1160}. Such a scenario could be confirmed by comparing observed radii with planetary formation models, such as those of \citet{2007ApJ...659.1661F}. { Note however that migration history can play a role in the observed radius of hot Jupiters when energy is deposited deep in the interior \citep{10.1093/mnras/staa1405}.}\\

In this work we have discussed how rotation alone can significantly impact and alter the dynamics of Hot Jupiters, ranging from the outer atmosphere wind and temperature structure, to the vertical transport of enthalpy. The latter effect may be of particular importance when trying to explain variations in hot Jupiter radii found at fixed irradiation, in particular the lack of inflation observed for some very-short-orbit hot Jupiters and hot Brown Dwarfs.
\\
One interesting discovery which might have significant implications for atmospheric dynamics and chemistry is the formation of a night-side hot-spot (via divergent day-night winds) in the mid-atmosphere of our slowly and synchronously rotating regime models. If such a night-side hot-spot is a robust feature of more slowly rotating hot/warm Jupiters, something that we wish to investigate using a next-generation GCM (such as expeRT/MITgcm) that robustly models radiative heating and cooling at all pressure levels, the associated temperature inversion may significantly alter atmospheric chemistry by acting, for example, as a cold trap and hence reshaping the vertical distribution of key observable models. Understanding such an effect would likely require modelling the atmospheres of slowly rotating hot/warm Jupiters with a GCM that includes both robust radiative transport and dis-equilibrium chemistry, something that is currently far too computationally expensive to robustly investigate. {Although work is ongoing to solve this issue; for example, \citet{2023A&A...672A.110L}, have coupled a reduced chemical network (`mini-chem') to the GCM Exo-FMS, in order to study the effects of non-equilibrium chemistry on the atmospheres of WASP-39b and HD189733b in a computationally efficient manner. As might be expected, this model showed that transport can have a significant affect on the chemistry and local atmospheric composition. However this comes at the cost of calculating a significantly reduced chemical network (12 species without photochemistry) and the need for $\sim73$ weeks of wall-clock time using 200 processors to model the 200 Earth-years of simulation time considered here.}  \\

\begin{acknowledgements}
F. Sainsbury-Martinez and P. Tremblin would like to acknowledge and thank the ERC for funding this work under the Horizon 2020 program project ATMO (ID: 757858). F. Sainsbury-Martinez would also like to thank UK Research and Innovation for additional support under grant number MR/T040726/1.\\
The authors also wish to thank Idris, CNRS, University Paris-Salcay, and Mdls for access to the supercomputers Poincare, and Ruche, without which the long time-scale calculations featured in this work would not have been possible. Additionally this work was granted access to the HPC resources of IDRIS (Jean-Zay) and CEA-TGCC (Irene/Joliot-Curie) under the 2021/2022 allocation - A0100410870 made as part of the GENCI Dari A10 call. \\
\end{acknowledgements}

\bibliographystyle{aa} 
\interlinepenalty=10000
\bibliography{papers}

\end{document}